\documentclass[journal abbreviation, manuscript]{copernicus}

\usepackage{rotating}
\usepackage{amsmath}

\usepackage[printonlyused]{acronym}

\newcommand{\sst}{\ac{SST}}
\newcommand{\ssta}{\ac{SSTa}}
\newcommand{\ogcm}{\ac{OGCM}}
\newcommand{\llc}{{LLC}}
\newcommand{\viirs}{\ac{VIIRS}}
\newcommand{\pae}{\ac{PAE}}
\newcommand{\healpix}{\ac{HEALPix}}
\newcommand{\acc}{\ac{ACC}}

\newcommand{\ulmo}{{\sc Ulmo}}
\newcommand{\LL}{\ac{LL}}
\newcommand{\dT}{\ensuremath{dT}}
\newcommand{\tildell}{$\widetilde{\rm LL}$}
\newcommand{\tildellviirs}{$\widetilde{\rm LL}_{\rm VIIRS}$}
\newcommand{\tildellllc}{$\widetilde{\rm LL}_{\rm LLC}$}
\newcommand{\tildellllcprime}{$\widetilde{\rm LL}'_{\rm LLC}$}
\newcommand{\tildellhead}{$\widetilde{\rm LL}_{\rm 2012-2015}$}
\newcommand{\tildelltail}{$\widetilde{\rm LL}_{\rm 2017-2020}$}
\newcommand{\viirsllcdifference}{{\tildellviirs}$-${\tildellllc}}

\newcommand{\headtaildifference}{{\tildellhead}$-${\tildelltail}}
\newcommand{\sigmathreshold}{197}

\newcommand{\nscans}{$\sim$500,000} % Number of unique VIIRS files
\newcommand{\pixsq}{pixel$^2$} % pixles squared
\newcommand{\nviirs}{2,851,702} % Number of VIIRS cutouts
\newcommand{\DT}{\ensuremath{\Delta T}}

\begin{document}

\nolinenumbers

\title{An evaluation of the LLC4320 global ocean simulation based on the submesoscale structure of modeled sea surface temperature fields}

% \Author[affil]{given_name}{surname}

\Author[1, $\star$]{Katharina}{Gallmeier}
\Author[2,3,4,5, $\star$]{J. Xavier}{Prochaska}
\Author[6, $\star$]{Peter}{Cornillon}
\Author[7]{Dimitris}{Menemenlis}
\Author[8]{Madolyn}{Kelm}

\affil[1]{Institute for Defense Analyses, Alexandria, VA, 22305, USA}
\affil[2]{Affiliate of the Department of Ocean Sciences, University of California, Santa Cruz, CA, 95064, USA}
\affil[3]{Department of Astronomy and Astrophysics, University of California, Santa Cruz, CA, 95064, USA}
\affil[4]{Kavli Institute for the Physics and Mathematics of the Universe (Kavli IPMU), 5-1-5 Kashiwanoha, Kashiwa, 277-8583, Japan}
\affil[5]{Simons Pivot Fellow}
\affil[6]{Professor Emeritus, Graduate School of Oceanography,University of Rhode
Island, Narragansett, RI, 02882, USA}
\affil[7]{Jet Propulsion Laboratory, California Institute of Technology, Pasadena, CA, 91109, USA}
\affil[8]{Earth System Science Department, University of California, Irvine, Irvine, CA, 92697, USA}
\affil[$\star$]{These authors contributed equally to this work.}

%% The [] brackets identify the author with the corresponding affiliation. 1, 2, 3, etc. should be inserted.

%% If authors contributed equally, please mark the respective author names with an asterisk, e.g., "\Author[2,*]{Anton}{Smith}" and "\Author[3,*]{Bradley}{Miller}" and add a further affiliation: "\affil[*]{These authors contributed equally to this work.}".

\correspondence{J. Xavier Prochaska (jxp@ucsc.edu)}

\runningtitle{TEXT}

\runningauthor{TEXT}

\received{}
\pubdiscuss{} %% only important for two-stage journals
\revised{}
\accepted{}
\published{}

%% These dates will be inserted by Copernicus Publications during the typesetting process.

\firstpage{1}

\maketitle

\begin{abstract}
We have assembled \nviirs\ nearly cloud-free
cutout images (sized 144$\times$144\,km$^2$)
of \ac{SST} data from the entire 2012--2020 Level-2
{\viirs} dataset to perform a quantitative comparison to the
ocean model output from the \ac{MITgcm}.
Specifically, we evaluate outputs from the LCC4320 $\frac{1}{48}^\circ$ global-ocean simulation for a one-year period starting on November 17, 2011 but
 otherwise matched in geography and day-of-year
to the {\viirs} observations.
In lieu of simple (e.g., mean, standard deviation) or
complex (e.g., power spectrum) statistics, we analyze 
the cutouts of \ac{SST} anomalies with an unsupervised \ac{PAE}
trained to learn the distribution of structures in {\sst} anomaly (\acs{SSTa})\acused{SSTa} on 
$\sim$10-to-$80$-km scales (i.e., submesoscale-to-mesoscale).
A principal finding is that the LLC4320 simulation
 reproduces well, over a large fraction of the ocean, the observed distribution of \ac{SST}
patterns, both globally and regionally. 
%{\color{red}I'd get rid of the next sentence, which I've colored} {\color{blue}blue. After accounting for a large-scale offset between the 
%data and model,  we find that 
Globally, the medians of the structure
distributions match to within $2\sigma$
for $65\%$ of the ocean, despite 
a modest, latitude-dependent offset.
Regionally, the model outputs reproduce mesoscale
variations in \ssta\ patterns revealed by the \ac{PAE}
in the \viirs\ data, including
subtle features imprinted by variations in bathymetry.
We also identify significant differences in the 
distribution of \ssta\ patterns in several regions: (1)
in the vicinity of the point at which western boundary currents separate from the continental margin, (2) in the \ac{ACC}, especially in the eastern half of the Indian Ocean, and (3) in an equatorial band equatorward of $15^\circ$.
It is clear that (1) is a result of premature separation in the simulated western boundary currents. The model output in (2), the Southern Indian Ocean, tends to predict more structure than observed, perhaps arising from a misrepresentation of the mixed layer or of energy dissipation and stirring in the simulation.
%, however the reasons for this are not clear. 
The differences in (3), the equatorial band, are also likely due to model errors, perhaps
arising from the shortness of the simulation or from the lack of high-frequency/wavenumber atmospheric forcing.
%also not clear and may be partly attributable to unresolved clouds and errors in the retrievals of {\sst} in the satellite-data.
Although we do not yet know the exact causes for these model-data \ssta\ differences, we expect that this type of comparison will help guide future developments of high-resolution global-ocean simulations.

\end{abstract}

\copyrightstatement{TEXT} %% This section is optional and can be used for copyright transfers.

\acresetall

\introduction  %% \introduction[modified heading if necessary]

\acp{OGCM} are an attempt to reproduce the physics and thermodynamics associated with large-scale oceanic processes. The first global implementation of an OGCM with ‘realistic’ coastlines and bathymetry was undertaken in the early 1970s on a 2$^\circ\times$2$^\circ$ grid with 12 vertical levels \citep{Cox1975}. A subjective evaluation of this model's performance compared the dynamic topography `patterns' of large scale (basin-wide) gyres determined from ship surveys with those obtained from the model. A more quantitative evaluation was also performed by comparing the transport through the Drake Passage determined from hydrographic sections with those obtained from the model---an excellent overview of the early work on \acsp{OGCM} is provided in K.\ Bryan's tribute to M.\ Cox's work \citep{bryan1991}.
These spatially coarse comparisons made clear that the model reproduced some of the general features of the large-scale circulation but missed others; often those it had missed were off by significant fractions when quantitative comparisons were made. Given the coarse resolution---grid spacing often twice that of the width of major ocean currents such as the Gulf Stream---and the representation for subgrid-scale processes, which attempt to incorporate the physical contribution of processes on scales smaller than the grid spacing, it is not surprising that this model missed some features of large-scale circulation. 

In the fifty years since Cox's work, the processing capacity of computers has increased dramatically from $\sim$1 megaflop, for the Univac 1108 used by Cox, to $\mathcal{O}(10^9)$ megaflops. Likewise, storage capacities have seen similar increases, more efficient codes have been introduced, and the observational data needed to constrain and force the models have seen staggering increases in volume as well as accuracy. Today, the highest-resolution global \acp{OGCM} are run on grids ranging from $\frac{1}{12}^\circ$ to $\frac{1}{48}^\circ$ with 100 or more vertical levels \citep[see, e.g.,][]{Arbic2018,Uchida2022}. As a result, these models resolve many mesoscale processes that earlier models had missed. 
They reproduce quite well most of the large-scale patterns in the global ocean as well as the currents associated with these patterns, 
offering confidence in studies that use them to
predict the evolution of the ocean and atmosphere on a warming planet. 

The evaluation methodology described in this manuscript is meant to be applied to unconstrained \acp{OGCM} of sufficient resolution to develop vigorous mesoscale and, to some extent, submesoscale variability. At the moment, we lack the observations and estimation tools that are needed to constrain the amplitude and phase of individual mesoscale (and submesoscale) eddies globally and in a dynamically-consistent manner. Therefore, the simulated mesoscale and submesoscale features of free-running models are not expected to match, one-to-one, the observations. While comparing the predicted and modeled fields at an instant in time works for constrained models, the evaluation of free-running models must be performed statistically, since the mesoscale and submesoscale details inevitably differ within the observed field and the one modeled for the same timestamp. %, assuming, of course, that there are observed fields with which to evaluate the model. In this study we 
Furthermore, to the best of our knowledge, evaluations of the highest-resolution global, free-running \acp{OGCM} have, to date, focused on scales substantially larger (one and a half to two orders of magnitude larger) than the horizontal grid spacing of the model \citep[e.g.,][]{Fox-Kemper2019}. 
As such, these evaluations do not assess
the capability of such models to reproduce statistically valid measures of the submesoscale structure of their output. The objective of the work presented herein is to address this deficiency for one of the highest-resolution, global, free-running \acp{OGCM} available, specifically, the $\frac{1}{48}^\circ$, 90-level simulation known as LLC4320. This simulation was developed as part of the \ac{ECCO} project in a collaborative effort between the \ac{MIT}, the \ac{JPL}, and the NASA Ames Research Center (ARC).

To perform the desired evaluation requires a dataset with global coverage
spanning at least the LLC4320 time period
with a daily cadence (or higher).
In addition, it requires observations with spatial
sampling comparable to LLC4320 horizontal grid spacing, which ranges
from $\sim$2\,km at the equator to $\sim$1\,km 
at $70^\circ$ latitude. \ac{SST} fields obtained from several different satellite-borne sensors meet these requirements. The selected \sst\ dataset and the {\llc4320} simulation are described in more detail in the next section. As discussed in \S\ref{sec:methods}, we use an unsupervised machine learning algorithm applied to approximately 150$\times$150\,km$^2$ regions, which we refer to as {\bf cutouts}, to capture a measure of the structure of the \sst\ fields on 
such scales. 
By adopting this algorithm, we are intentionally agnostic to specific 
structures or patterns.  The algorithm ``learns'' the structures that
are dominant in the data and, equally as important, their distribution.
Furthermore, it can be applied in the same fashion to observational
data and model output.
This measure of field structure for the satellite-derived \sst\ fields is then compared statistically with that obtained from similar-sized squares of the model output. The results of these comparisons are discussed in \S\ref{results_discussion}.

\section{Data}

\subsection{Satellite-Derived \sst\ Data}

The {\viirs} instrument carried on the \ac{NPP} satellite provided the highest spatial-resolution global \sst\ products, 750\,m at nadir (degrading to 1700\,m at the swath edge, %Fig.\,\ref{SepFig} {\color{red}Not sure that we need this figure; can pull if too much detail.}), 
with at least daily coverage for the period covered by the {\llc4320} simulation. {\viirs} is a multi-detector instrument for which variations in the gain from detector-to-detector introduces striping in the resulting fields. In addition, geometric distortions in pixel location arise as the distance from nadir increases, referred to as the bow-tie effect, and render regions more than approximately 500\,km from nadir useless for our analysis unless they are corrected. These two issues, striping and inconstant pixel size, significantly impact the structure of the retrieved \sst\ fields. For this reason, we elected to use the \acl{NOAA}'s \ac{L2P}, \acl{RAN2} (\acsu{RAN2}) of the {\viirs} data \citep{10.1117/12.2518908}, the only product we are aware of that addresses both of these issues \footnote{The \ac{AVHRR} is not a multi-detector instrument so it does not suffer from the striping and geometric distortion issues associated with data from {\viirs} but the coarser spatial resolution, 1.1\,km at nadir, introduces greater degradation in resolution with distance from nadir and the noisier instrument results in a product with at least twice the noise than that obtained from {\viirs} \citep{rs9090877}. The \ac{SLSTR} instrument carried on the European Sentinel satellites provides an interesting alternative dataset but, given our lack of detailed familiarity with these data, we elected to continue using {\viirs}.}.

We downloaded all of the \ac{RAN2} \ac{L2P} files for the years 2012--2020, inclusive, from the \ac{JPL} \acl{PO.DAAC} (\acs{PO.DAAC},  https://podaac.jpl.nasa.gov). Each file contains the retrievals from 10 minutes of satellite data, approximately 5400 scans with 3200 pixels per scan. There are approximately 500,000 files for the period studied.  These \nscans~files total $\sim$90\,Tb and form the basis of our observational analysis.

% \begin{figure}[ht]
% \includegraphics[width=\linewidth]{Figures/pcc_pixel_size}
% \caption{Associated with the lower x-axis: along-scan and along-track dimensions of {\viirs} (black solid and dashed) and \acs{MODIS} (red solid and dashed) pixels as a function of distance from nadir. In \cite{prochaska2021deep} \acs{MODIS} only data between the cyan lines were used. For this study {\viirs} data are used across the entire swath. Associated with the upper x-axis: the along track separation of grid points for the {\llc4320} model output as a function of latitude. For this study only cutouts south of $57^\circ$ N, the vertical green line, were used.}
% \label{SepFig}
% \end{figure}

\subsection{\sst\ Output from the LLC4320 Simulation}

The LLC4320 simulation was completed in 2015 by coauthor Menemenlis with help from collaborators at \ac{MIT} and NASA ARC \citep[see, e.g.,][]{Rocha2016a,Rocha2016b,Arbic2018}. The LLC4320 simulation is a global-ocean and sea-ice simulation that represents full-depth ocean processes. The simulation is based on a \ac{LLC} configuration of the \acl{MITgcm} \citep[\acs{MITgcm};][]{https://doi.org/10.1029/96JC02775,hill2007investigating}. The LLC4320 grid has 13 square tiles with 4320 grid points on each side and 90 vertical levels for a total grid-cell count of 2.2$\times$10$^{10}$. Nominal horizontal grid spacing is $\frac{1}{48}^\circ$, ranging from 0.75\,km near Antarctica to 2.2\,km at the Equator, and vertical levels have $\sim$1-m thickness near surface to better resolve the diurnal cycle. The simulation is initialized from an \ac{ECCO}, data-constrained, global-ocean and sea-ice solution with nominal $\frac{1}{6}^\circ$ horizontal grid spacing \citep{Menemenlis2008}. From there, model resolution is gradually increased to \acs{LLC1080} ($\frac{1}{12}^\circ$ grid), \acs{LLC2160} ($\frac{1}{24}^\circ$ grid), and finally \acs{LLC4320} ($\frac{1}{48}^\circ$ grid). Configuration details are similar to those of the $\frac{1}{6}^\circ$ \ac{ECCO} solution except that the {\llc4320} simulation includes atmospheric pressure and tidal forcing. The inclusion of tides allows successful shelf-slope dynamics, water mass modification, and their contribution to the global-ocean circulation \citep{https://doi.org/10.1002/grl.50825}. Surface boundary conditions are from the $0.14^\circ$ \ac{ECMWF} atmospheric operational model analysis, starting in 2011. Another unique feature of this simulation is that hourly output of full 3-dimensional model prognostic variables were saved, making it a remarkable tool for the study of ocean and air-sea exchange processes and for the simulation of satellite observations.
The 0000-GMT and 1200-GMT global \sst\ fields for the uppermost 1-m level of the LLC4320 output were downloaded 
using the {\tt xmitgcm} package
for the 365-day period starting on November 17, 2011 yielding 730 files
totaling $\sim$0.5\,Tb. Hereinbelow, we will use LLC and LLC4320 interchangeably to refer to the $\frac{1}{48}^\circ$ MITgcm simulation.
Although the entire model domain was downloaded, only the region from the southern extreme to $57^\circ$N was considered in the geographic analysis to avoid the change in grid geometry occurring at $57^\circ$N. %{\color{red}Kat, does the geometry change at 57 or a bit higher?}

% %%%%%%%%%%%%%%%%%%%%%%%%%%%%%%%%%%%%%%%%%%%%%%%
% %%%%%%%%%%%%%%%%%%%%%%%%%%%%%%%%%%%%%%%%%%%%%%%
\section{Methods}
\label{sec:methods} 

\subsection{Creation of Comparable SSTa Cutouts}

Following our previous study on \sst\ patterns \citep{prochaska2021deep},
we chose to analyze cutouts, approximately
$\sim$150$\times$150\,km$^2$ regions extracted from the parent observational and modeled \sst\ fields.
The size of these samples was chosen in part to focus on features at scales of
$\sim$30\,km or smaller (i.e., submesoscale).
Using these modest-sized cutouts also yields
a massive number of cutouts---O($\sim$10$^6$)---with array dimensions that are
easily tractable to machine learning techniques
($\sim$100$\times$100 pixels).
In the following subsections, we detail
the procedures to generate such cutouts from the
{\viirs} data and {\llc4320} outputs that have nearly 
equal dimension and geographical coverage.

\subsubsection{{\viirs} Cutouts \label{section:viirs_cutouts}}

In each {\viirs} parent field ({\it image} hereinafter), we identified every 
192$\times$192\,\pixsq\ subarray that has 
fewer than 2\%\ of its pixels masked 
(quality\_level~$< 5$) because of land, corrupted pixels usually associated with cloud cover, or missing data. 
This $2\%$ threshold was 
established after sampling at a wider
range of thresholds (as high as $5\%$)
and assessing the outputs.
For thresholds greater than $2\%$, we found that clouds significantly bias estimates of the degree of structure obtained by the machine learning algorithm (discussed in \S\ref{MLalgorithm}) applied to the datasets of cutouts.

We then divided each image into a grid with cell
size of 96$\times$96\,\pixsq\
and selected the closest cutout to the center of each grid cell 
that satisfies the 2\% threshold (or 0 if none satisfy). This approach randomized the sampling thus avoiding possible biases that may have emerged from cutouts sampled on a regular grid. In a well-sampled image, this implies each cutout has $\sim$50\%
overlap with its 4 nearest neighbors and $\sim$25\% with
the next 4 nearest neighbors.
Unlike our analysis of \ac{MODIS} data, we did not place a 
restriction on distance from nadir because
the physical size of the {\viirs} pixels 
varies by approximately a factor of two from nadir to swath edge \citep{rs14143476} compared with a variation of approximately a factor of five for \ac{MODIS}. % Fig.\,\ref{SepFig}. 

\begin{figure}[ht]
\includegraphics[width=\linewidth]{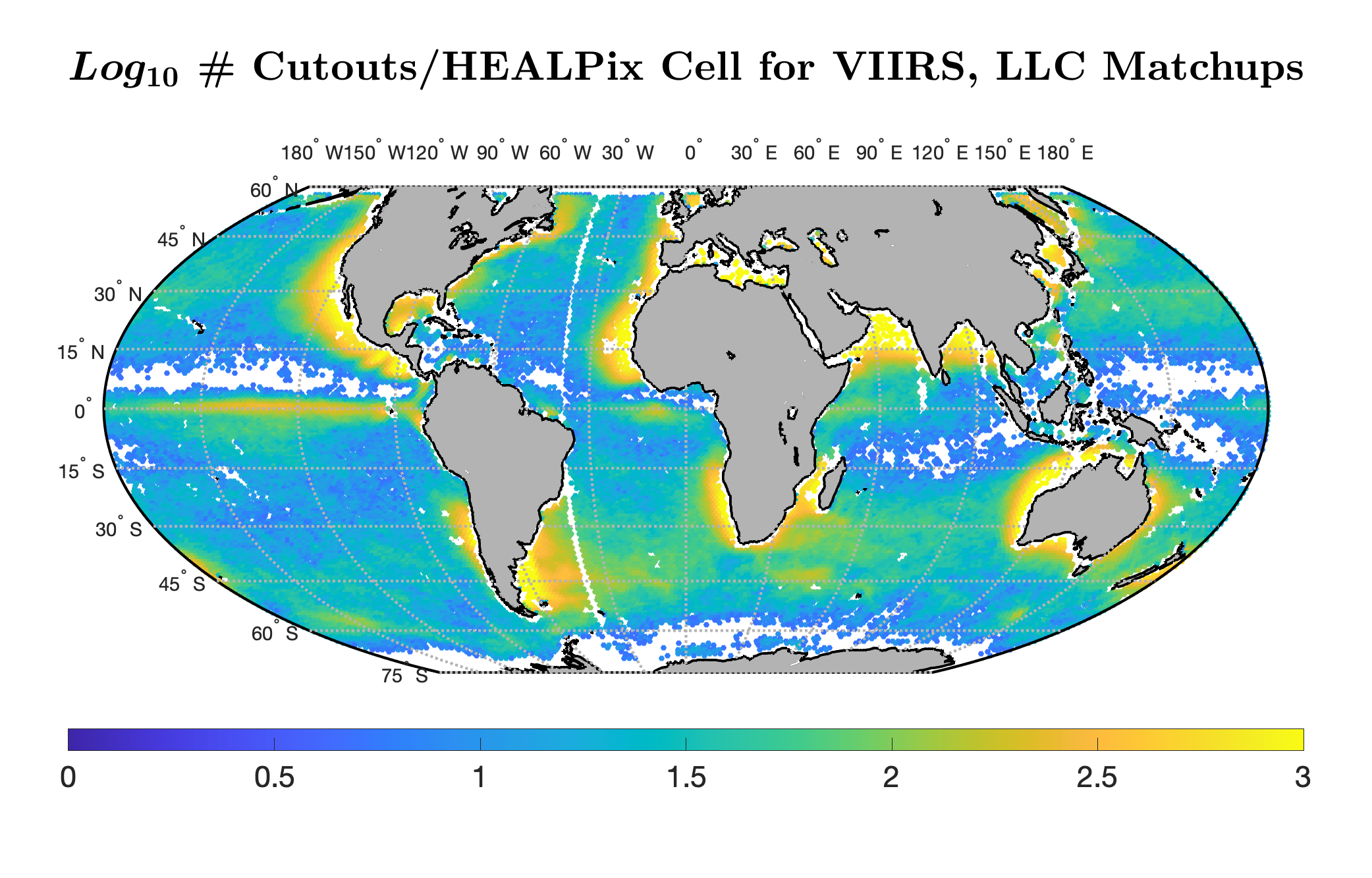}
\caption{
%Concentration of cutouts extracted from the 
Geographic distribution of cutouts in our
98\%-clear {\viirs} dataset, shown 
using a log$_{10}$-based scale of color intensity.
Each equal-sized ({\healpix}) spatial cell, plotted as dots here, covers approximately
10,000~km$^2$. {\healpix} cells with less than five {\viirs} cutouts or less than five {\llc4320} cutouts are shown in white. Land is shown as light gray. Meridional white line at $\sim$35$^\circ$W is due to an {\llc4320} sampling artifact.
}
\label{fig:viirs_geo_p}
\end{figure}

From the full parent dataset, 
we extracted \nviirs\ cutouts (limiting to $<57$\degree\ N;
see \ref{sec:ogcm_cutouts}).
The geographical distribution of these is shown 
in Fig.\,\ref{fig:viirs_geo_p}, which highlights
the regions of the ocean that are preferentially cloud
free (e.g., the equatorial Pacific ocean and coastal regions). For this and all subsequent geographic plots, we used the {\healpix}\footnote{
https://healpix.sourceforge.io} 
schema 
\citep{healpix},
which tesselates the surface into equal-area curvilinear quadrilaterals
and was introduced for all-sky analysis of astronomical data. 
Here and throughout the manuscript we
adopted nside=64, which yields a {\healpix}
cell with approximately 100$\times$100\,km$^2$ area.
Values presented are numbers associated with each {\healpix} cell; in this case the number of cutouts in the cell. 

% \begin{figure}[ht]
% \includegraphics[width=\linewidth]{Figures/viirs_concentration.png}
% \caption{
% %Concentration of cutouts extracted from the 
% Geographic distribution of cutouts in our
% 98\%\ clear {\viirs} dataset shown 
% on a log-based scale of color intensity.
% Each equal-sized spatial cell covers approximately
% 10,000~km$^2$ of area.
% }
% \label{fig:viirs_geo}
% \end{figure}

For cutouts with one or more masked pixels, we ``inpainted''
them using the biharmonic algorithm provided in 
the {\tt scikit-} {\tt image} software package \citep{scikit-image}; the algorithm, we found, performs
well even for data with steep gradients \citep{prochaska2021deep}.
We then downscaled the arrays with a local mean 
to 64$\times$64\,\pixsq\ or approximately
144$\times$144\,km$^2$ at
$\sim$2.1\,km sampling, which approximately matches the coarsest sampling
of the \ogcm\ outputs. Note that the actual size of cutouts is a function of distance from nadir; % (Fig.\,\ref{SepFig}); 
we did not resample the observed data to account for these changes.
Last, we demeaned each cutout to produce \ssta\ fields.
This defines the final, pre-processed dataset that we
use for all {\viirs} analyses to follow.

% %%%%%%%%%%%%%%%%%%%%%%%%%%%%%%%%%%%%%%%%%%%%%%%%%%%%%%
% OGCM 
\subsubsection{{\llc4320} Cutouts \label{llc_cutout_selection}}
\label{sec:ogcm_cutouts}

Roughly nine years (8 yrs 11 months) of {\viirs} data comprise the $\sim$3 million (2,932,452) {\viirs} cutouts, whereas the \llc4320 simulation output used for this study spans only one year. 
%Alone restricting the {\viirs} cutouts to be 98\% clear eliminates more than two thirds of the data. 
To further restrict the {\viirs} cutouts to one year would leave too little data for a full-globe comparison between satellite observations and the model outputs. For this reason, we compare $\sim$9 years of {\viirs} data to the one-year model simulation. 

Our approach to constructing cutouts from the {\llc4320} outputs
intentionally paralleled the methodology and outputs for \viirs.
We first identified all 64$\times$64\,\pixsq\ regions that have a
valid \sst\ value (i.e., avoiding land). 
The geographic location (lat, lon) of each of these was recorded.
Second, we considered the full year of {\llc4320} outputs taken every
12~hours from 2011-11-17 to 2012-11-15 inclusive.
For each cutout in the {\viirs} sample, we matched in location
to the closest valid 64$\times$64\,\pixsq\ 
region in the {\llc4320} dataset.  We then
identified the {\llc4320} timestamp closest in time from
the start of the given year (with the {\llc4320} in 12~hour intervals).
This is akin to matching on day-of-year and then time-of-day
to the nearest 12 hours. This `climatological' matchup of cutouts was performed to avoid seasonal and regional biases in the sampling.

Much of the analysis presented in subsequent sections of this paper compares the statistics of cutouts in a given {\healpix} cell. Ideally, the cutouts associated with a {\viirs}-{\llc4320} match-up lie in the same {\healpix} cell. This, however, is not always the case; the closest {\llc4320} cutout to a {\viirs} cutout may lie in an adjacent {\healpix} cell when the {\viirs} cutout lies outside of the {\llc4320} grid, generally in coast waters. 

We wished to maintain an approximately-constant sampling size 
of 2.25\,km matched to {\viirs} or 144$\times$144\,km$^2$ 
total for  a 64$\times$64\,\pixsq\ cutout.
Therefore, we sized the array extracted from 
the {\llc4320} outputs according to the local size of the
grid, which varies as approximately $\cos(\rm lat)$ for
latitudes $\leq 57^\circ$\,N.
At latitudes $>57^\circ$\,N, {\llc4320} horizontal grid spacing asymptotes to $\sim$1 km in the polar cap. To avoid the complications brought by different grid characteristics, we constrained our analysis to south of $57^\circ$\,N.

Each extracted {\llc4320} array is downscaled to a 64$\times$64\,\pixsq\ cutout using the local mean. We then injected random noise using a Gaussian deviate with a standard deviation $\sigma = 0.04$\,K based on an analysis of the noise properties of the {\viirs} data \citep{rs9090877}. Lastly, we demeaned each cutout to generate
\ssta\ arrays.

\subsection{Characterization of the \ssta\ Cutouts}

Each cutout was assigned a {\LL} metric by the machine learning algorithm, \ulmo, used for this work. This metric describes the frequency of occurrence of the cutout within the full set. The {\LL} metric tends to correlate with the {\sst} structure, at least at the spatial scales of the fields under consideration here, with structure increasing with decreasing {\LL} \citep{prochaska2021deep}.
This simply follows from the fact that the parent sample is dominated
by cutouts with little inherent structure.
Comparing the distribution of {\LL} values across the global ocean thus identifies geographic regions where
the structure of the model output at submesoscale-to-mesoscale matches (or fails to match) 
the observations.

% %%%%%%%%%%%%%%%%%%%%%%%%%%%%%%%%%%%%%%%%%%%%%%
\subsubsection{Brief Overview of \textbf{\textsc{Ulmo}}\label{MLalgorithm}}

The \ulmo\ machine learning algorithm is a \acl{PAE} \citep[\acs{PAE};][]{pae}
designed to assign a relative probability of occurrence to each
cutout in a large dataset.  It is an unsupervised method,
which learns representations of the diversity of \ssta\ 
patterns without human assessment.
The \pae\ combines two deep learning algorithms to perform
its analysis.  
The first is an autoencoder that generates a reduced
dimensionality representation (aka, a latent vector) for
each cutout in a complex latent space.
The second step is a normalizing flow \citep{Papamakarios}, %{\color{red}X, check this reference}
which transforms the autoencoder latent space
into a Gaussian manifold with the same dimensionality.  
One can then calculate
the relative probability of any cutout occurring
within the Gaussian manifold with standard statistics.
We refer to this relative probability as the Log-Likelihood
(\LL) metric.

In the following section, we compare distributions of the
{\LL} metric for the {\viirs} and {\llc4320} cutouts in discrete
geographical regions across the global ocean.
This provides a quantitative technique to compare
the \sst\ patterns predicted by the \ogcm\ against those
observed in the real ocean.  We note that because the {\LL} metric
is only a scalar description of a given pattern's frequency of
occurrence, it is possible---in principle---to have similar
{\LL} distributions despite qualitative differences in the
\sst\ patterns.  This would, however, require a remarkable
coincidence, and our visual inspection of regions with consistent
{\LL} distributions have not revealed any such examples.
We also emphasize that the opposite is not true:  regions
with significantly different {\LL} distributions do have
qualitatively differing distributions of \sst\ patterns.

\begin{figure}[ht]
\includegraphics[width=\linewidth]{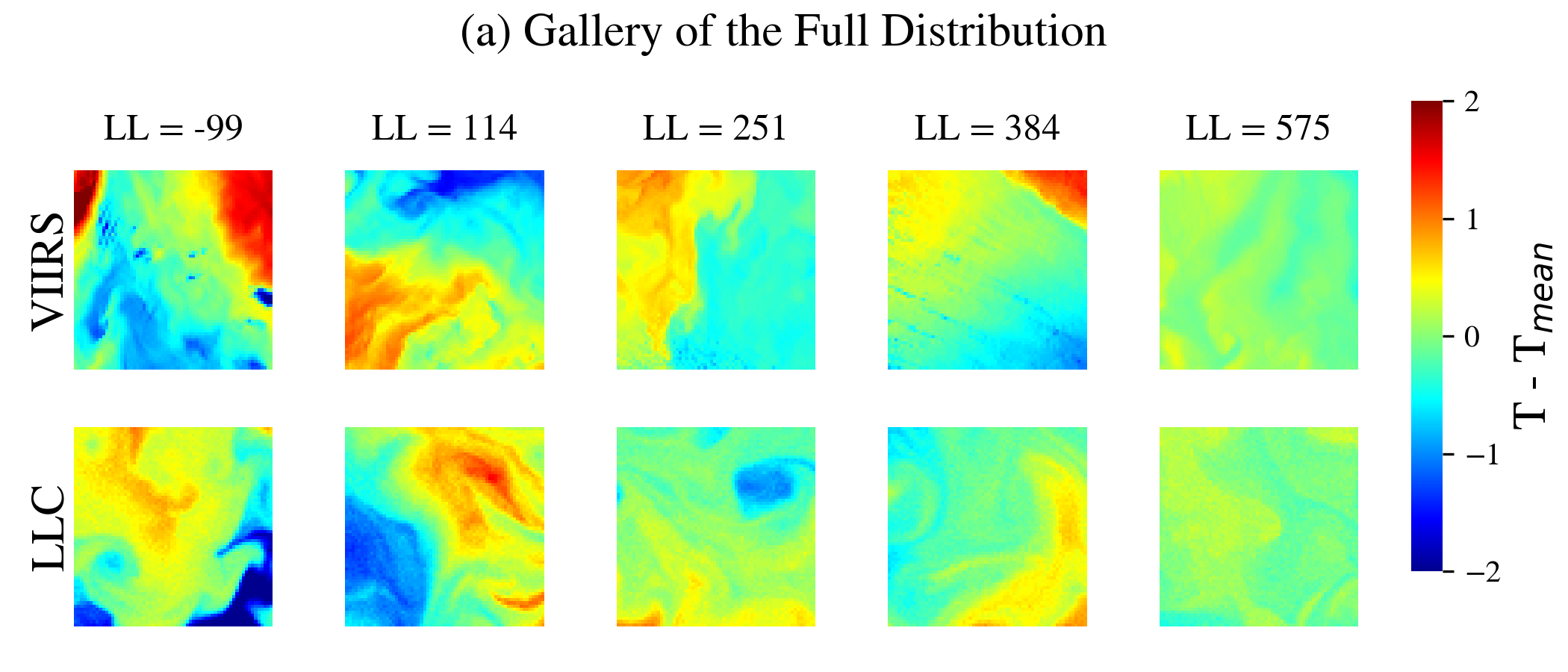}
\includegraphics[width=\linewidth]{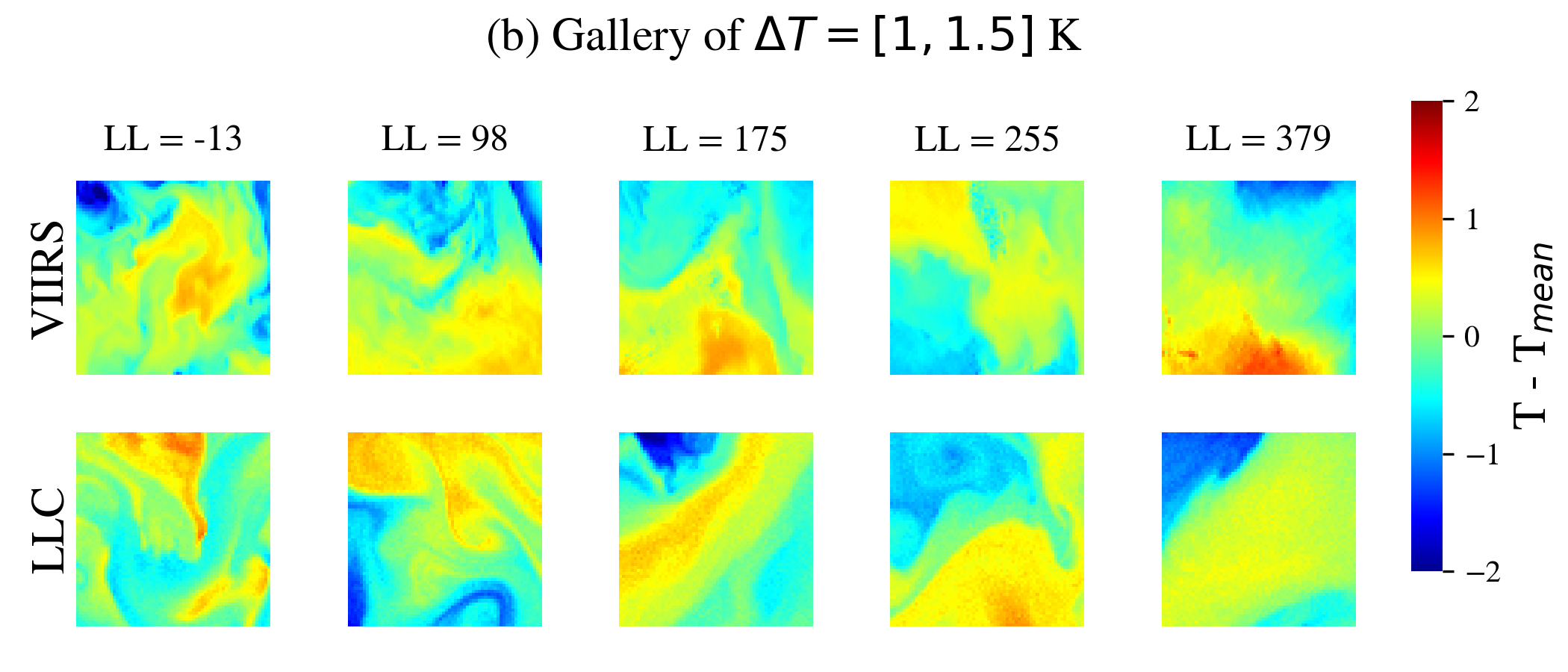}
\caption{
Galleries of cutouts showing the progression of structure of the associated \sst\ fields as a function of \LL. (a) Galleries for the entire set of cutouts. Upper row for \viirs\ cutouts, lower row for \llc\ cutouts. Each cutout is randomly selected from the 50 cutouts with \LL\ nearest to the median \viirs\ \LL\ value for the associated \LL\ quintile. The \LL\ of the median is shown above the \viirs\ cutout. (b) Similar galleries for all cutouts with $1<\DT <1.5$\,K where $\DT \equiv T_{90}-T_{10}$ with $T_{N}$ denoting the $N^{th}$-percentile.
}
\label{fig:ULMO_gallery}
\end{figure}

Fig.~\ref{fig:ULMO_gallery} presents galleries of \viirs\ and \llc\ cutouts designed to show how the structure of the cutouts vary as a function of \LL. For Fig.~\ref{fig:ULMO_gallery}a the entire \LL\ \viirs\ population is divided into quintiles. For each quintile, one \viirs\ cutout and one \llc\ cutout is randomly selected from the 50 \LL\ values nearest to the median of the \viirs\ \LL\ distribution. The \LL\ of the median values are shown above the \viirs\ cutouts. These galleries show a well defined progression from fields with a large temperature range and accompanying gradients to fields with a smaller temperature range, weaker gradients and less
complex patterns. %Of importance for the work presented herein, the two galleries also track each other quite well--the characteristics of \viirs\ and \llc\ fields are similar for similar \LL\ values. 

The correlation between temperature range and \LL\ seen in Fig.~\ref{fig:ULMO_gallery}a was noted by \citet{prochaska2021deep} in their discussion of the \ac{MODIS} dataset. In the analysis to follow (Section 4.2), we compare \LL\ values of \viirs\ cutouts with those for \llc\ cutouts at the same geographic location, which suggests that the temperature range of the cutouts we compare will be similar. (Admittedly, there are a few regions where this is not the case, and we address this when it occurs.) Important from the perspective of the work presented herein is rather how the characteristics of cutouts vary with \LL\ when the temperature range is bound to a small range. 
To further provide a sense for how cutout characteristics other than temperature range evolve with \LL, we present a second gallery of \viirs\ and \llc\ cutouts in Fig.~\ref{fig:ULMO_gallery}b.
These cutouts are restricted to have $1<\DT<1.5$\,K 
where \DT\ is defined as the difference in $90^{th}$ and $10^{th}$ percentiles of the \sst\ distribution: $\DT \equiv T_{90} - T_{10}$ where $T_{N}$ denotes the $N^{th}$-percentile. These galleries were constructed in the same fashion as those for Fig.~\ref{fig:ULMO_gallery}a. In this case, the progression is from cutouts for which \sst\ contours are relatively convoluted to cutouts with relatively straighter contours; in other words, the `structure' of the cutouts decreases as \LL\ increases. As with the galleries in \ref{fig:ULMO_gallery}a, the characteristics of the \llc\ cutouts in this gallery track those of the cutouts in the \viirs\ gallery.

%[Try using KS test in addition to median LL?]

% %%%%%%%%%%%%%%%%%%%%%%%%%%%%%%%%
\subsubsection{Training on {\viirs} and Evaluation}
\label{sec:ulmo_training}

In \cite{prochaska2021deep} we introduced the \ulmo\ algorithm
and trained a model with \ac{L2} \ac{MODIS} data sampled
at $\sim$2\,km and using 64$\times$64\,\pixsq\ cutouts.
While this model could have been applied here to the {\viirs} and
{\llc4320} cutouts,
%\footnote{We did perform this experiment and
%recovered similar results but with a few notable exceptions. {\color{red}I think that I would remove this footnote; it's a distraction at this point.}},
we have generated a new \ulmo\ model from the {\viirs} dataset.
The same hyperparameters derived for \ulmo\ in \cite{prochaska2021deep} were adopted here.

We trained the \ac{PAE} on 150,000 random {\viirs} 
cutouts from 2013 and used the remainder of the data from 
that year (181,184 cutouts) for evaluation.
We trained the autoencoder for 10 epochs with a batch
size of 256 and achieved good learning loss
convergence.
We then trained the normalizing flow for 10 epochs
(batch size of 64) and also achieved good convergence.
The \viirs-trained \ulmo\ model was then applied to all 
of the {\viirs} and {\llc4320} cutouts to calculate {\LL} for each.

\section{Results/Discussion\label{results_discussion}}

We divide the comparison of the {\viirs} {\sst} dataset with the {\llc4320} model output, in the context of the patterns learned by {\ulmo}, into those related to the shapes of the two {\LL} probability distributions and those related to their geographic distributions.

% [JXP: Am pointing out here we now can do analysis with seasonal cuts,
% e.g., isolate the 3 summer months.  Not sure we should or will
% but just wanted to keep it in mind. {\color{red}Much as I would love to do a more detailed analysis I think that we need to push this one out without this level of detail. I would like to do a seasonal analysis of {\viirs} along in a subsequent paper.}]

% [JXP: This section is quite short.  I say we move
% what is in here into the other sections;  some of
% it is Results some is closer to methodology]

\subsection{Overall Statistical Comparison of {\viirs} and {\llc4320} {\LL} Values}

\begin{figure}[ht]
\includegraphics[width=\linewidth]{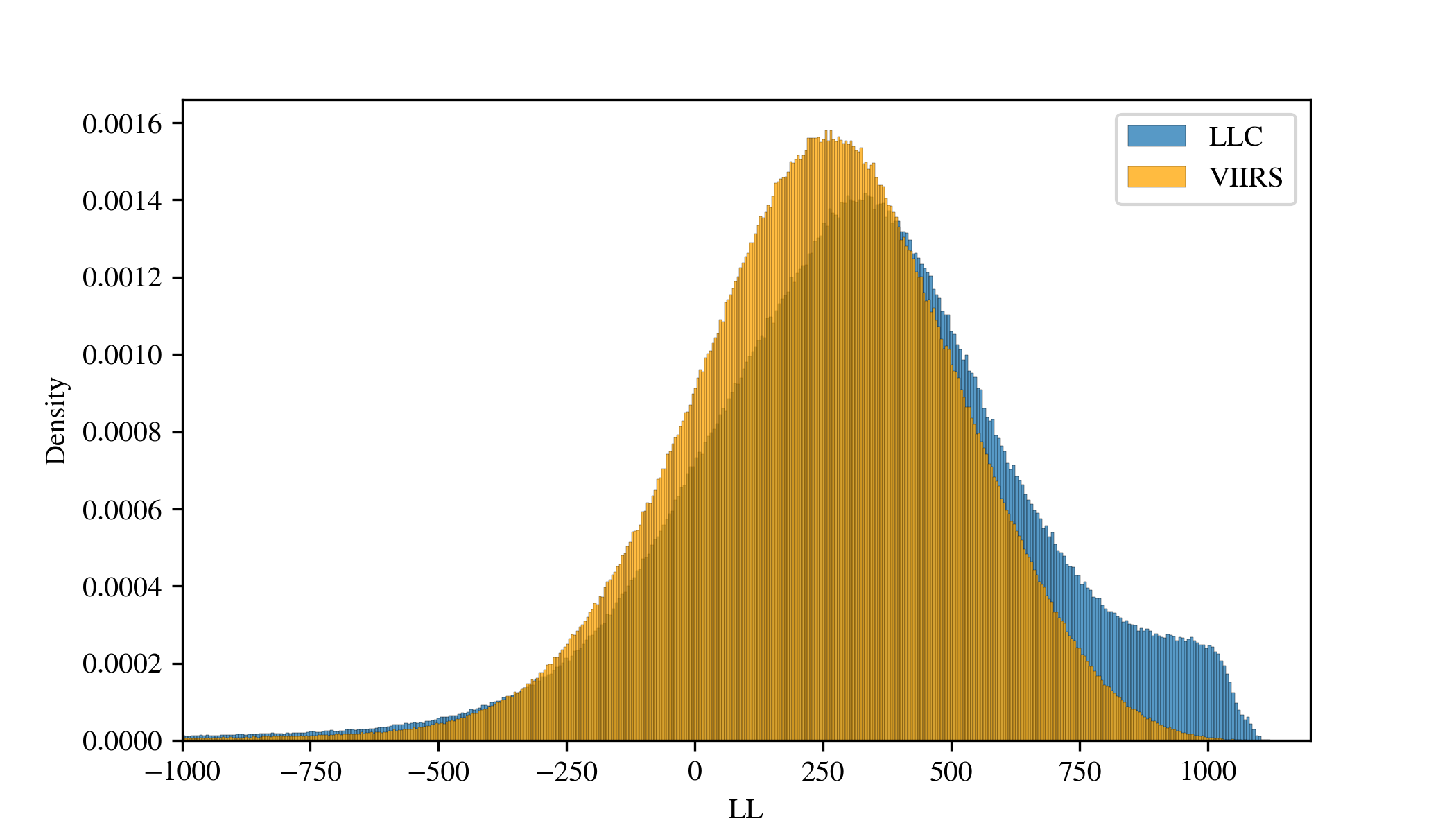}
\caption{Histograms of the {\LL} metric
for the full sample of {\viirs} and {\llc4320} cutouts.}
\label{fig:LL_hist}
\end{figure}

Figure \ref{fig:LL_hist} compares the distribution of 
the {\LL} metric for the 9~years of
{\viirs} observations against 
the {\llc4320} results matched in
space and day-of-year.
The two distributions are quite similar, suggesting that, on average, the {\sst} pattern distribution learned by {\ulmo} from {\viirs} 
is close to the same distribution derived from {\llc4320}. 
There are, however, differences, subtle for much of the range of {\LL} values and not so subtle for $LL>800$:
\begin{itemize}
    \item[$\bullet$] $LL<-375$: This range corresponds to relatively energetic regions, generally associated with strong currents and large {\sst} gradients \citep{prochaska2021deep}. In these regions, the probability of finding a {\viirs} cutout with a given {\LL} value is lower than that of finding an {\llc4320} cutout with the same {\LL} value. This suggests that in dynamic regions, {\llc4320} fields tend to have slightly more structure than {\viirs} fields, i.e., that the fields are possibly more energetic. An example of cutouts in this {\LL} range is discussed in the paragraphs following the bolded text {\bf Gulf Stream} in  \S\ref{Differences}.
    \item[$\bullet$] $-375<LL<375$: This corresponds to mid-range fields, generally found at mid-latitudes (Fig.\,\ref{fig:viirs_and_llc_ll}a), away from eastern and western boundary currents. The probability of finding a {\viirs} cutout with a given structure ({\LL} value) in this {\LL} range is increasingly higher, as {\LL} increases from -375 to 375, than that of finding an {\llc4320} cutout within this {\LL} range. An example of cutouts in this {\LL} range is discussed in  the paragraphs following the bolded text {\bf Southern Ocean} in \S\ref{Differences}.
    \item[$\bullet$] $375<LL<800$: In this {\LL} range, the probability for
    the {\llc4320} of finding a given {\LL} value is higher---less structure---than for {\viirs}. {\sst} fields with these {\LL} values have relatively little structure with retrieval and instrument noise associated with the satellite-derived fields having a relatively larger impact on their observed structure. As noted in \S\ref{llc_cutout_selection}, white noise was added to the {\llc4320} fields in an attempt to remove the importance of noise in determining the {\LL} value but, in retrospect, the level of noise may not have been sufficient to address this, hence the higher probability of finding {\llc4320} cutouts in this range. Cutouts with $LL>375$ tend to be found equatorward of approximately $15^\circ$. 
    \item[$\bullet$] $LL>800$: The probability distribution for {\llc4320} {\LL} values levels off at 800 falling rapidly to zero for values larger than 1100 or so. By contrast, few {\viirs} cutouts have {\LL} values greater than 800. This is likely due to noise in the satellite-derived {\sst} cutouts as well as unresolved clouds, discussed in  the paragraphs following the bolded text {\bf Equatorial Band} in \S\ref{Differences}.
\end{itemize}

\subsection{Geographic Comparison of {\tildellviirs} and {\tildellllc} Values}

\begin{figure}[ht]
\includegraphics[width=\linewidth]{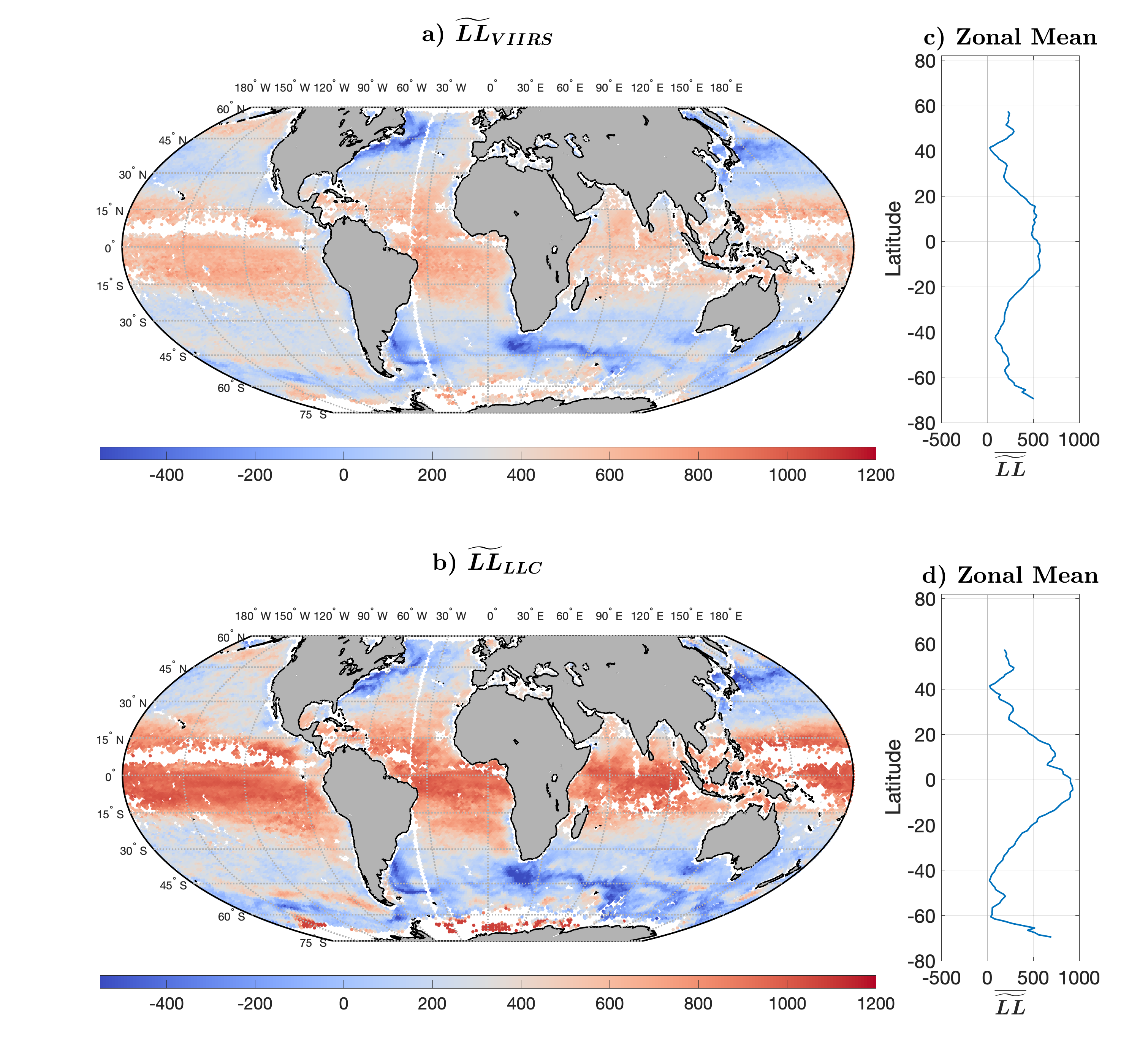}
\caption{(a) {\healpix} median {\LL} for {\viirs} cutouts ({\tildellviirs}), (b) {\tildellllc}, (c) Zonal average of {\tildellviirs}, and (d) Zonal average of {\tildellllc}. Each dot in a and b is associated with an equal area {\healpix} cell. 
White dots correspond to locations with less than five {\viirs} cutouts and less than five {\llc4320} cutouts in the {\healpix} cell. Land is shown in gray.}
\label{fig:viirs_and_llc_ll}
\end{figure}

\begin{figure}[ht]
\includegraphics[width=\linewidth]{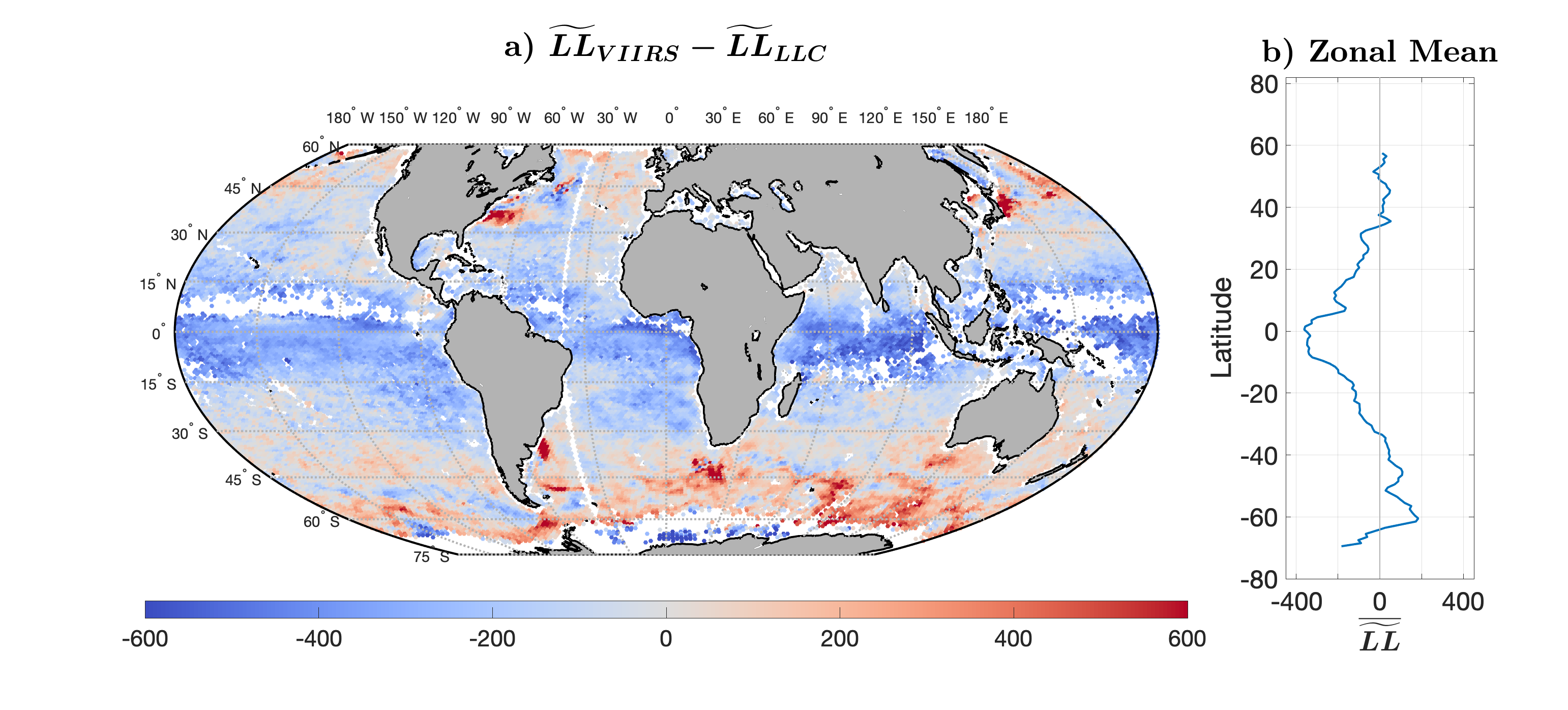}
\caption{(a) {\viirsllcdifference}. (b) Zonal mean of {\viirsllcdifference}.
It is apparent that the {\llc4320} model has {\sst} patterns
with less structure in Equatorial regions.
In contrast, the dynamic regions of the global ocean
(e.g., Western boundary currents) exhibit lower {\tildellllc} 
indicating a higher degree of structure within these areas
in the model output.
}
\label{fig:viirs_minus_llc_ll}
\end{figure}

% In this section, we examine geographic 
% similarities and differences between the 
% {\llc4320} model output and \viirs.
As mentioned in \S\ref{section:viirs_cutouts}, we compare the geographic distribution of {\viirs} {\LL}s with that of {\llc4320}, based on the {\healpix} tesselated surface covering the entire Earth: 41,952 equal-area spatial cells, each of $\sim$100$\times$100\,km$^2$ size.
Within each cell, we consider the distribution
of {\LL} values from the {\viirs} data and {\llc4320} output. To minimize the effects of outliers within the
distributions, we utilize the
median {\LL} value (designated {\tildell} hereinafter) as a characteristic metric of
the structure in \sst\ at a given location. Furthermore, we only consider {\healpix} cells containing a least five {\viirs} cutouts and five {\llc4320} cutouts. The distributions for {\tildellviirs} and {\tildellllc} are shown in Fig.\,\ref{fig:viirs_and_llc_ll} and their difference, $\Delta_{LL} = ${\viirsllcdifference}, in Fig.\,\ref{fig:viirs_minus_llc_ll}. Recall that {\LL} tends to increase with decreasing structure in the {\sst} field, hence positive values of {$\Delta_{LL}$} suggest less structure in the satellite-derived cutouts than in the cutouts of the model output and negative values the contrary. 

Figure \ref{fig:viirs_minus_llc_ll} suggests that there are significant differences between the submesoscale-to-mesoscale structure of the simulation and that of the satellite-derived dataset. However, closer examination of the two plots in Fig.\,\ref{fig:viirs_and_llc_ll} suggests that there is significant similarity in the larger scale ($>$ several hundred kilometers) of the two fields; our eyes are drawn to the differences, the large red equatorial regions and blue regions at higher latitudes, not the similarities. We therefore begin by examining the similarities of the fields and then consider their differences.

\subsubsection{Similarities}

\begin{figure}[ht]
\includegraphics[width=0.9\linewidth]{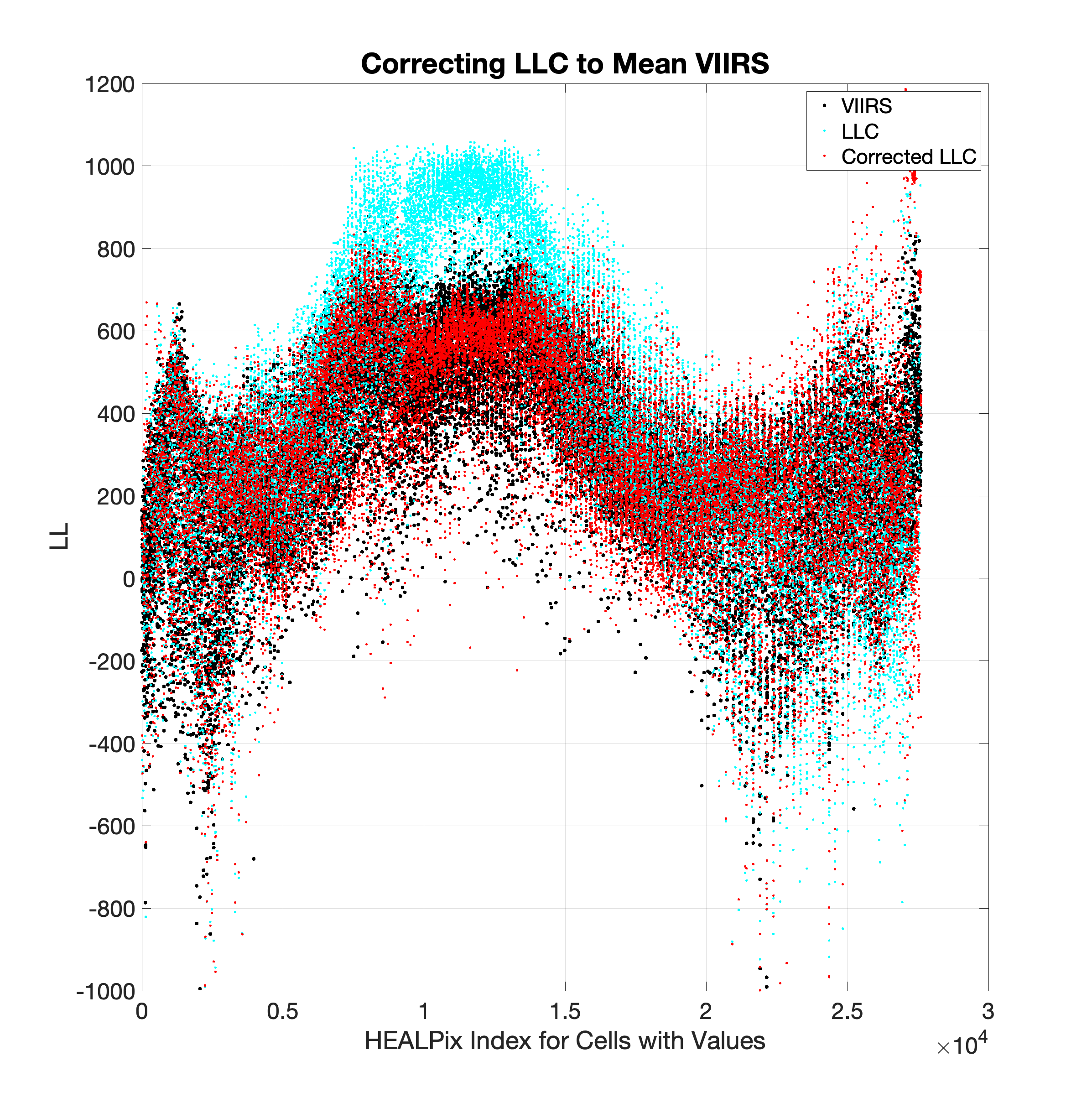}
\caption{ {\tildell} versus {\healpix} cell index: {\tildellviirs} values are black dots, {\tildellllc} are cyan, and {\tildellllcprime} are red.
}
\label{fig:correctingllc}
\end{figure}

To highlight the similarities in the {\tildell} fields on
smaller scales ($\mathcal{O}(100\,\rm km)$), 
we remove from {\tildellllc} the large regional differences 
apparent between them (Fig.\,\ref{fig:viirs_and_llc_ll}).
%In order to highlight the similarities in the {\tildell} fields we remove the large regional differences between them. 
This is done by averaging {\tildellviirs} and {\tildellllc} over 100 sequential values based on the {\healpix} index. Since the vector of {\healpix} cells is arranged geographically, the indicies over which the average is performed tend to correspond to a relatively tight geographical region. {\tildellviirs} (black) and {\tildellllc} (cyan) are shown in Fig.\,\ref{fig:correctingllc}. A `corrected' {\tildellllc}, designated as {\tildellllcprime}, is then determined from:

$$ \widetilde{LL}'_{LLC_i} =  \widetilde{LL}_{LLC_i} - \frac{1}{100}\displaystyle\sum_{j\in B}{(\widetilde{LL}}_{LLC_j} - \widetilde{LL}_{VIIRS_j}) $$
where $i \in $ all {\healpix} cells with at least 5 cutouts in the {\viirs} dataset and 5 cutouts in the {\llc4320} dataset,  
$$ B: \left[\lfloor \frac{i-1}{100} \rfloor*100+1, \lfloor \frac{i-1}{100} \rfloor*100+100\right] $$
and $\lfloor x \rfloor$ dnotes the largest integer less than or equal to x. 
{\tildellllcprime} values are shown in red in Fig.\,\ref{fig:correctingllc}. 

\begin{figure}[ht]
\includegraphics[width=\linewidth]{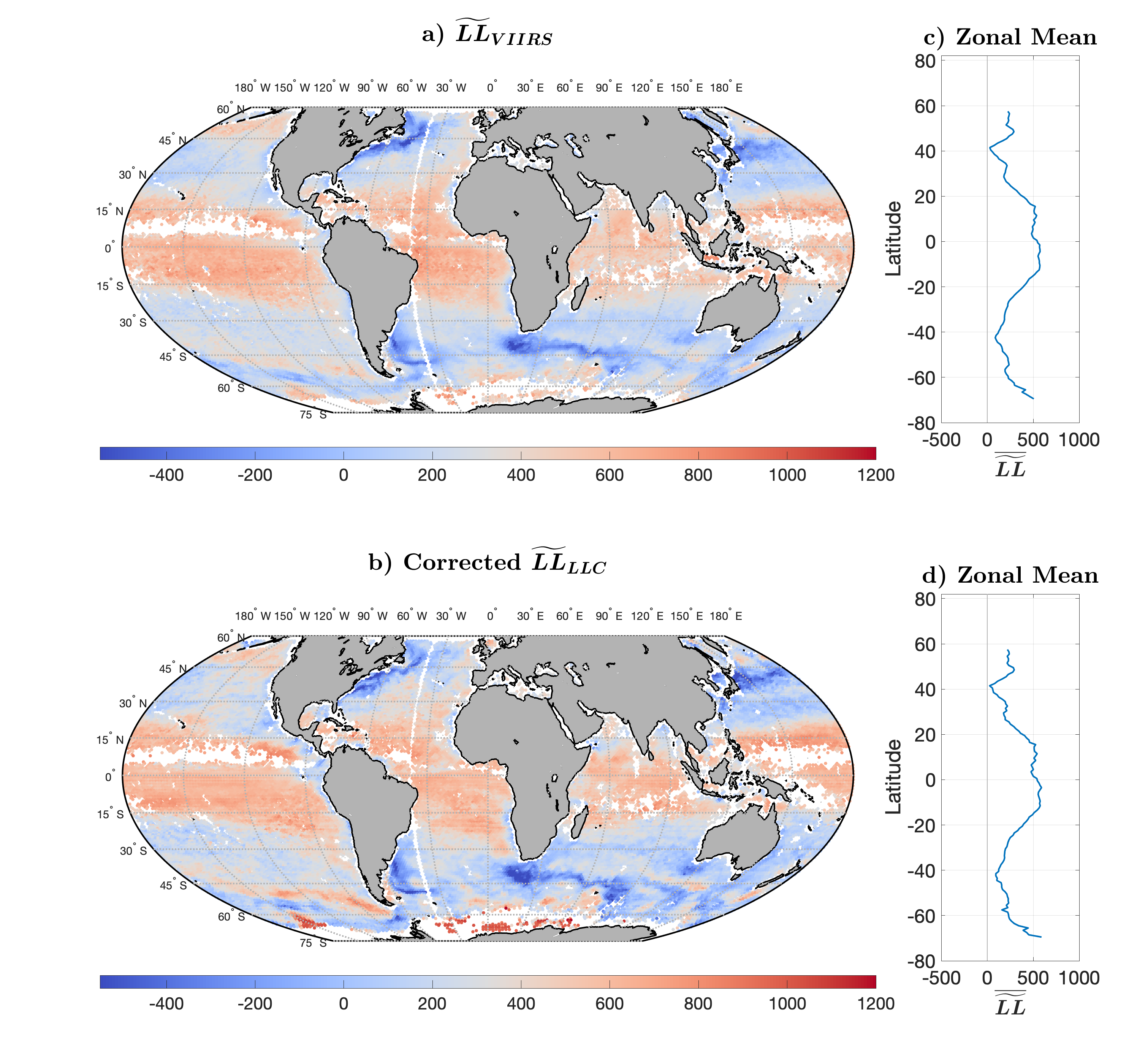}
\caption{As in Fig.\,\ref{fig:viirs_and_llc_ll} except {\tildellllcprime} is shown in (b).
}
\label{fig:correctedfield}
\end{figure}

The {\tildellllc} geographic distribution is shown  with that of {\tildellllcprime} in Fig.\,\ref{fig:correctedfield}. The large scale similarities in the general shape of the distributions is much more evident in this figure. Note also that the zonal mean for {\tildellviirs} and that for {\tildellllcprime} are now quite similar for latitudes equatorward of $60^\circ$. 

\begin{figure}[ht]
\includegraphics[width=\linewidth]{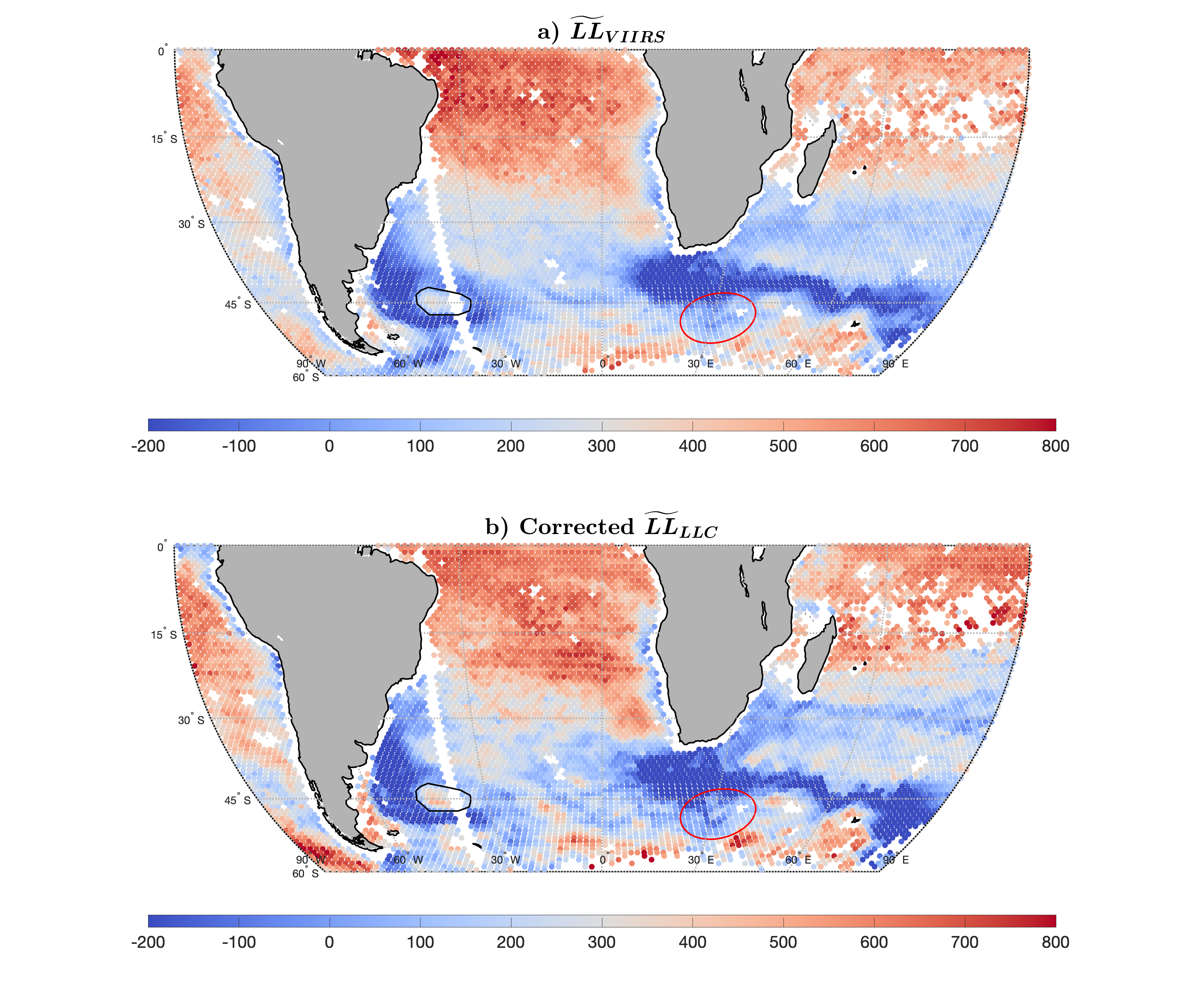}
\caption{As in Fig.\,\ref{fig:correctedfield} but palette constrained to better show features in the two focus area: black polygon at $\sim$35$^\circ$W, 45$^\circ$S and red polygon at $\sim$30$^\circ$E, 50$^\circ$S.
}
\label{fig:zoomed_correctedfield}
\end{figure}

\begin{figure}[ht]
\includegraphics[width=\linewidth]{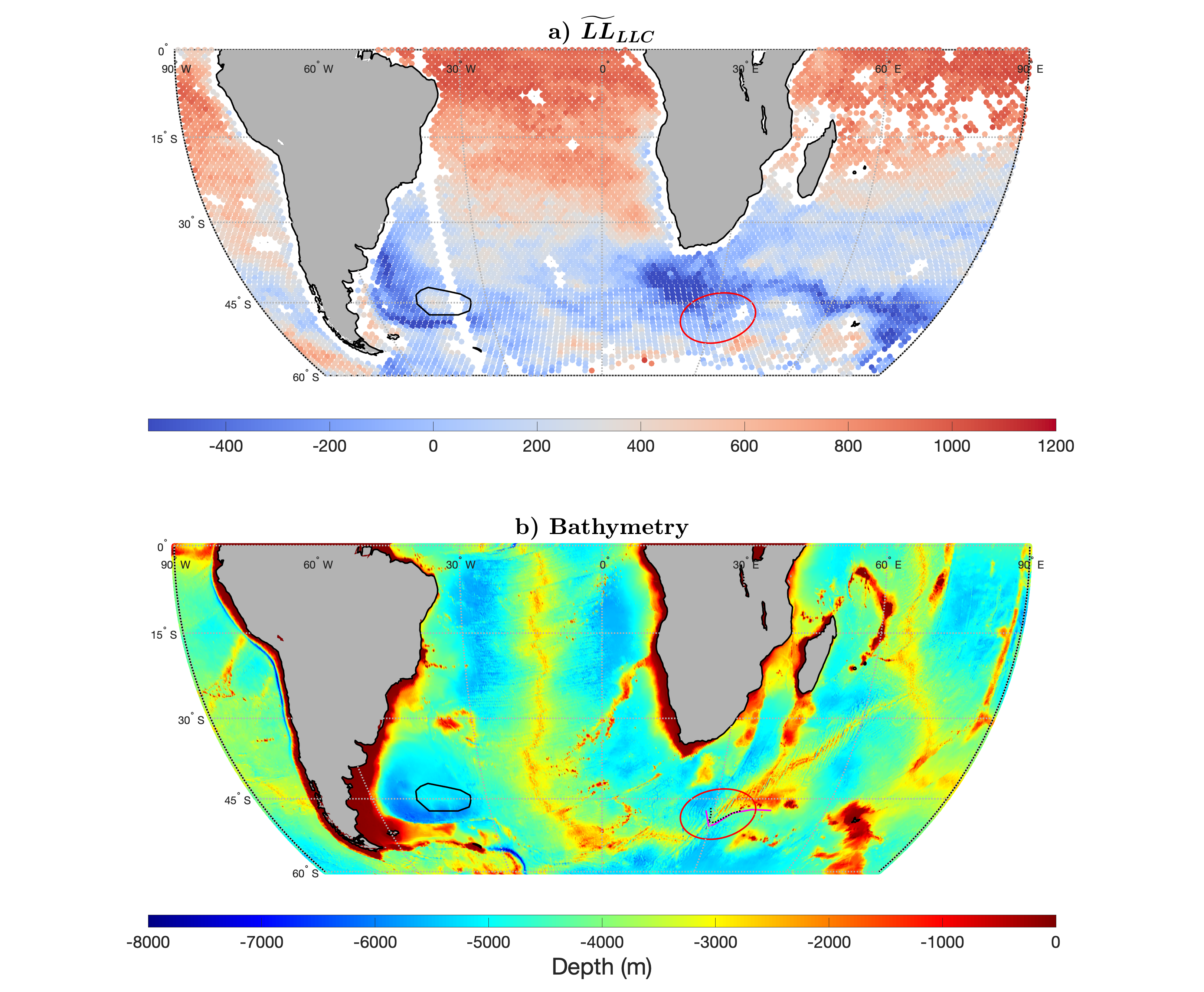}
\caption{(a) Uncorrected {\tildellllc}. (b) Bathymetry. Focus regions: black polygon at $\sim$35$^\circ$W, 45$^\circ$S and red polygon at $\sim$30$^\circ$E, 50$^\circ$S. Dotted black and solid magenta lines in the focus area encircled with the red polygon were manually digitized from Fig.\,\ref{fig:viirs_and_llc_ll}a and Fig.\,\ref{fig:correctedllc_with_bathymetry}a. for {\viirs} and {\llc4320}, respectively.
}
\label{fig:correctedllc_with_bathymetry}
\end{figure}

It is not just the similarities in the large-scale distribution that we find intriguing but also
a number of small-scale features. Consider, 
for example, the {\tildell} fields at approximately $45^\circ$S in the black and red polygons of Fig.\,\ref{fig:zoomed_correctedfield}. For both {\tildellviirs} and {\tildellllc} there is a local maximum in {\tildell} corresponding to a minimum in the structure of the {\sst} cutouts in the black polygon. This feature is associated with a zonal bathymetric ridge at approximately $45^\circ$S crossed by two meridional ridges, one at $44^\circ6'$W and the other at $39^\circ7'$W (Fig.\,\ref{fig:correctedllc_with_bathymetry}b, although the ridges are difficult to see in this figure). The peaks of these ridges are at depths of approximately 5000\,m in a basin extending to depths of 6000\,m.  Both {\LL} fields show a band of negative {\tildell} values to the west and south of the feature. The {\tildellviirs} field also shows a well-defined band to the north and east while the {\tildellllc} fields only shows a suggestion of such a band but the local minimum is still well-defined in both. 

The second feature of interest is the thin, hooked band of low {\tildellviirs} values---corresponding to relatively more structure in the cutouts---in the red polygon south of South Africa (Fig.\,\ref{fig:zoomed_correctedfield}a). This band is well reproduced in the {\llc4320} output (Fig.\,\ref{fig:zoomed_correctedfield}b). Again, referring to the bathymetric image, (Fig.\,\ref{fig:correctedllc_with_bathymetry}b) it is clear that the shape of this feature is a consequence of the underlying bathymetry. Specifically, it appears that a tendril of the retroflected Agulhas Current has flowed to the south before turning toward the east to pass through a gap in the southwest-northeast ridge, partially blocking the main part of the retroflected current. The top of the ridge is found at depths of approximately 2000\,m while the relatively wide gap, through which the tendril passes, is as deep as 3500\,m. The manually digitized center of the feature is shown with the dotted black and solid magenta lines in Fig.\,\ref{fig:correctedllc_with_bathymetry}b; the model appears to reproduce quite accurately this subtle feature in the circulation.

\subsubsection{Differences\label{Differences}}

\begin{figure}[ht]
\includegraphics[width=\linewidth]{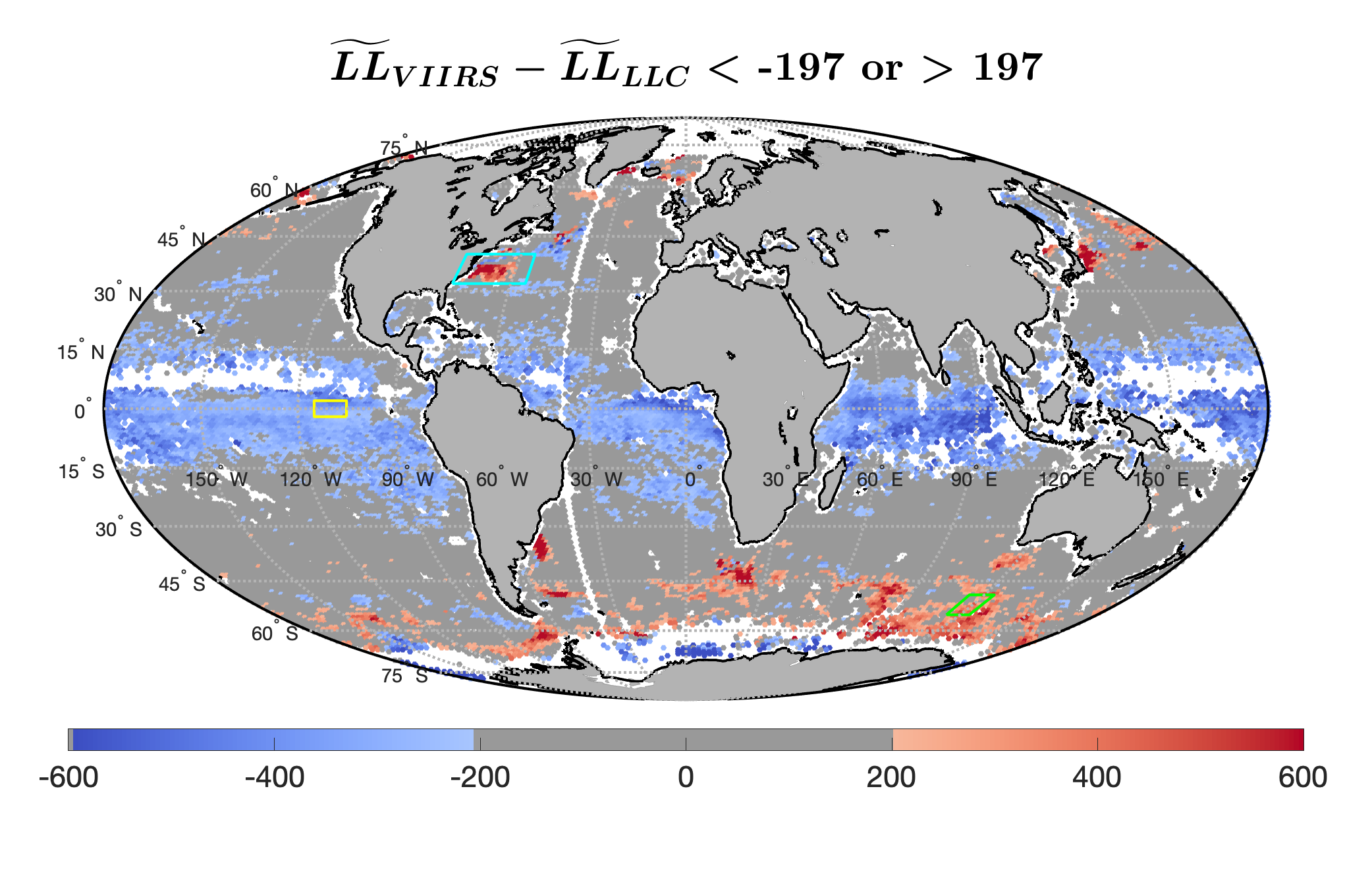}
\caption{As in Fig.\,\ref{fig:viirs_minus_llc_ll}a with $|\widetilde{LL}_{VIIRS} - \widetilde{LL}_{LLC}|<{\sigmathreshold}$ masked darker gray to highlight the significant differences between 
{\viirs} observations and the {\llc4320} model outputs. 
The yellow rectangle ($\sim$110$^\circ$W, 0$^\circ$N) designates the focus area for galleries shown in Fig.~\ref{fig:above_equator_galleries} highlighting Equatorial differences, the green rectangle ($\sim$120$^\circ$E, 55$^\circ$S) for galleries shown in Fig.~\ref{fig:acc_galleries} highlighting Southern Ocean differences and the cyan rectangle ($\sim$70$^\circ$W, 35$^\circ$N) for galleries shown in Fig.~\ref{fig:four_galleries} highlighting western boundary current differences.
}
\label{fig:viirsllmasked}
\end{figure}

To highlight the regional distributions of significant differences between the structure of satellite-derived {\sst} cutouts and that of {\sst} cutouts produced by {\llc4320}, we replot in Fig.\,\ref{fig:viirsllmasked} the data of Fig.\,\ref{fig:viirs_minus_llc_ll}, masking---dark gray---all values between $-${\sigmathreshold} and {\sigmathreshold}. These thresholds are based on differences between the {\tildell} distribution obtained from the first four years, 2012--2015, of the {\viirs} dataset and the last four years, 2017--2020, {\headtaildifference} (see the Appendix).  {\healpix} cells with {\tildell} values beyond these thresholds correspond either to regions for which the retrieved {\viirs} cutouts are not good measures of the {\sst} or that the {\llc4320} output is associated with deficiencies 
in the simulation. 

{\viirsllcdifference} differences evident in Fig.\,\ref{fig:viirsllmasked} fall into three general groupings: the wide band of negative values (more structure in the {\viirs} cutouts than in the {\llc4320} cutouts) centered on the Equator, the less continuous band of positive values in the Southern Ocean, and the very positive (much more {\llc4320} structure than {\viirs} structure) patches in the vicinity of the separation of western boundary currents from the continental margin, specifically, the Gulf Stream in the western North Atlantic, the Kuroshio in the western North Pacific, the Brazil Current in the South Atlantic, and the Agulhas current where it retroflects south of South Africa. The Equatorial and Southern Ocean bands are also evident in the zonal mean {\viirsllcdifference} of the \underline{un}masked field shown in Fig.\,\ref{fig:viirs_minus_llc_ll}b; the zonal mean is roughly flat at -350 from $5^\circ$S to the Equator, rises rapidly poleward of these two points for about $5^\circ$ of latitude to -150 and then continues to increase approximately linearly, but more slowly, from there ($10^\circ$S and $5^\circ$N) to a value of zero at $30^\circ$N and $30^\circ$S. 

In the following we address each of the three regions primarily in the context of {\sst} galleries constructed from two small and geographically close regions (three colored rectangles shown in Fig.\,\ref{fig:viirsllmasked}), which exemplify characteristics of the differences that we find to be of interest.

\bigskip
\noindent{\bf Equatorial Band: \label{equatorial_band}}
In general, there are four possibilities for low values of {\viirsllcdifference} in the equatorial region ($15^\circ$S to $15^\circ$N):

\begin{enumerate}
    \item {\it The year simulated, 2012, is atypical, differing from the 2012--2020 mean.} This is very unlikely given the magnitude of the differences, -350 at the Equator, as well as the fact that the zonal mean of {\headtaildifference} is essentially flat at 0 equatorward of $50^\circ$ (Fig.\,\ref{fig:head_tail_ll}) suggesting little interannual variability.
    
    \item {\it Unresolved clouds in the {\viirs} cutouts.} Unresolved clouds tend to add structure to the cutouts; the `quieter' the field, the more significant the impact noise has on the structure, and the greater the decrease will be in {\tildell}. To address this potential problem, the 194 cutouts of the {\healpix} cell at $36^\circ$N, $112^\circ30'$W were examined for unresolved clouds. Thirty percent were found to be of high quality, 50\% were found to be significantly contaminated, and there was uncertainty as to how to classify the remaining 20\%. Calculating mean {\viirsllcdifference} for the high quality only cutouts increased the mean value by 43, small compared to the difference of $-350$ for this {\healpix} cell, 
    so this is not likely the primary explanation for the differences. 
    We further address this issue below in the context of 
    galleries of {\sst} fields.
   
    \item {\it Noise in the {\viirs} {\sst} fields.} Cutouts in this part of the ocean tend to have relatively high {\tildell} values (Fig.\,\ref{fig:viirs_and_llc_ll}a), corresponding to relatively less structure. This means that noise in the field, which is assumed to change slowly if at all with latitude, will become relatively more important, decreasing {\tildell}. Noise has been added to the {\llc4320} {\sst} fields in an attempt to address this but possibly not enough resulting in more negative values of {\viirsllcdifference}.
 
    \item {\it {\llc4320} does not reproduce the submesoscale-to-mesoscale structure well in the Equatorial regions.} There is a suggestion based on the examination of a small patch of this region, discussed below, that the model is missing structure in at least some parts of this region.
\end{enumerate}

\begin{figure}[ht]
\includegraphics[width=\linewidth]{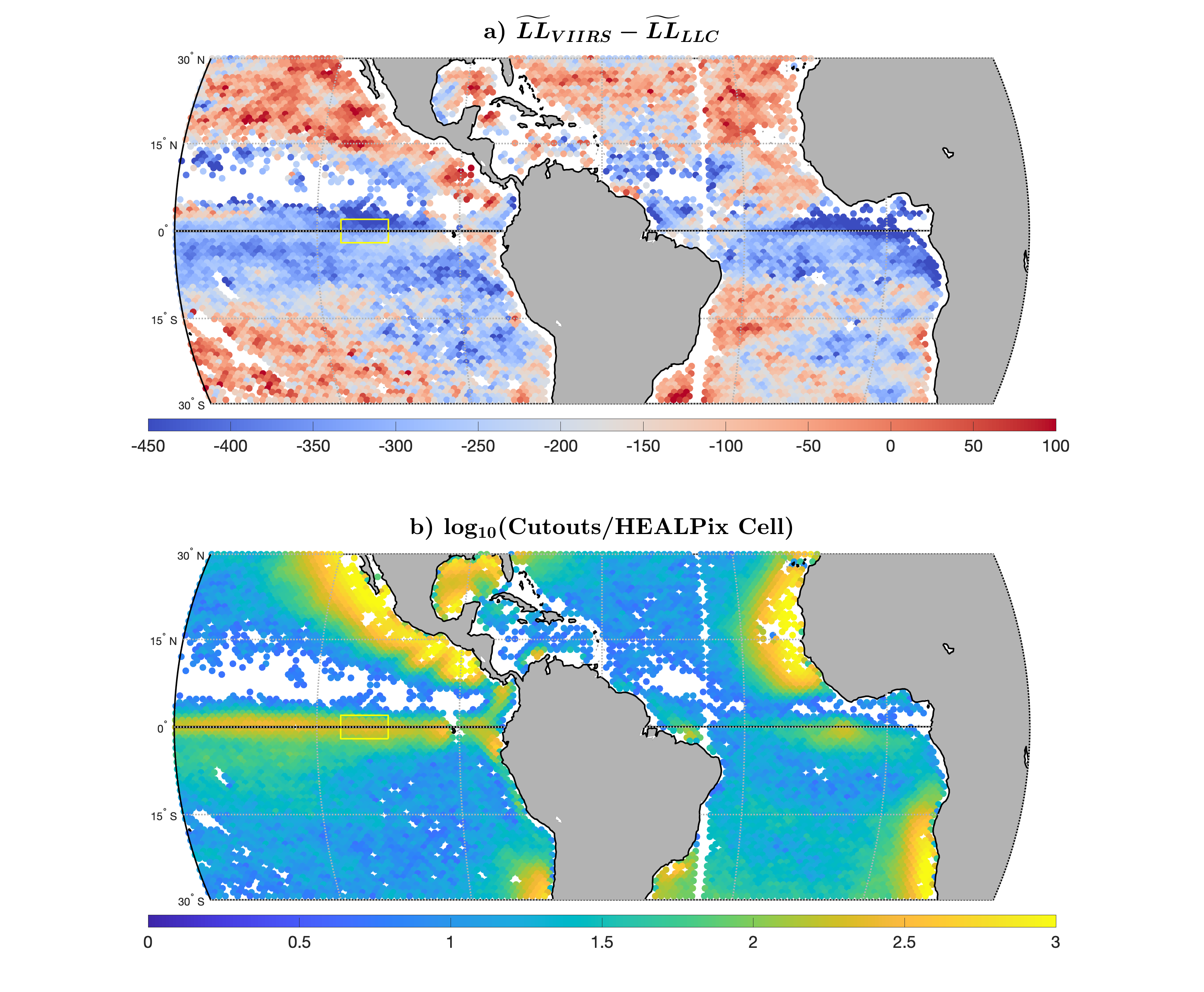}
\caption{(a) {\viirsllcdifference} for the Equatorial Pacific and Atlantic. (b) $log_{10}$ of the \# of cutouts/{\healpix} cell. Thick horizontal black line is the Equator. Yellow rectangle is the focus area. }
\label{fig:zoomed-eq-pacific}
\end{figure}

\begin{figure}
    \includegraphics[width=\linewidth]{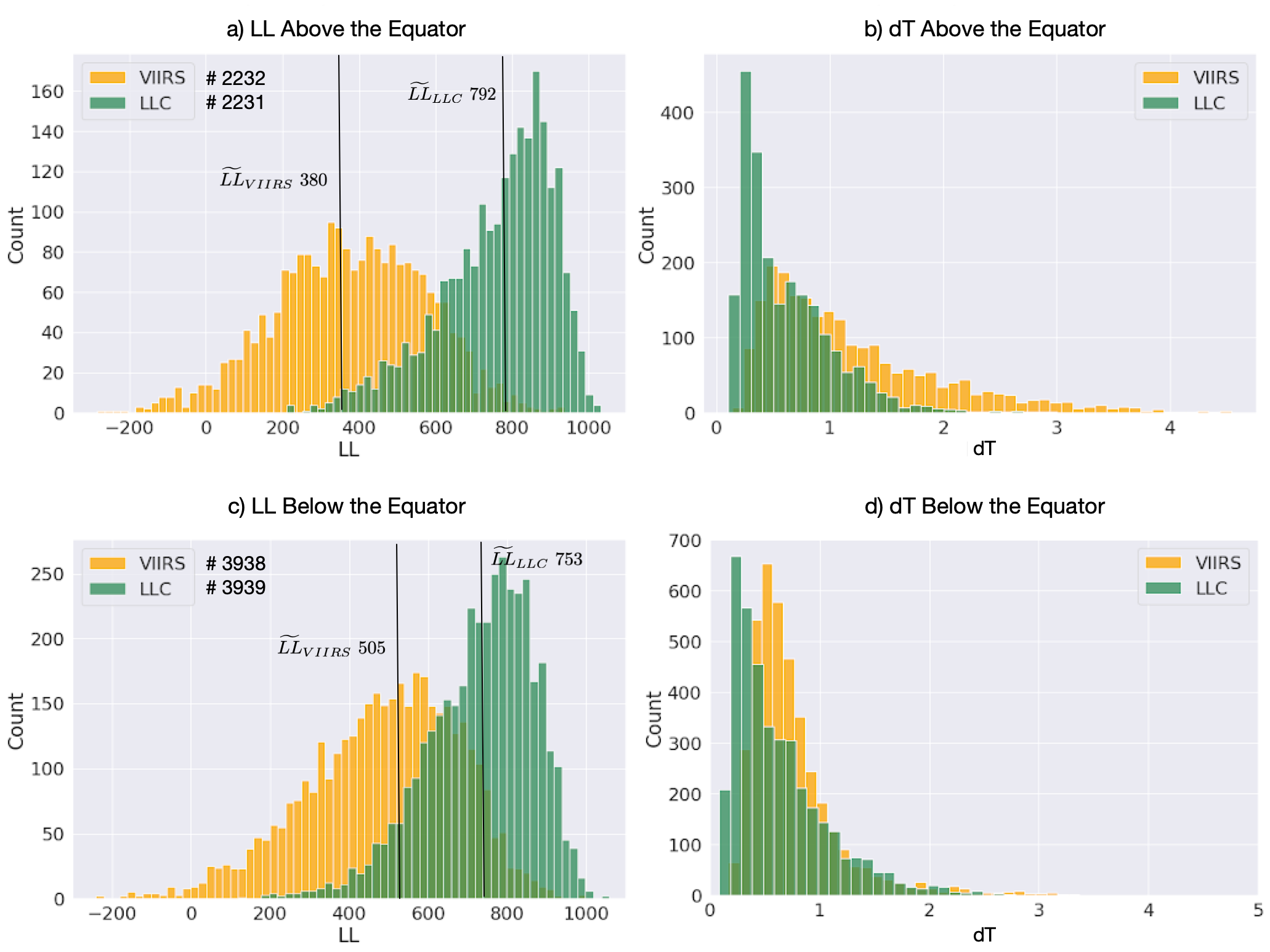} \hfill
    \caption{(a) Histograms of {\LL} for all {\viirs} (yellow) and {\llc4320} (green) cutouts above the Equator in the focus area. (b) Histograms of \dT\ for the same region as (a). (c) As for (a) except below the Equator. (d) As for (b) except below the Equator. The median values of the distributions are indicated for (a) and (c), as are the number of cutouts contributing to each. The number of cutouts apply to the corresponding \dT\ frames.}
    \label{fig:equator_histograms}
\end{figure}

The rectangular patch [($2^\circ$S, $2^\circ$N), ($105^\circ$W, $95^\circ$W)] west of the Galapagos Island in the equatorial Pacific (the yellow rectangle in Figs.\,\ref{fig:viirsllmasked} and \ref{fig:zoomed-eq-pacific}) stands out because of the significant step of {\viirsllcdifference} along the Equator. The {\tildell} distribution for the region in the rectangle above the Equator is provided in Fig.\,\ref{fig:equator_histograms}a and that below the Equator in Fig.\,\ref{fig:equator_histograms}c. Consistent with the geographic distributions of {\tildell} in Fig.\,\ref{fig:viirs_and_llc_ll}, 
the median {\tildellllc} value increases from south to north across the Equator 
while the median {\tildellviirs} value decreases, both contributing to a larger structural difference between {\viirs} cutouts and {\llc4320} cutouts to the north of the equator than that to the south. 
Histograms of \dT\ are provided because there is a correlation, although weak, between {\LL} and \dT, while \dT\ is a more readily understood measure of similarity and differences between {\viirs} and {\llc4320} {\sst} fields.  The distribution of 
$\dT_{LLC}$ is similar across the Equator, while that of {\viirs} shows a much longer tail consistent with more variability and lower {\LL} values.

\begin{figure}
    \includegraphics[width=\textwidth]{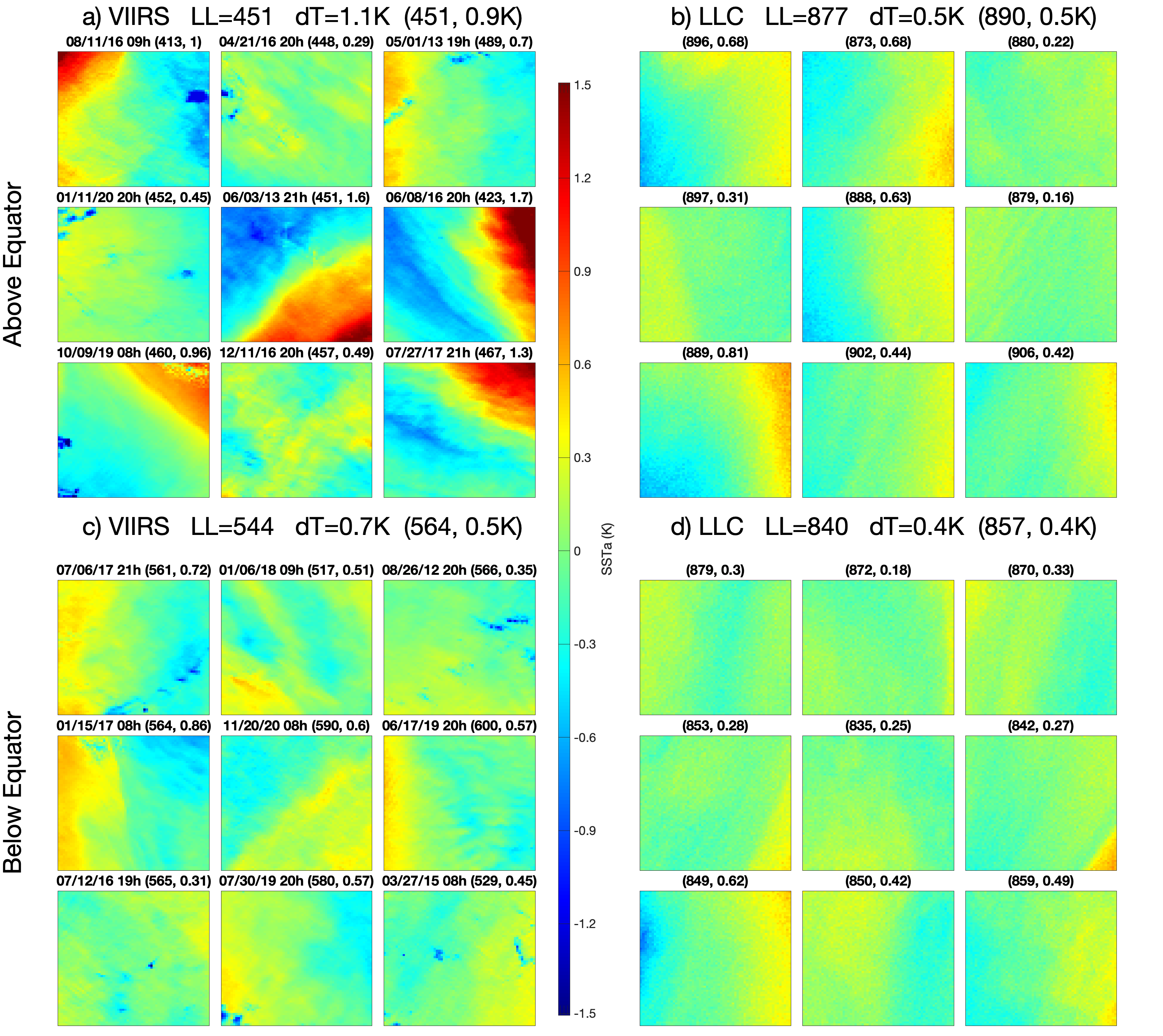} 
    \caption{(a) Gallery for {\viirs} cutouts above the Equator in the yellow rectangle of Fig.\,\ref{fig:zoomed-eq-pacific}a. (b) As in (a) but for {\llc4320} cutouts. (c) {\viirs} cutouts below the Equator in the yellow rectangle of Fig.\,\ref{fig:zoomed-eq-pacific}a. (d)  As in (c) but for {\llc4320} cutouts. 
    The mean {\tildell} and {\dT} for all cutouts in the given region (e.g., $0^\circ$ to $2^\circ$N, $105^\circ$ to $95^\circ$W for gallery a) follow the dataset name and the numbers following in parentheses are the mean {\tildell} and {\dT} of the cutouts in the gallery. The date and time of each {\viirs} cutout is shown above it and {\tildell} and {\dT} for that cutout follow in parentheses. The date of the corresponding {\llc4320} cutout (same position in the {\llc4320} gallery) is the same as that of the {\viirs} cutout. The time is the closest of 0 and 1200 to the {\viirs} time. It is evident that the gallery from the {\viirs} observations above the Equator shows greater structure than any of the other subsets. 
    }
    \label{fig:above_equator_galleries}
\end{figure}

Visual examination of the {\sst} fields supports the above conclusions. Consider {\sst} fields of the {\viirs} and {\llc4320} cutouts in the yellow rectangle of Fig.\,\ref{fig:zoomed-eq-pacific}, again separating them into those above the Equator and those below. Galleries of nine cutouts each are shown in Fig.\,\ref{fig:above_equator_galleries}. As previously noted each {\viirs} cutout was matched in space and day-of-year with an {\llc4320} cutout. To generate the galleries shown in Fig.\,\ref{fig:above_equator_galleries}, two sets of cutout pairs were formed. One set consisted of 40\% of all pairs in the region for which the {\viirs} {\LL} values were closest to the median {\viirs} {\LL} value for the region. The second set was similarly constructed except based on {\llc4320} {\LL} values. The intersection of the two sets defined the pool from which nine cutout pairs were randomly drawn. These pairs for the region above the equator, $0^\circ$ to $2^\circ$N and $105^\circ$ to $95^\circ$W, are shown in Figs.\,\ref{fig:above_equator_galleries}a and b. For each {\viirs} cutout in gallery a, its {\llc4320} partner is shown in the same location in gallery b. Similarly, Figs.\,\ref{fig:above_equator_galleries}c and d show pairs for the region below the Equator.
(Remember that because the {\llc4320} simulation is free running, there is no reason to expect the features in the pairs to be identical or even similar, it is the `structure' of the fields that is of interest.) %The cutouts for each gallery were randomly selected from those in the corresponding distribution with {\LL} falling within $\pm$ 20 of the distribution's median value. 
Visually, the {\sst} fields of the {\llc4320} cutouts are very similar in both regions showing very little structure with correspondingly large {\LL} values. The {\viirs} fields below the Equator show a little more structure than the {\llc4320} fields consistent with the smaller {\LL} values. By contrast, most of the {\viirs} fields above the Equator have significantly more structure than any of those in the other three galleries, with correspondingly lower {\LL} values. Also evident in the {\viirs} galleries are blemishes in the fields, scattered regions of colder temperatures. We believe these to be unresolved clouds, i.e., clouds not detected by the retrieval algorithm. Because they add structure to the fields, we believe that they decrease the {\LL} value of the cutout.
This would reduce the magnitude of {\viirsllcdifference} both above and below the Equator but is not likely to significantly impact the observed {$\widetilde{\rm LL}_{\rm VIIRS(above)} - \widetilde{\rm LL}_{\rm VIIRS(below)}$}.
%This would reduce the magnitude of {\viirsllcdifference} both above and below the Equator but it possible that differences in cloud contamination above and below the Equator are partially to blame for differences in the observed {$\widetilde{\rm LL}_{\rm VIIRS(above)} - \widetilde{\rm LL}_{\rm VIIRS(below)}$}.
Also, note that the distribution of the number of cutouts per {\healpix} cell is symmetric about the Equator in the area of interest (Fig.\,\ref{fig:zoomed-eq-pacific}b) suggesting that the contribution of clouds to corruption of the {\LL} values is also symmetric about the Equator in the rectangle of interest.

\bigskip
\noindent{\bf Southern Ocean: \label{southern_ocean}} We examine the same four possibilities for high {\viirsllcdifference} values in the Southern Ocean as we did for low values in the Equatorial region:

\begin{enumerate}
    \item {\it The year simulated, 2012, is atypical, differing from the 2012--2020 mean.} Possible but unlikely given the extent of the region covered by anomalously high differences and that there were few differences in this region for which {\headtaildifference} exceeded the $2\sigma$ threshold.
    
    \item {\it Unresolved clouds in the {\viirs} fields.} Very unlikely because clouds in the {\viirs} fields would tend to increase the structure, which would decrease {\tildell}. Hence {\viirsllcdifference} would become more positive than if clouds are not present in the {\viirs} fields, rendering the difference more anomalous, not less so.
   
    \item {\it Noise in the {\viirs} {\sst} fields.} Unlikely because the geophysical variability of {\sst} in these regions tends to overwhelm noise in the {\viirs} cutouts. Furthermore, noise in the {\viirs} cutouts would tend to reduce the associated {\tildell} rendering the differences between uncontaminated {\tildell} {\viirs} values and those obtained from the LLC4320 simulation even larger.
 
    \item {\it {\llc4320} does not reproduce the submesoscale-to-mesoscale structure well in the Southern Ocean.} There is a suggestion based on the examination of small patches of this region, which we discuss in more detail below, that the model has more structure in this region than is recorded in actual observations. This indicates that the mixed layer or energy dissipation and stirring due to subgrid-scale physics are not represented with sufficient accuracy in these regions.
\end{enumerate}

\begin{figure}
    \includegraphics[width=0.9\textwidth]{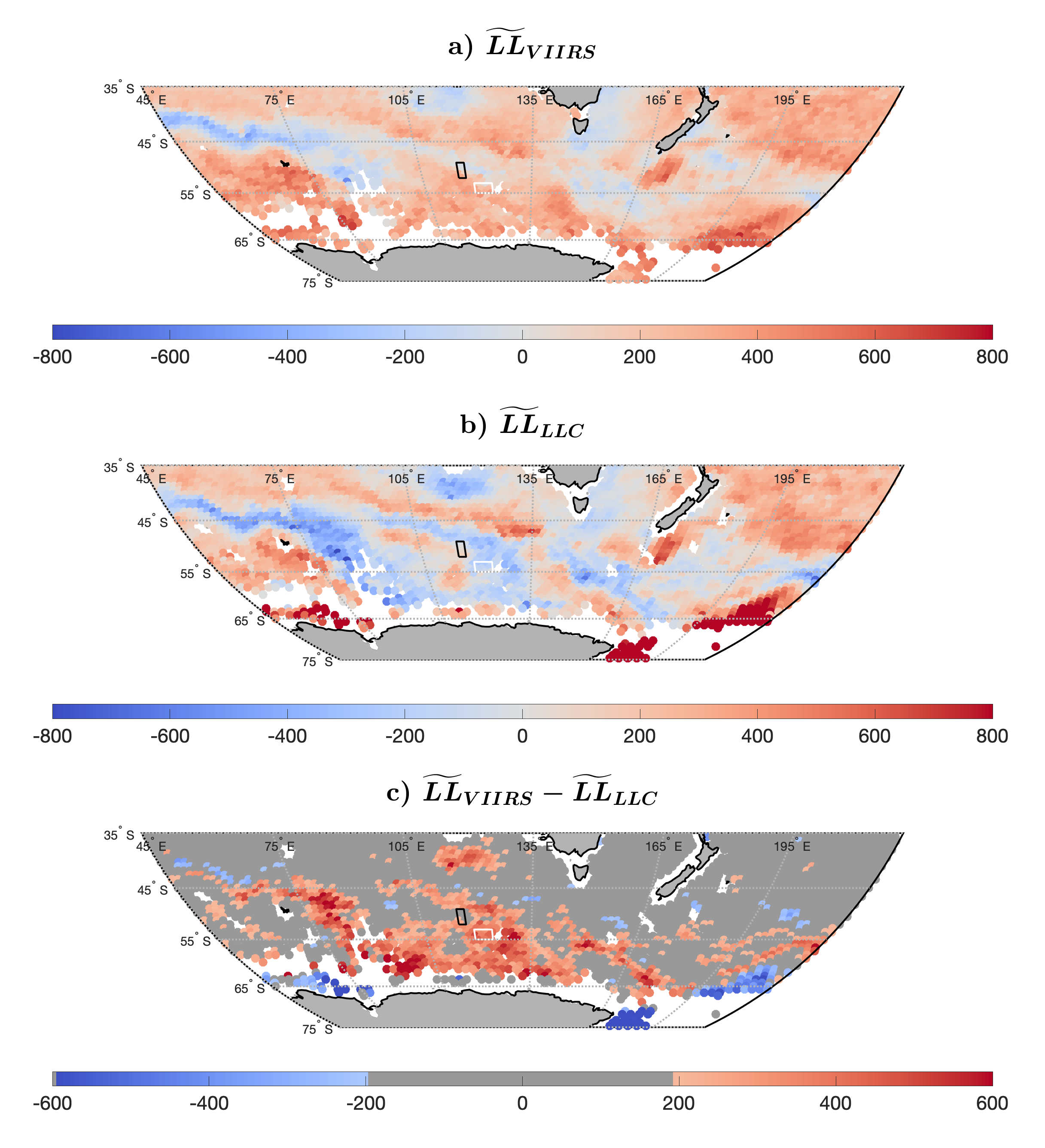} 
    \caption{(a) {\tildellviirs} for the focus area of the Southern Ocean, (b) {\tildellllc}, and (c) masked {\viirsllcdifference}. 
    The black rectangle ($\sim$116$^\circ$W, -50$^\circ$S) 
    indicates a region of agreement, $\overline{LL_{VIIRS} - LL_{LLC}}=-63$. 
    The white rectangle ($\sim$120$^\circ$W, -54$^\circ$S) 
    is an anomalous region with $\overline{LL_{VIIRS} - LL_{LLC}}=335$.} 
    \label{fig:ACC-zoom}
\end{figure}

In Fig. \ref{fig:ACC-zoom}c, we replot the masked $\widetilde{LL}_{VIIRS} - \widetilde{LL}_{LLC}$ field of Fig.\,\ref{fig:correctedllc_with_bathymetry} for the Southern Ocean south of Australia with {\tildellviirs} and {\tildellllc} for the same region in the upper two panels. The band of low {\tildellviirs} and {\tildellllc} values from $45\ {\rm to}\ 70^\circ$E at about $41^\circ$S corresponds to the significant structure in the field associated with the {\acc}. (Note that this band originates south of South Africa where the Agulhas retroflection joins the {\acc}, the band of negative values of {\viirsllcdifference} in Fig.\,\ref{fig:zoomed_correctedfield}.) It appears that the {\acc} as modeled by {\llc4320} for 2012 is slightly to the south of the {\viirs} {\acc} for 2012--2020; the positive values of {\viirsllcdifference} south of the band and the corresponding negative values to the north of the band. Although a small shift, the fact that the width of the bands for both {\viirs} and {\llc4320} are virtually identical suggests that the modeled {\acc} is a bit to the south of the envelope of paths in this period, i.e., the slight shift may be significant in the context of the modeled processes. Apart from the slight shift to the south, the model appears to have reproduced the current quite well in this region. East of about $70^\circ$E {\tildell} for the modeled {\acc} is substantially more negative than that of the observed values, resulting in the anomalous {\viirsllcdifference} values in this region. Interestingly, the {\viirs} field shows a positive band north and south of the northern branch of the stream as does the {\llc4320} field. In fact, the general pattern of the {\tildellllc} has the same general shape as that of {\tildellviirs}. 
Therefore it appears that the model has the correct general structure for the flow in this region but, as we now emphasize,
is too energetic. 

\begin{figure}
    \includegraphics[width=\textwidth]{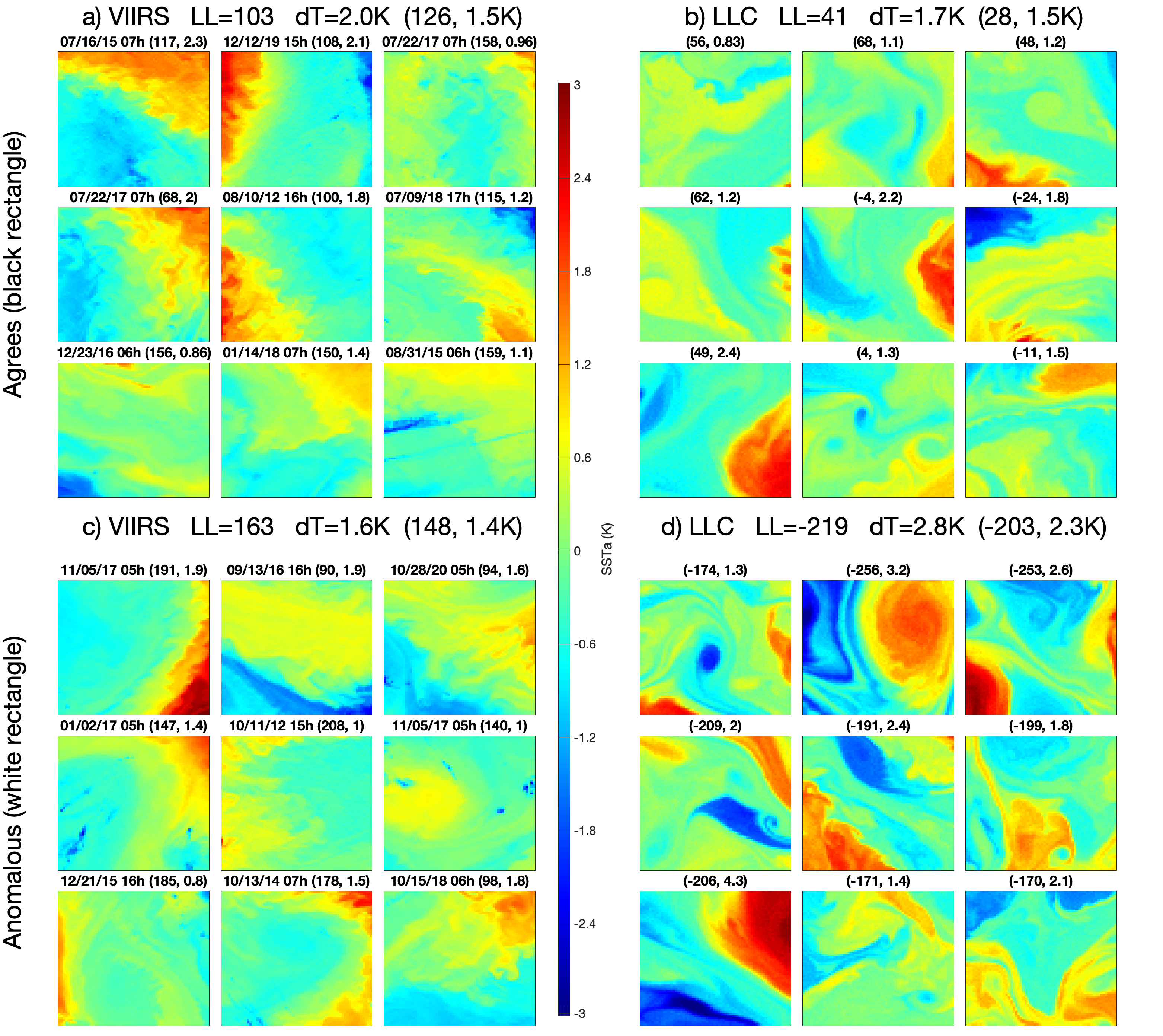} 
    \caption{(a) Gallery of 9 randomly selected {\viirs} cutouts within the black rectangle of Fig.\,\ref{fig:ACC-zoom},
    the region of `agreement'. 
    (b) Similarly for {\llc4320} cutouts in the same region of `agreement'. 
    (c) {\viirs} cutouts in the anomalous region, the white rectangle in Fig.\,\ref{fig:ACC-zoom}. 
    (d) Same as (c) but for {\llc4320} cutouts. Dates, times, {\LL} and {\dT} as in Fig.~\ref{fig:above_equator_galleries}.}
    \label{fig:acc_galleries}
\end{figure}

Galleries of {\ssta} cutouts in the black rectangles of Fig.\,\ref{fig:ACC-zoom} (a region for which the model {\tildell} values are in general agreement with the {\viirs} values) are shown in Fig.\,\ref{fig:acc_galleries}a for {\viirs} and \ref{fig:acc_galleries}b for {\llc4320}. Galleries for the anomalous region, the white rectangles in Fig.\,\ref{fig:ACC-zoom}, are shown in Fig.\,\ref{fig:acc_galleries}c for {\viirs} and \ref{fig:acc_galleries}d for {\llc4320}. Cutouts for these galleries were randomly selected as described for the generation of galleries for the equatorial region (Fig.\,\ref{fig:above_equator_galleries}). The anomalous behavior is clear in the lower two panels; the {\llc4320} fields are much bolder with substantially lower $\overline{LL}$ values and larger \dT\ values. The {\llc4320} field is clearly more energetic. By contrast, {\llc4320} cutouts in the region for which there appears to be agreement are more similar to {\viirs} cutouts. 

\bigskip
\noindent{\bf Gulf Stream: \label{gulf_stream}} The conclusions for the four possibilities of high values in the Gulf Stream are similar to those for the Southern Ocean;

\begin{enumerate}
    \item {\it The year simulated, 2012, is atypical, differing from the 2012--2020 mean.} Very unlikely given, as will be shown below, that the modeled Gulf Stream is south of the most extreme southern positions of paths of the Gulf Stream from a number of observational sources. 
    
    \item {\it Unresolved clouds in the {\viirs} fields.} Very unlikely because clouds in the {\viirs} fields would tend to increase the structure, i.e., decrease {\tildell}. Hence cloud-free fields would tend to increase {\viirsllcdifference}, rendering it more anomalous.
   
    \item {\it Noise in the {\viirs} {\sst} fields.} Unlikely because the geophysical variability of {\sst} in these regions overwhelms noise in the {\viirs} fields but, if it were to contribute, it would again increase the {\viirsllcdifference}, rendering it more anomalous.
 
    \item {\it {\llc4320} does not reproduce the submesoscale-to-mesoscale structure well in the Gulf Stream region.} As will be shown below, this is likely the cause of the differences but, unlike the differences in the Southern Ocean, we believe that these differences are due to premature separation of the Gulf Stream from the continental margin, i.e., that the Gulf Stream is in the wrong place as opposed to it being in the correct location but too energetic as appears to the case in the \ac{ACC} south of Australia. 
\end{enumerate}

\begin{figure}[ht]
\includegraphics[width=\linewidth]{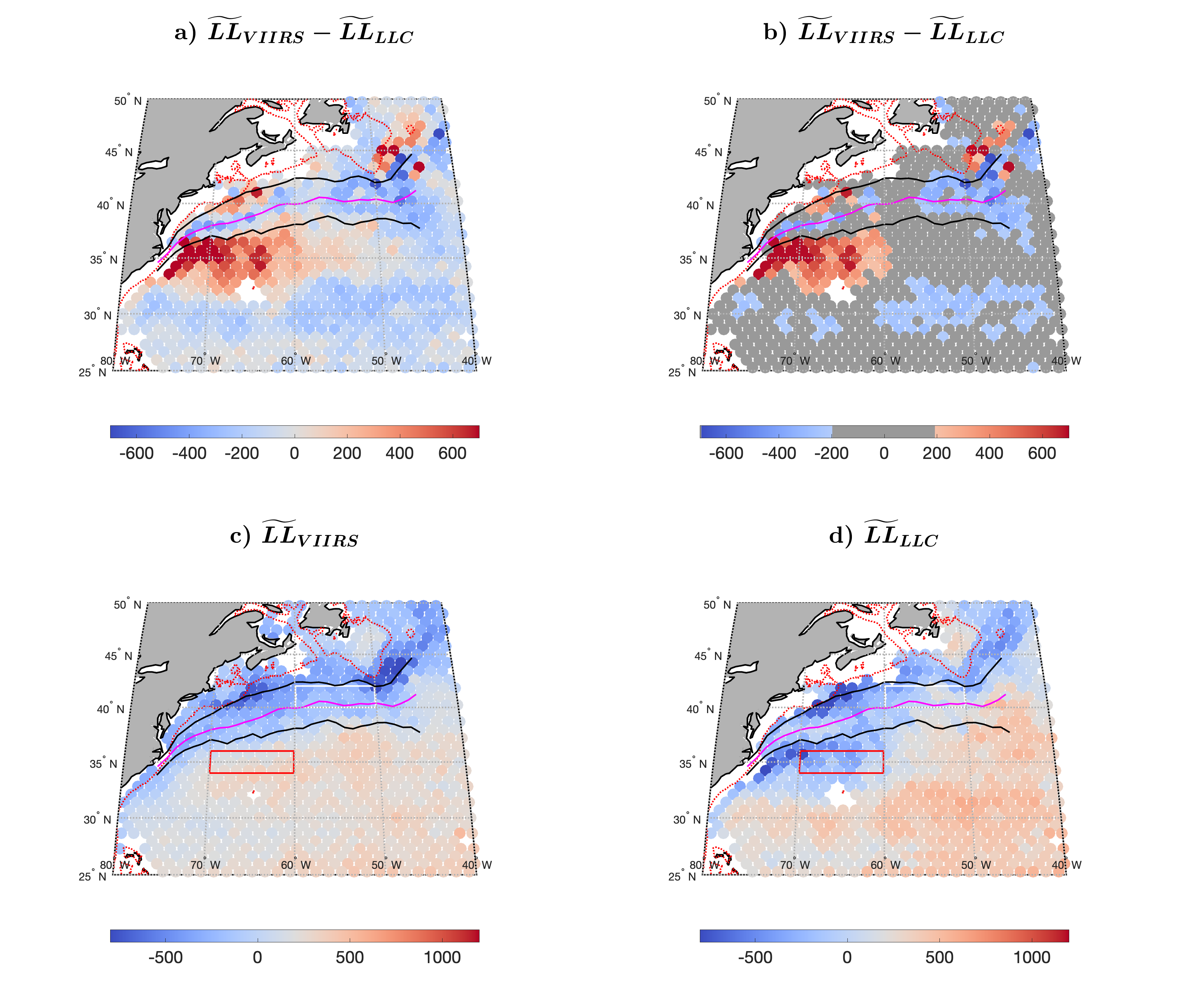}
\caption{(a) {\viirsllcdifference} for the Gulf Stream region after separation from the continental margin. (b) Masked {\viirsllcdifference} for same area. (c) {\tildellviirs}. (d) {\tildellllc}. Thick magenta line shows the mean Gulf Stream path digitized from 2-day \ac{AVHRR} {\sst} composites for 1982--1999. Thick black lines show the envelope containing 98\% of these paths for the same period, 1\% beyond the limits on each side of the envelope. Dotted red line is the 200-m isobath. Red and white rectangles show the focus areas from which galleries of cutouts are selected for Fig.\,\ref{fig:four_galleries}.}
\label{fig:gulf_stream-Peters-version}
\end{figure}

Figure \ref{fig:gulf_stream-Peters-version}a shows {\viirsllcdifference} in the Gulf Stream region downstream of the point at which it separates from the continental margin. Figure \ref{fig:gulf_stream-Peters-version}b shows the same data but masked, showing only the {\healpix} cells with values exceeding the thresholds identified in the Appendix, $\pm$\sigmathreshold. Also shown in these plots is the mean path of the Gulf Stream (magenta line) and its northern and southern extent (black lines). These were determined from manual digitizations of the path of the stream---defined as the maximum cross-stream {\sst} gradient in the vicinity of the stream---in warmest-pixel composites of all \ac{AVHRR} 1-km  {\sst} fields in contiguous 2-day intervals \citep{Lee1996a}. The mean path of the stream was determined by averaging, over all 2-day composites between 1982 and 1999, the point at which these paths intersected integral degrees of longitude. The northern and southern extents are the latitudes for which 99\% of the paths lie to the south---the northern extent---and 99\% lie to the north---southern extent---for 1982--1999. Large positive differences south of the southern extreme suggest that the {\llc4320} output contains more structure in its cutouts than {\viirs}, whereas {\viirs} fields show more structure within the bounds of the northern and southern extremes.

The large patch of positive values west of $60^\circ$W corresponds to the premature separation of the modeled Gulf Stream from the continental margin reported by \citet{cornillon-2018} based on the mean path and extreme envelope of paths shown in Figs.~\ref{fig:gulf_stream-Peters-version} and on the \ac{OSCAR} surface currents (https://podaac.jpl.nasa.gov/ datasetlist, search Keywords: Oceans/Ocean Circulation, Projects: OSCAR) for 2012. Simply put, the modeled Gulf Stream is found some 250\,km to the south of the mean observed stream at $70^\circ$W and approximately 100\,km to the south of the southern extreme observed between 1982 and 1986.

The cause of the positive and negative anomalous values of Fig.\,\ref{fig:gulf_stream-Peters-version}b become more clear from the individual plots of {\tildellviirs} and {\tildellllc}, Fig.\,\ref{fig:gulf_stream-Peters-version}c and d, respectively. The most negative values of {\tildell} are seen in the satellite-derived fields along the edge of the continental slope and, in particular, south of Georges Bank, south of the eastern side of the Gulf of Maine and south and east of the Grand Banks. (Note that the relatively sharp gradient in {\tildell} values follows the 200-m isobath, shown as dotted red lines in Fig.\,\ref{fig:gulf_stream-Peters-version}.) Values remain low to the southern extreme of the Gulf Stream. Recall, that these {\healpix} values are medians obtained from all cutouts in the 8+year interval. During this period the Gulf Stream meanders in the envelope with regions of significant structure at some times and regions of less structure at others, resulting in less structure on the average than is found near the shelf break in very active regions that are topographically constrained---south of Georges Bank and south and east of the Grand Banks. The {\llc4320} output also shows the most negative values south of Georges Bank but less so south of the Grand Banks. This is likely associated with the premature separation of the stream from the continental margin, the negative values of {\LL} between $65^\circ$ and $75^\circ$W and south of about $36^\circ30'{N}$. Two aspects of interest associated with the model field after separating are: 1) the relatively smaller width (meridional extent) of the region covered by the Gulf Stream immediately after separation when compared with the broader distribution associated with the {\viirs} data and the rapid increase in {\LL} values---decrease in structure---at approximately $62^\circ$; the stream appears to die at that point and 2) the positive {\LL} values east of $60^\circ$W and south of $33^\circ$N, the cause of the statistically significant negative differences when compared with the {\viirs} results. The former may be due to the fact that model simulation is only for one year while the {\viirs} data cover 8+ years. The reasons for the rapid die-off of the stream and the relatively quieter (larger values of {\LL}) east of $62^\circ$ are not obvious.

\begin{figure}
    \includegraphics[width=\textwidth]{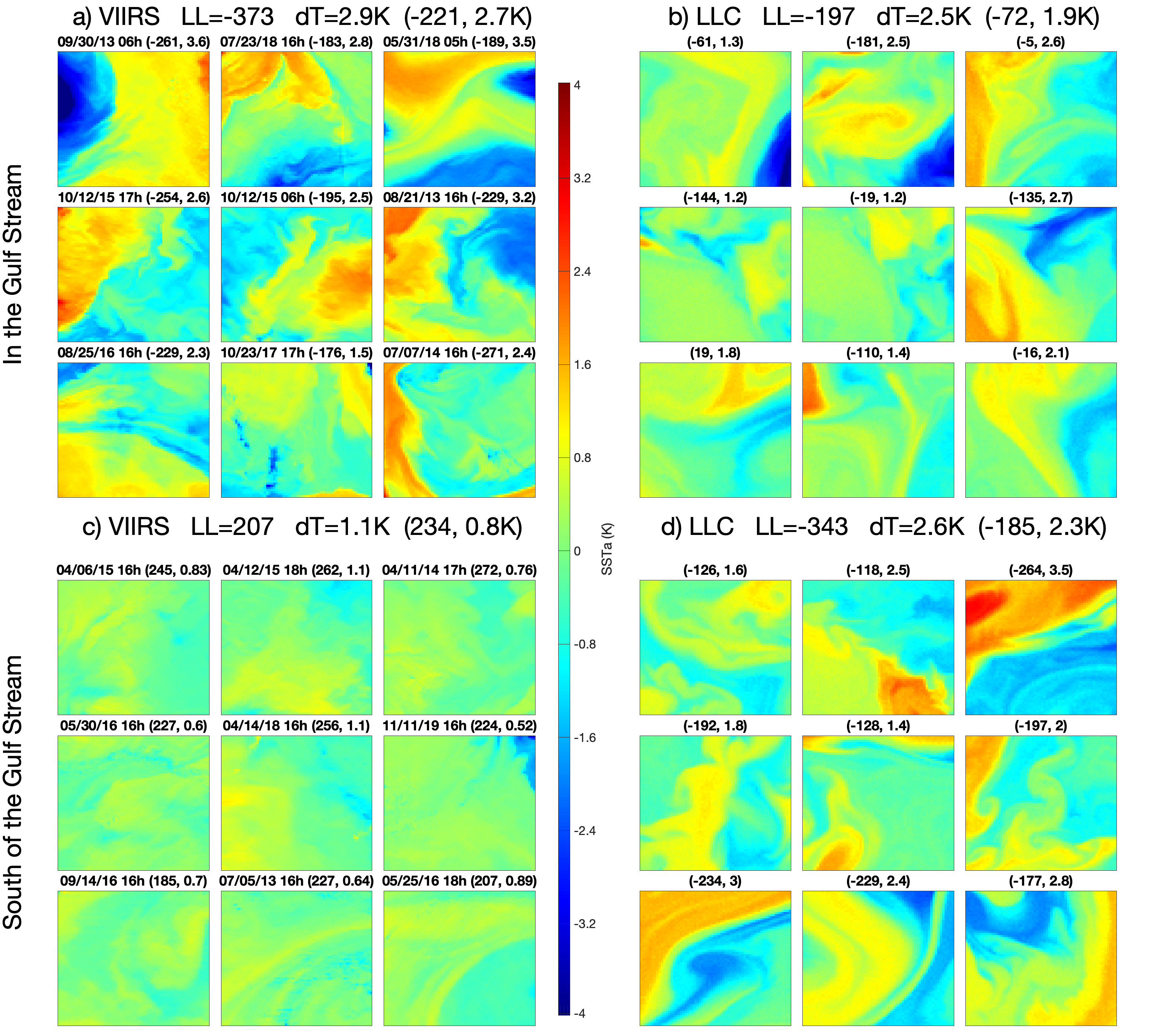} 
    % \\[\smallskipamount]
    % \includegraphics[width=\textwidth, trim= 0 0 4 0,clip]{Figures/real gulf stream.png}
    \caption{(a) Gallery for {\viirs} cutouts in red rectangle of Fig.\,\ref{fig:gulf_stream-Peters-version}c. (b) Same for {\llc4320} cutouts in red rectangle of Fig.\,\ref{fig:gulf_stream-Peters-version}d. (c) {\viirs} cutouts in white rectangle of Fig.\,\ref{fig:gulf_stream-Peters-version}c. (d) {\llc4320} cutouts in white rectangle of Fig.\,\ref{fig:gulf_stream-Peters-version}d. Dates, times, {\LL} and {\dT} as in Fig.~\ref{fig:above_equator_galleries}.}
    \label{fig:four_galleries}
\end{figure}

Next, we examine {\viirs} and {\llc4320} cutouts inside the Gulf Stream envelope (the white rectangles in Figs.~\ref{fig:gulf_stream-Peters-version}c and d at [($40^\circ$N, $42^\circ$N), ($60^\circ$W, $50^\circ$W)]) and outside the observed Gulf Stream in the region of anomalously low {\llc4320} values associated with the premature separation of the Gulf Stream (the red rectangles in Figs.~\ref{fig:gulf_stream-Peters-version}c and d at [($34^\circ$N, $36^\circ$N), ($70^\circ$W, $60^\circ$W)]). Galleries of nine cutouts each for both {\viirs} and {\llc4320} for both regions are shown in  Fig.\,\ref{fig:four_galleries}. Cutouts for these galleries were randomly selected as described for the generation of galleries for the equatorial region (Fig.\,\ref{fig:above_equator_galleries}). The characteristics of the {\viirs} cutouts outside of the Gulf Stream (Fig.\,\ref{fig:four_galleries}c) differ substantially from those of the other three galleries as does the mean {\LL}. {\viirs} cutouts within the stream (Fig.\,\ref{fig:four_galleries}a) are similar to those in the {\llc4320} gallery of cutouts south of the Gulf Stream (Fig.\,\ref{fig:four_galleries}d), consistent the suggestion that the modeled stream separates prematurely from the continental margin. %except for the short wavelength features which are not evident in the {\llc4320} cutouts, e.g., the apparent instabilities on the southwestern side of the warm region in the (1,3) cutout of  Fig.\,\ref{fig:four_galleries}c, which are separated by $\sim$25\,km. We suspect that the lack of such small scale features in the {\llc4320} cutouts relates to the actual resolution of the model, which is coarser than the grid on which the model operates. 
The structure of {\llc4320} cutouts in the Gulf Stream (Fig.\,\ref{fig:four_galleries}b) lies between that of {\viirs} cutouts in the stream (Fig.\,\ref{fig:four_galleries}a) and {\llc4320} cutouts south of the stream (Fig.\,\ref{fig:four_galleries}d) as do the {\LL} values. This is consistent with the modeled stream being to the south of the observed stream. 

% %%%%%%%%%%%%%%%%%%%%%%%%%%%%%%%%%%%%%%%%%%%%%%%%%%%%%%%%%%%%%%%%%%%%%%%%%%
\conclusions

In this manuscript we set out to confront outputs from the well-adopted {\llc4320} simulation with a large dataset of global observations.
Specifically, we have focused on the submesoscale dynamics traced by
\sst, an observable with decades of global coverage provided by a series of 
sensors on remote sensing satellites.
This manuscript used \ac{L2} data from \viirs, restricted to nearly-clear ($\ge 98\%$ cloud free) cutouts with dimension
$\sim$150$\times$150\,km$^2$ selected across the ocean.

Our approach for quantitative comparison between data and model is
unconventional.  We trained a deep-learning \acf{PAE} on the \viirs\ 
data to learn the distribution of \ssta\ patterns
observed in the ocean and then applied this \ac{PAE} to  
geographically and seasonally matched \ssta\ cutouts from
the {\llc4320} model.
An advantage of this approach is that it is intentionally unsupervised;
the network learned from the data the features most characteristic
of ocean dynamics traced by \ssta.
On the flip side, the results---especially any differences between
data and model---are more difficult to interpret.  
The \LL\ metric calculated from the \ac{PAE} is known to 
correlate with \DT\ and other physical measures of \ssta\, 
yet with significant scatter \citep{prochaska2021deep}.
And uncertainties are not inherently calculated;
instead we have estimated them by
applying \ulmo\ to two independent subsets of \viirs\ data. 

Proceeding in this manner, we found that, in general, the distribution
of \ssta\ patterns present in the \viirs\ observations are 
well-predicted by the {\llc4320} model
(e.g., Fig.\,\ref{fig:LL_hist}).
Globally, the medians of the \LL\ distributions from {\viirs} and 
{\llc4320} agree within $2\sigma$ for 65\%\ of the ocean (Fig.\,\ref{fig:viirsllmasked}).
However, there is a modest but significant and latitude-dependent
offset between data and model
with the latter exhibiting less structure in the \ssta\ cutouts
near the Equator and greater structure towards the poles.
%{\color{blue}[are we certain this is real, i.e., a "flaw" in the model? if so, say so} {\color{red}I'm not certain that this is a flaw in the model so would prefer not to say so.}]
After correcting for this latitude-dependent offset, we
find that the model frequently recovers mesoscale features
imprinted in the \LL\ distributions and seen in the {\tildellviirs} field.
This includes the reproduction of detailed
mesoscale dynamics often forced by deep bathymetric features. We emphasize here that the {\viirs}--{\llc4320} comparison is being performed on spatial scales of $\mathcal{O}(50$\,km) and less and that it is changes in the structure at these scales that is informing the large scale patterns observed, i.e., the submesoscale structure of cutouts  appears to be tied to larger scale processes. 
One may conclude that the {\llc4320} model has captured salient
mesoscale dynamics across the majority of the ocean.

There are, however, a few notable exceptions.  One of these is the
location of the Gulf Stream, a previously known failure of
the LLC4320 simulation \citep{cornillon-2018}. Giving confidence to the approach taken in this manuscript to evaluate the performance of the \llc4320 simulation is the fact that a known region of concern is clearly identified as problematic.

A more subtle difference occurs at the Equator.
We have shown from the \viirs\ data that the structure in the \ssta\
cutouts just north of the Equator exceeds that of its southern
counterpart (Fig.\,\ref{fig:above_equator_galleries}).
%{\color{blue}We believe that this arises from differences in wind and ocean circulation between the two hemispheres stemming from  our rotating planet \citep{CITE}. [True??]} {\color{red}As I mentioned in somewhere above, a fair amount of work has been done here and it is known that some models do not reproduce {\sst} well in this region. However, I don't know much about equatorial dynamics and would have to do a literature review to sort it out. I'd prefer not to do that right now.}
This difference in \ssta\ in \viirs, however,  is not
reproduced by the {\llc4320} simulation and there is no reason to believe that cross-Equator differences in the {\viirs} data are spurious.

Third, we highlighted inconsistencies between the {\llc4320} model
and \viirs\ observations in the \ac{ACC}, where the former
frequently exhibits a higher degree of structure in \ssta\ 
and, presumably, more energetic surface currents. We attribute this increased structure to a misrepresentation of the mixed layer and of subgrid-scale processes that are responsible for energy dissipation and stirring. %{\color{blue}[why??]} {\color{red}My sense is that it has something to do with the modeled \ac{ACC} being slightly off track but I really don't know why nor how this would result in more structure..} 

We hope that our analysis will inspire similar investigations of both
global and regional models.
With the construction of large, well-curated datasets,
as we have done with \viirs, one may construct
a series of tests.
The construction of such a dataset requires intentional
decisions on how to extract and preprocess cutouts for
direct comparison to model outputs.
Furthermore, the data volume $\mathcal{O}$(100\,Tb)
is sufficiently large to require best practices with storage,
databasing, and computing.
The authors provide their code (including workflow)
with the manuscript and encourage discussion
with parties interested in building their own similar analyses.

We also wish to emphasize several of the weaknesses of our
methodology and identify paths for improvement in future work.
First, and perhaps foremost, we have not accounted for the 
mismatch in effective spatial resolution between model and 
observations.  The pixelization of the \ac{L2} \viirs\ 
product is $\sim$750\,m at nadir hence can resolve features in the few kilometer range.
The {\llc4320} simulation, meanwhile, has a finest cell size of $\sim$1\,km
but the formulation is not expected to properly resolve features 
on scales less than $\mathcal{O}(10)$\,km \citep[see, e.g.,][]{Su2018}.
We considered smoothing (i.e., degrading) the \viirs\ data
to better match the model outputs but were not confident that we
could do so with high accuracy.
Furthermore, the \ac{PAE} itself is effectively smoothing 
the data by passing each cutout through a 512-dimension
bottleneck \citep[e.g., see Fig.\,3 of][]{prochaska2021deep}.

Related to the above discussion, the initial analysis 
ignored latitude dependence in \ssta\ structure, despite the
predicted and observed dynamical differences driven by geophysical
fluid dynamics.  Further work might, for example, vary the size of the cutouts
proportional to the Rossby Radius of Deformation.
Similarly, if we were to expand the cutout size to larger
scales to better assess mesoscale features, 
it may become necessary to match the orientation of the data
(here dictated by the satellite path) with the model
(fixed with rows/columns parallel to longitude/latitude).

Another weakness of our implementation is the lack of
any error estimation from the \ac{PAE} outputs (i.e., for 
individual \LL\ values).
This is a general weakness of deep learning algorithms
\citep[but see Bayesian Neural Nets; e.g.\ ][]{bayes_NN}.
Therefore, we approached uncertainty estimation using
an empirical estimate generated from subsets of
the data (see the Appendix).
While effective, it is approximate and relies on the central limit
theorem to assume a Gaussian deviate.
Related, the \LL\ metric of \ulmo\ has no intrinsically
physical, mathematical or statistical (despite the name) meaning!
Future work focused on comparisons of \ssta\ or other
patterns may consider the scattering transform 
\citep{mallat2012,cheng2021},
which has sound mathematical underpinning and may allow
for proper statistical tests.

Last, but far from least, are the significant ``blemishes''
in the data that are absent in the model outputs.
Foremost are clouds.
The mitigation for clouds adopted here was to 
(1) limit to cutouts with fewer than 2\%\ of the pixels 
masked by the retrieval algorithm \citep{rs14143476} and (2) inpaint these masked pixels.
The latter step was required for the \ac{PAE} and is, in general,
required for convolutional neural nets, which expect `complete' fields.
In our exploration of the cutouts, however, we identified 
a high incidence of clouds that were not masked in the \viirs\ data.
These ranged from minor blemishes in otherwise uniform fields 
where the clouds generate non-negligible structure (especially evident in Figs.\,\ref{fig:above_equator_galleries}a and c) to, in a few cases, corruption of the entire field in the cutout.
Another negative consequence, perhaps the most serious,
is the terrific reduction of potential data and the resultant
geographic biases of the dataset that follow from the
98\%\ clear criterion (Fig.\,\ref{fig:viirs_geo_p}).
To the greatest extent possible, future work must continue to
identify and mitigate clouds;  our own efforts are well underway.

As we conclude, we emphasize that perhaps the greatest value
of this manuscript was the construction and now dissemination
of the large dataset of cutouts for comparison with ocean models as well as with other satellite-derived {\sst} datasets.
This includes the software to generate and analyze them.
All of these products are publicly available as described below.

\codedataavailability{
All of the data generated and analyzed in this manuscript is publicly
available as {\it parquet} tables and {\it hdf5} files at Dryad (LINK TO APPEAR).
The code developed throughout the project is provided at 
https://doi.org/10.5281/zenodo.7763845 
(doi: 10.5281/zenodo.7763845) or at https://github.com/AI-for-Ocean-Science/ulmo.
} %% use this section when having data sets and software code available

\appendix

%\section{Correcting for Large {\llc4320} Differences \label{appendix:correcting_llc}}
%
%In order to highlight the similarities in the {\tildell} fields we remove the large regional differences between them. This is done by averaging {\tildellviirs} and {\tildellllc} over 100 sequential values based on the {\healpix} index. Since the vector of {\healpix} cells is arranged geographically, the indicies over which the average is performed tend to correspond to a relatively tight geographical region. {\tildellviirs} (black) and {\tildellllc} (cyan) are shown in Fig.~\ref{fig:correctingllc}. A `corrected' {\tildellllc} is then determined from:
%
%$$ \widetilde{LL}'_{LLC_i} =  \widetilde{LL}_{LLC_i} - \frac{1}{100}\displaystyle\sum_{j\in B}{(\widetilde{LL}}_{LLC_j} - \widetilde{LL}_{VIIRS_j}) $$
%where $i \in $ all {\healpix} cells with at least 5 cutouts in the {\viirs} dataset and 5 cutouts in the {\llc4320} dataset and 
%$$ B: \left[\lfloor \frac{i-1}{100} \rfloor*100+1, \lfloor \frac{i-1}{100} \rfloor*100+100\right] $$
%`Corrected' {\tildellllc} values are shown in red in Fig.~\ref{fig:correctingllc}. 
%
% \begin{figure}[ht]
% \includegraphics[width=0.9\linewidth]{Figures/pcc_1x1_correcting_LLC_to_mean_VIIRS.png}
% \caption{ {\tildell} versus {\healpix} cell index. Black {\tildellviirs}, cyan {\tildellllc} and red corrected {\tildellllc}.
% }
% \label{fig:correctingllc}
% \end{figure}

\section{{\healpix} Uncertainty 
\label{appendix:healpix_uncertainty}}

\begin{figure}[ht]
\includegraphics[width=\linewidth]{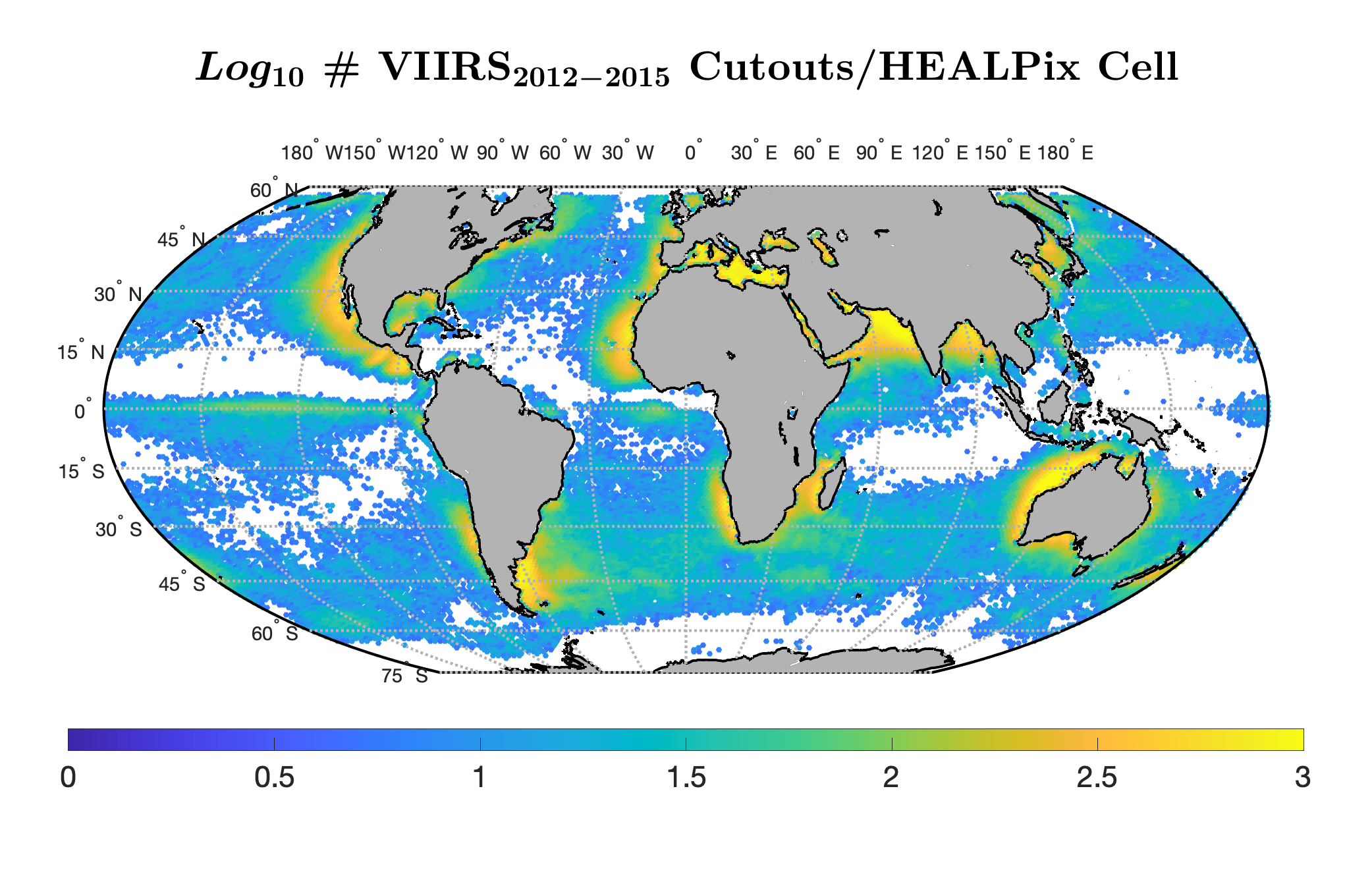}
\caption{Number of {\viirs} cutouts per {\healpix} cell for the period 2012--2015. {\healpix} cells with less than five cutouts for 2012--2015 or less than five cutouts for 2017--2020 are shown in white.
}
\label{fig:head_tail_counts}
\end{figure}

A question, which arises naturally in the context of Fig.\,\ref{fig:viirs_minus_llc_ll}, is what constitutes a 
statistically significant difference in {\tildell} between the model output and the {\viirs} fields. 
To address this, the {\viirs} dataset is divided into two 4-year segments, 1 February 2012 through 31 January 2016 (referred to as 2012--2015 hereafter) and 1 January 2017 through 31 December 2020 (2017--2020). Subsequently, {\tildell} is calculated for each {\healpix} cell for each of the two periods. Figure \ref{fig:head_tail_counts} shows the distribution of cutouts for the first of the two periods. Because the periods for which these data are being calculated are substantially shorter than that of the dataset from which they are drawn (Fig.\,\ref{fig:viirs_geo_p}), the number of {\healpix} cells with less than 5 cutouts (white in Fig.\,\ref{fig:head_tail_counts}) is substantially larger. 

\begin{figure}[ht]
\includegraphics[width=\linewidth]{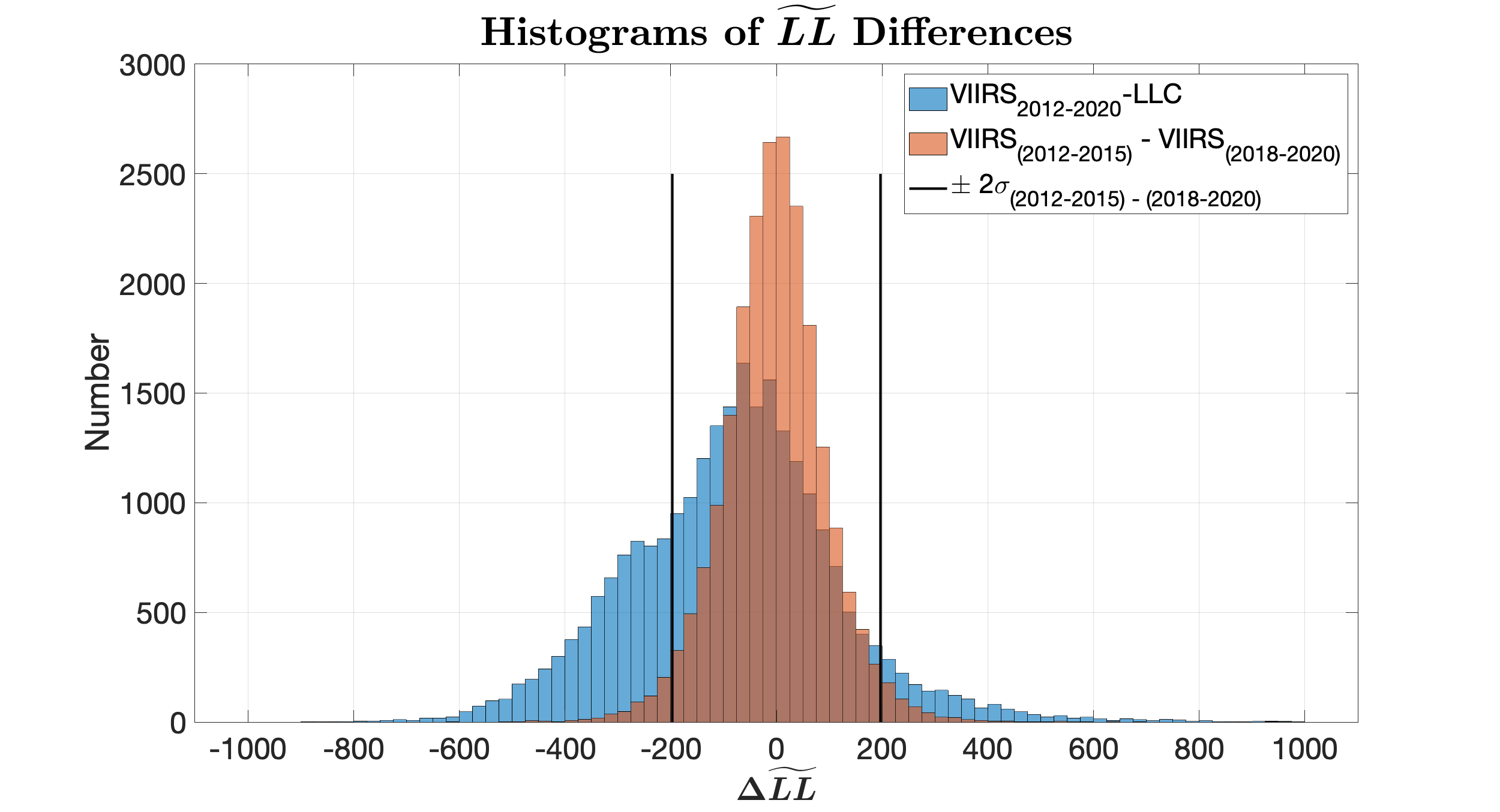}
\caption{Histograms of {\viirsllcdifference} (blue) and {\headtaildifference} (light brown). Vertical black lines are $\pm2\sigma$ of {\headtaildifference}.
}
\label{fig:histogram_ll_differences}
\end{figure}

Figure \ref{fig:histogram_ll_differences} shows a histogram of the differences of the two {\tildell} fields {\headtaildifference}. Also shown in Fig. \ref{fig:histogram_ll_differences} is the histogram of the differences of the {\viirs} and {\llc} {\tildell} fields, {\viirsllcdifference}. %This histogram is discussed in \S\ref{section:discussion}; in this section we focus on {\headtaildifference}. 
The two vertical black lines denote $\pm2\sigma$ of the {\headtaildifference} distribution. We use these in the body of the manuscript to identify significant outliers in the fields. There are three primary contributors to the variance of the {\headtaildifference} distribution. First, there is the uncertainty associated with the assignment of an {\LL} value by the machine learning algorithm to each cutout within cell; think of this as instrument noise. Second, cutouts in each cell are being sampled from a three-dimensional space-time region, the spatial extent defined by the cell boundaries (approximately 100\,km on a side) and the temporal extent defined by the four-year period from which each distribution is drawn; think of this as the uncertainty of estimated values based on the finite sample size. Third, difference in the {\tildell} value between the period covered by the two datasets, i.e., the true geophysical difference. For {\headtaildifference}, the latter is the difference between 2012--2015 and 2017--2020. For {\viirsllcdifference} this would be the difference between the simulated period, 2012, and the period from which the {\viirs} data is sampled, 2012--2020. This means that the variability of {\headtaildifference} places an upper bound on uncertainty in the {\LL} values, variability due to position within the cell and the period from which cutouts contributing to the cell are drawn. As shown in the next paragraph, we believe that the geophysical contribution of uncertainty to the {\headtaildifference} differences is small hence the variability of {\headtaildifference} is a good measure of what constitutes a significant deviation between two datasets. In light of this, we will use two standard deviations of the {\headtaildifference} distribution to identify regions in which the model output agrees/disagrees with the satellite-derived fields.

\begin{figure}[ht]
\includegraphics[width=\linewidth]{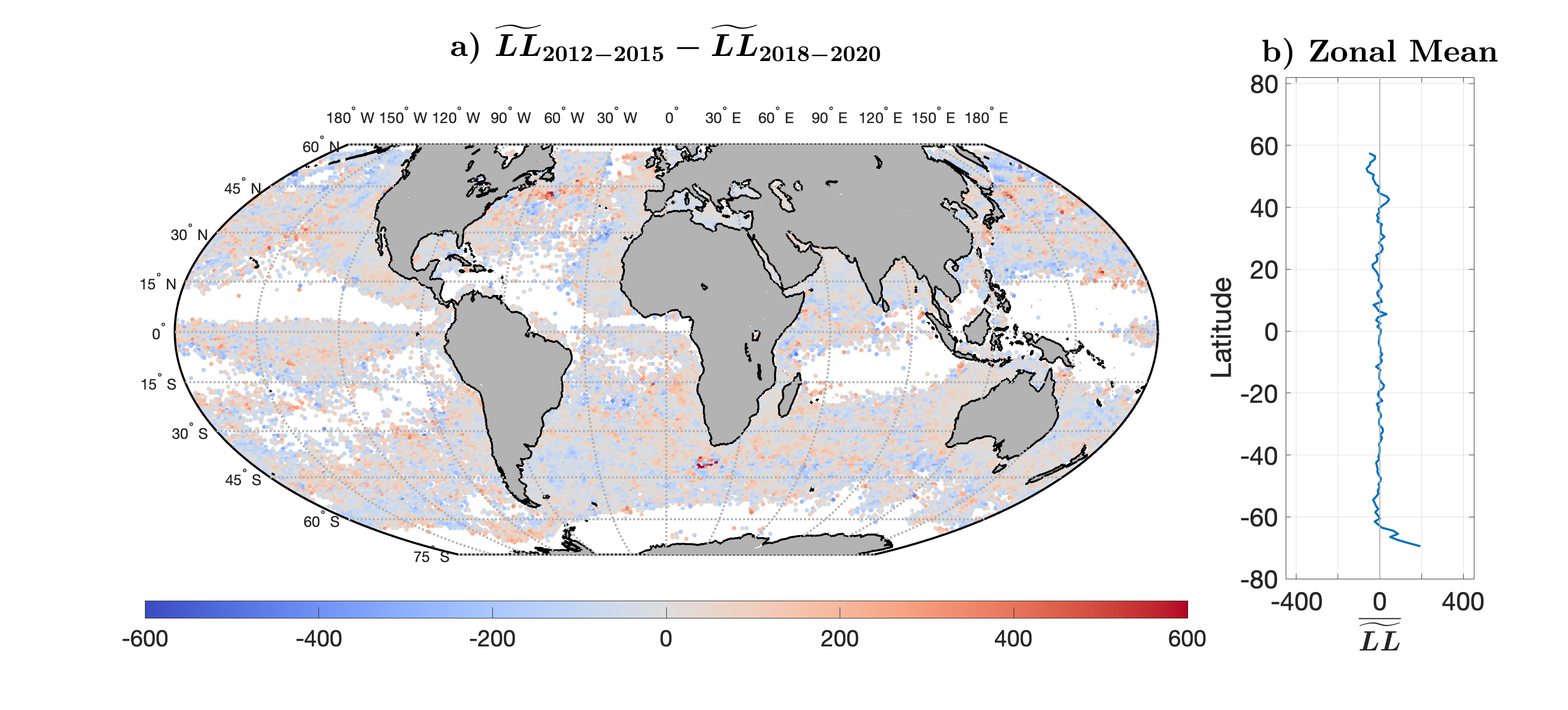}
\caption{(a) {\headtaildifference}. White areas - less than 5 cutouts in the corresponding {\healpix} cell in 2012--2015 and/or 2017--2020. The same color palette is used in this figure as in Fig.~\ref{fig:viirs_minus_llc_ll} to facilitate comparison as well as to emphasize the significant differences in {\viirsllcdifference}. (b) Zonal mean of a. 
}
\label{fig:head_tail_ll}
\end{figure}

Also of importance in understanding the significance of differences between {\viirsllcdifference} is the degree to which these differences are distributed geographically. Specifically, shifts in major ocean currents as well as changes in forcing from one period to another could result in different structures in the submesoscale-to-mesoscale range, which would display as geographic regions of positive or negative differences between two periods.  Figure \ref{fig:head_tail_ll} suggests that, with a few exceptions, the distribution of {\headtaildifference} is, in fact, quite random, i.e., at least for this pair of 4-year periods, the differences in submesoscale-to-mesoscale structure is relatively random. There are, however, some regions of more than a few {\healpix} cells, which stand out as significantly different between the two periods either in the positive or negative sense. A narrow negative band is evident along the northern edge of the {\acc} south of the Indian Ocean suggesting that the {\acc} may have shifted south between 2012--2015 and 2017--2020---more negative values of the difference correspond to less structure in the second period. 

\begin{figure}[ht]
\includegraphics[width=\linewidth]{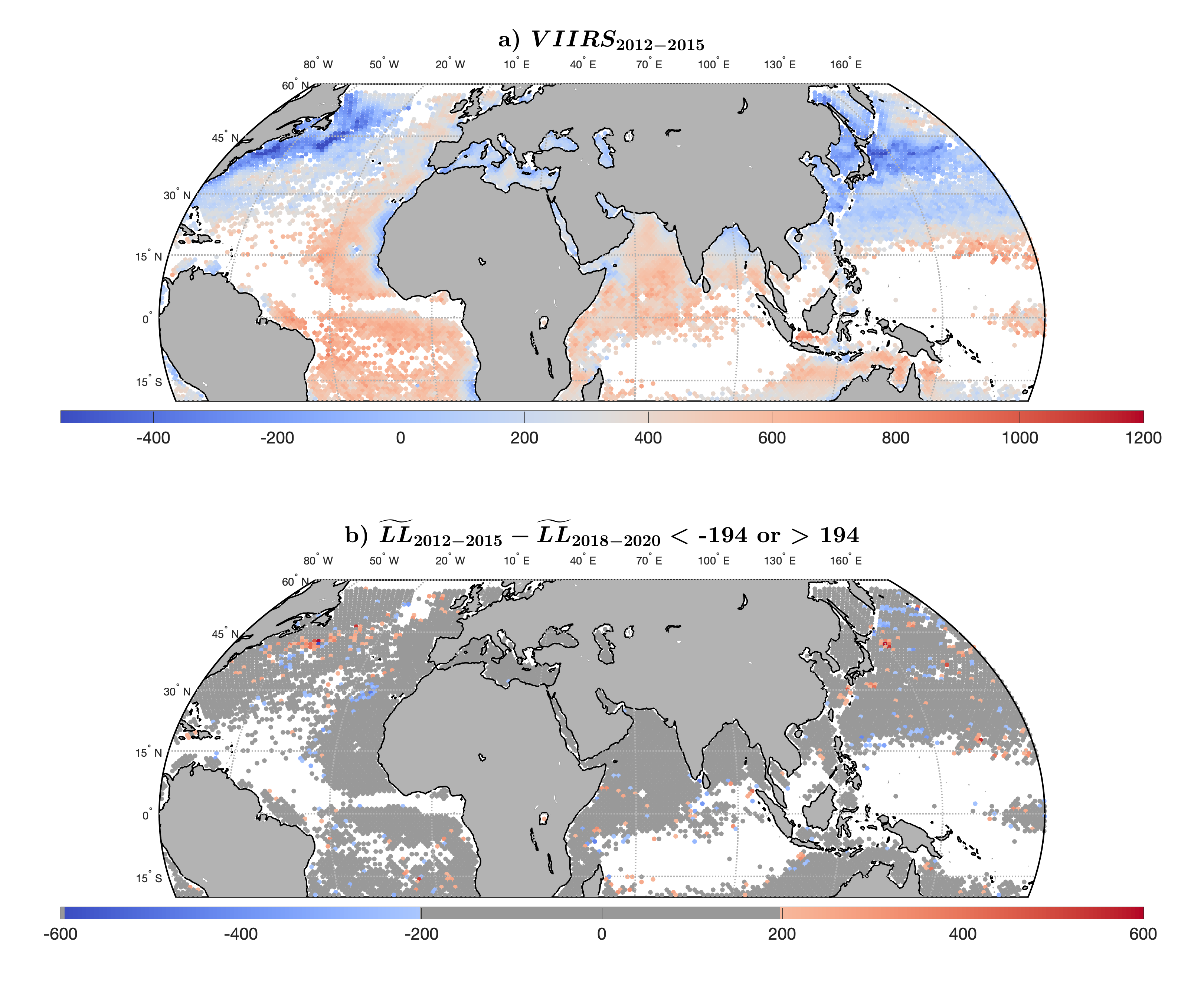}
\caption{ a) {\tildellhead} for the northern hemisphere. b) Masked {\headtaildifference}. Dark gray: $|${\headtaildifference}$|<2\sigma={\sigmathreshold}$. Light gray: land. White - less than 5 cutouts/{\healpix} cell for both 2012--2015 and 2017--2020.
}
\label{fig:northern_hemisphere_zoom}
\end{figure}

\begin{figure}[ht]
\includegraphics[width=0.55\linewidth]{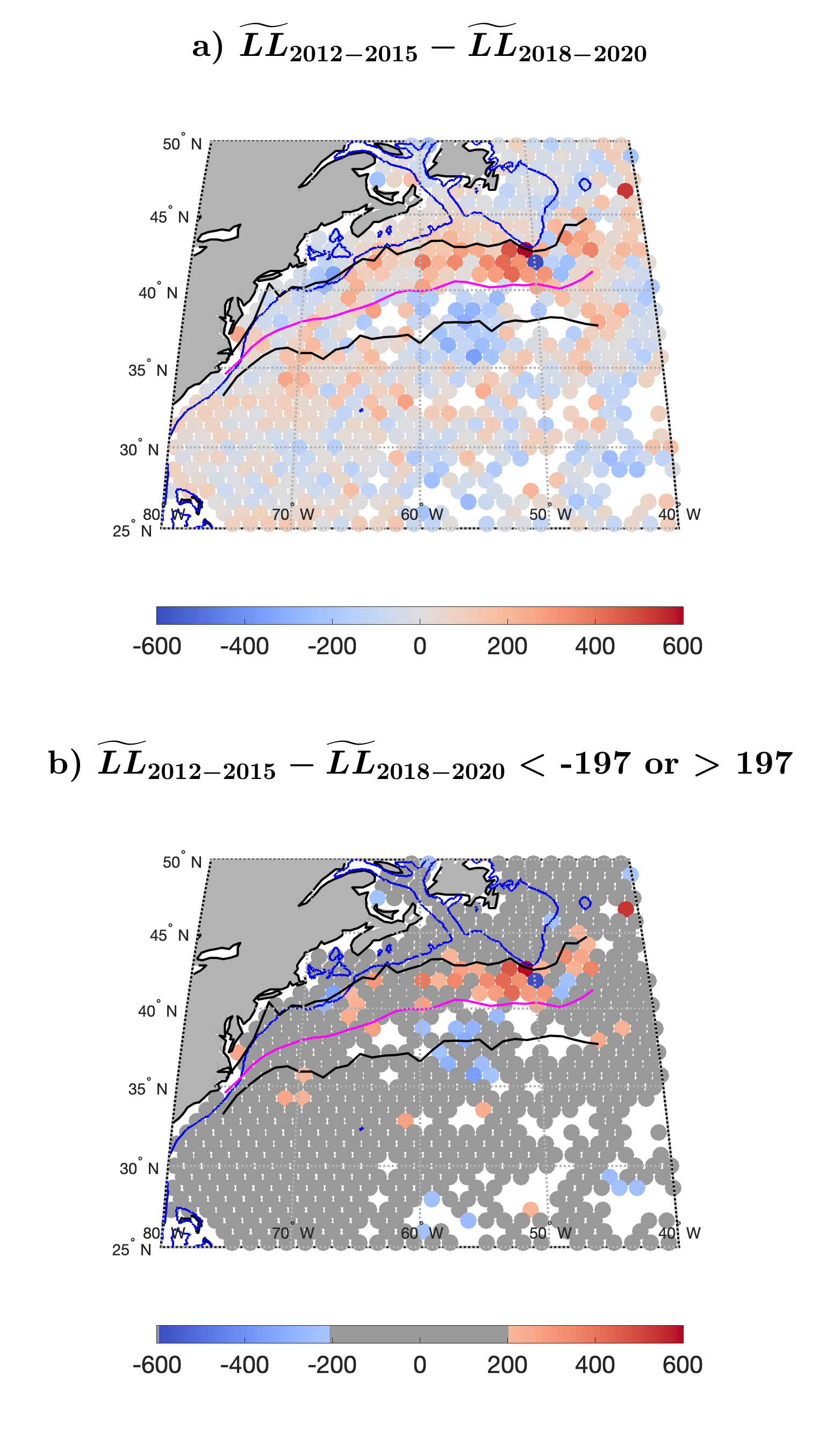}
\caption{(a) As in Figs.~\ref{fig:head_tail_ll} and \ref{fig:northern_hemisphere_zoom} but focused on Gulf Stream region. (b) Masked version of (a). Magenta line between approximate $75^\circ$ W and $45^\circ$ W is the mean path of the Gulf Stream digitized from two-day composites of 1-km \ac{AVHRR} {\sst} fields for 1982--1999. Black lines are for the northern most and southernmost extents of Gulf Stream paths digitized from the same dataset as (a) but for 1982--1986.
}
\label{fig:GS_zoom}
\end{figure}

Significant differences are also evident in the vicinity of the Gulf Stream and Kurshio (Fig.\,\ref{fig:northern_hemisphere_zoom}b masked to show only {\healpix} cells with values more than two standard deviations from the mean, i.e., significant outliers). Figure \ref{fig:northern_hemisphere_zoom}a shows the {\tildell} distribution for 2012--2015 (the geographic distribution of {\tildell} for 2017--2020 is virtually indistinguishable from that shown for 2012--2015 in this plot). The dark blue areas on the western side of the North Atlantic and North Pacific north of approximately $35^\circ$N correspond to significant structure in the {\sst} fields in and north of the associated western boundary currents---the Gulf Stream and Kuroshio, an observation documented in \citet{prochaska2021deep}. The region of enhanced differences between the two periods appears to be on the northern edge and to the north of these currents. Figure \ref{fig:GS_zoom} is a blow-up of the region in the vicinity of the Gulf Stream: Fig.\,\ref{fig:GS_zoom}a shows the unmasked {\headtaildifference} values and Fig.\,\ref{fig:GS_zoom}b shows the masked values. The more positive differences north of the Gulf Stream mean path (magenta line in the figure) suggest an increase in structure in 2017--2020 compared with that in 2012--2015. That much of the differences are north of the northernmost extent of the Gulf Stream (upper black line) argues that not only has the mean path of the stream likely moved to the north in this period but that this displacement resulted in more submesoscale-to-mesoscale turbulence in the region north of the stream. Of particular interest, although not the focus of this manuscript, is the similarity in the patterns in the vicinity of the Kuroshio suggesting that the phenomena is hemispheric as opposed to confined to one ocean basin. The point here is that, although there are some regions of significant differences, these tend to be relatively small and are, in general, associated with strong currents---the Gulf Stream, the Kuroshio, and the {\acc}.

\noappendix       %% use this to mark the end of the appendix section. Otherwise the figures might be numbered incorrectly (e.g., 10 instead of 1).

%% Regarding figures and tables in appendices, the following two options are possible depending on your general handling of figures and tables in the manuscript environment:

%% Option 1: If you sorted all figures and tables into the sections of the text, please also sort the appendix figures and appendix tables into the respective appendix sections.
%% They will be correctly named automatically.

%% Option 2: If you put all figures after the reference list, please insert appendix tables and figures after the normal tables and figures.
%% To rename them correctly to A1, A2, etc., please add the following commands in front of them:

\appendixfigures  %% needs to be added in front of appendix figures

\appendixtables   %% needs to be added in front of appendix tables

%% Please add \clearpage between each table and/or figure. Further guidelines on figures and tables can be found below.

\authorcontribution{ KG performed the initial studies on which this manuscript is based. KG, XP and PC contributed to the subsequent analyses of the data. All contributed to the writing of the manuscript, which was led by KG. Figures were generated by KG and PC. XP provided input on the machine learning portion of the project. PC and DM provided input related to physical oceanography. PC provided the satellite data expertise. DM provided the modeling expertise. MK and XP processed the satellite data.}

\competinginterests{The authors declare that they have no conflict of interest.} 

%\disclaimer{TEXT} %% optional section

\begin{acknowledgements}
JXP acknowledges future support from the Simons Foundation and the University of California, Santa
Cruz.
DM carried out research at the Jet Propulsion Laboratory, California Institute of Technology, under a contract with NASA, with support from the Physical Oceanography (PO) and Modeling, Analysis, and Prediction (MAP) programs.  KG was supported during the summer of 2021 through the \acl{URI}/\acl{SURFO}, \acs{NSF} award number OCE-1950586. Support for PC was provided by the Office of Naval Research: ONR N00014-17-1-2963, NASA: 80NSSC18K0837, NASA:80NSSC20K1728 and the State of Rhode Island.

The {\viirs} L2 {\sst} data were provided by the \acl{GHRSST} and the \acl{NOAA} and obtained from the \acl{NASA}/\acl{PO.DAAC}. The {\llc4320} {\sst} fields were obtained via the {\tt xmitgcm} package obtained at: https://xmitgcm.readthedocs.io/en/latest/.

Some of the results in this paper have been derived using the HEALPix (K.M. Górski et al., 2005, ApJ, 622, p759) package. The authors acknowledge use of the Nautilus cloud computing system which is supported by the following US National Science Foundation (NSF) awards: CNS-1456638, CNS-1730158, CNS-2100237, CNS-2120019, ACI-1540112, ACI-1541349, OAC-1826967, OAC-2112167. High-end computing for the LLC4320 simulation were provided by the NASA Advanced Supercomputing (NAS) Division at the Ames Research Center.
\end{acknowledgements}

\newpage
\noindent{\bf Acronyms}
{\begin{acronym}[12345678901234]
\acro{(A)ATSR}{one or all of ATSR, ATSR-2 and AATSR}
\acro{AATSR}{Advanced Along Track Scanning Radiometer}
\acro{ACC}{Antarctic Circumpolar Current}
\acro{ACCESS}{Advancing Collaborative Connections for Earth System Science}
\acro{ACL}{Access Control List}
\acro{ACSPO}{Advanced Clear Sky Processor for Oceans}
\acro{ADA}{Automatic Detection Algorithm}
\acro{ADCP}{Acoustic Doppler Current Profiler}
\acro{ADT}{absolute dynamic topography}
\acro{AESOP}{Assessing the Effects of Submesoscale Ocean Parameterizations}
\acro{AGU}{American Geophysical Union}
\acro{AI}{Artificial Intelligence}
\acro{AIRS}{Atmospheric Infrared Sounder}
\acro{AIS}{Ancillary Information Service}
\acro{AIST}{Advanced Information Systems Techonology}
\acro{AISR}{Applied Information Systems Research}
\acro{ADL}{Alexandria Digital Library}
\acro{API}{Application Program Interface}
\acro{APL}{Applied Physics Laboratory}
\acro{API}{Application Program Interface}
\acro{AMSR}{Advanced Microwave Scanning Radiometer}
\acro{AMSR2}{Advanced Microwave Scanning Radiometer 2}
\acro{AMSR-E}{Advanced Microwave Scanning Radiometer - EOS}
\acro{ANN}{ Artificial Neural Network}
\acro{AOOS}{Alaska Ocean Observing System}
\acro{APAC}{Australian Partnership for Advanced Computing}
\acro{APDRC}{Asia-Pacific Data-Research Center}
%\acro{ARC}{\acs{ATSR} Reprocessing for Climate}
\acro{ARC}{ATSR Reprocessing for Climate}
\acro{ASCII}{American Standard Code for Information Interchange}
\acro{AS}{Aggregation Server}
\acro{ASFA}{Aquatic Sciences and Fisheries Abstracts}
\acro{ASTER}{Advanced Spaceborne Thermal Emission and Reflection Radiometer}
%{ -- \it http://asterweb.jpl.nasa.gov}
\acro{ATBD}{Algorithm Theoretical Basis Document}
\acro{ATSR}{Along Track Scanning Radiometer}
\acro{ATSR-2}{Second ATSR}
\acro{AVISO}{Archiving, Validation and Interpretation of Satellite Oceanographic Data}
\acro{ANU}{Australian National University}
\acro{AVHRR}{Advanced Very High Resolution Radiometer}
\acro{AzC}{Azores Current}

\acro{BAA}{Broad Agency Announcement}
\acro{BAO}{bi-annual oscillation}
\acro{BES}{Back-End Server}
\acro{BMRC}{Bureau of Meteorology Research Centre}
\acro{BOM}{Bureau of Meteorology}
\acro{BT}{brightness temperature}
\acro{BUFR}{Binary Universal Format Representation}
%{ -- \it http://www.wmo.ch/web/www/WDM/Guides/Guide-on-DataMgt-1.htm}

\acro{CAN}{Cooperative Agreement Notice}
\acro{CAS}{Community Authorization Service}
\acro{CC}{cloud cover}
\acro{CCA}{Cayula-Cornillon Algoritm}
\acro{CCI}{Climate Change Initiative}
\acro{CCLRC}{Council for the Central Laboratory of the Research Councils}
%{ --- \it http://www.cclrc.ac.uk/}
\acro{CCMA}{Center for Coastal Monitoring and Assessment}
\acro{CCR}{cold core ring}
\acro{CCS}{California Current System }
\acro{CCSM}{Community Climate System Model}
\acro{CCSR}{Center for Climate System Research}
\acro{CCV}{Center for Computation and Visualization}
\acro{CDAT}{Climate Data Analysis Tools}
\acro{CDC}{Climate Diagnostics Center}
\acro{CDF}{Common Data Format}
\acro{CDR}{Common Data Representation}
\acro{CEDAR}{Coupled Energetic and Dynamics and Atmospheric Regions}
%{ -- \it http://cedarweb.hao.ucar.edu/}
\acro{CEOS}{Committee on Earth Observation Satellites}
\acro{CERT}{Computer Emergency Response Team}
\acro{CenCOOS}{Central \& Northern California Ocean Observing System}
% \acro{CF}{NetCDF Climate and Forecast Metadata Conventions}
\acro{CF}{clear fraction}
\acro{CGI}{Common Gateway Interface}
%\acro{CHAP}{\textsmaller{CISL} \ac{HPC} Advisory Panel}
\acro{CHAP}{CISL High Performance Computing Advisory Panel}
\acro{CIFS}{Common Internet File System}
\acro{CIMSS}{Cooperative Institute for Meteorological Satellite Studies}
\acro{CIRES}{Cooperative Institute for Research (in) Environmental Sciences}
\acro{CISL}{Computational \& Information Systems Laboratory}
\acro{CLASS}{Comprehensive Large Array-data Stewardship System}
\acro{CLIVAR}{Climate Variability and Predictability}
\acro{CLS}{Collecte Localisation Satellites}
\acro{CME}{Community Modeling Effort}
\acro{CMS}{Centre de M\'et\'eorologie Spatiale}
\acro{CNN}{Convolutional Neural Network}
\acro{COA}{Climate Observations and Analysis}
\acro{COARDS}{Cooperative Ocean-Atmosphere Research Data Standard}
\acro{COAPS}{Center for Ocean-Atmospheric Prediction Studies}
\acro{COBIT}{Control Objectives for Information and related Technology}
%\acro{COCO}{\acs{CCSR} Ocean Component model}
\acro{COCO}{CCSR Ocean Component model}
\acro{CODAR}{Coastal Ocean Dynamics Applications Radar}
\acro{CODMAC}{Committee on Data Management, Archiving, and Computing}
\acro{Co-I}{Co-Investigator}
\acro{CORBA}{Common Object Request Broker Architecture}
\acro{COLA}{Center for Ocean-Land-Atmosphere Studies}
%{ -- \it http://grads.iges.org/cola.html}
\acro{CPU}{Central Processor Unit}
\acro{CRS}{Coordinate Reference System}
\acro{CSA}{Cambridge Scientific Abstracts}
\acro{CSC}{Coastal Services Center}
\acro{CSIS}{Center for Strategic and International Studies}
\acro{CSL}{Constraint Specification Language}
\acro{CSP}{Chermayeff, Sollogub and Poole, Inc.}
%{ -- \it http://csp-architects.com/contact.htm}
\acro{CSDGM}{Content Standard for Digital Geospatial Metadata}
%\acro{CTD}{Conductivity, Temperature and Salinity probes}
\acro{CSV}{Comma Separated Values}
\acro{CTD}{Conductivity, Temperature and Salinity}
\acro{CVSS}{Common Vulnerability Scoring System}
\acro{CZCS}{Coastal Zone Color Scanner}

\acro{DAAC}{Distribute Active Archive Center}
\acro{DAARWG}{Data Archiving and Access Requirements Working Group}
\acro{DAP}{Data Access Protocol}
\acro{DAS}{Data set Attribute Structure}
\acro{DBMS}{Data Base Management System}
\acro{DBDB2}{Digital Bathymetric Data Base} 
\acro{DChart}{Dapper Data Viewer}
\acro{DDS}{Data Descriptor Structure}
\acro{DDX}{XML version of the combined DAS and DDS}
\acro{DFT}{Discrete Fourier Transform}
\acro{DIF}{Directory Interchange Format}
\acro{DISC}{Data and Information Services Center}
\acro{DIMES}{Diapycnal and Isopycnal Mixing Experiment:  Southern Ocean}
\acro{DMAC}{Data Management and Communications committee}
\acro{DMR}{Department of Marine Resources}
\acro{DMSP}{Defense Meteorological Satellite Program}
\acro{DoD}{Department of Defense}
\acro{DODS}{Distributed Oceanographic Data System}
%{ -- \it http://www.unidata.ucar.edu/packages/dods}
\acro{DOE}{Department of Energy}
%{ -- \it http://www.energy.gov}
\acro{DSP}{U. Miami satellite data processing software}
\acro{DSS}{direct statistical simulation}

\acro{EASy}{Environmental Analysis System}
%{ -- \it http://members.cox.net/fjobrien/global/layout/people.htm}
\acro{ECCO}{Estimating the Circulation and Climate of the Ocean}
\acro{ECCO2}{Estimating the Circulation and Climate of the Ocean, Phase II}
\acro{ECS}{EOSDIS Core System}
\acro{ECHO}{Earth Observing System Clearinghouse}
\acro{ECMWF}{European Centre for Medium-range Weather Forecasting}
\acro{ECV}{Essential Climate Variable}
\acro{EDC}{Environmental Data Connector}
\acro{EDJ}{Equatorial Deep Jet}
\acro{EDFT}{Extended Discrete Fourier Transform}
\acro{EDMI}{Earth Data Multi-media Instrument}
%{ -- \newline \it http://www.newmediastudio.org/Homepage/TNMSHomeFramset.htm}
\acro{EEJ}{Extra-Equatorial Jet}
\acro{EIC}{Equatorial Intermediate Current}
\acro{EICS}{Equatorial Intermediate Current System}
\acro{EJ}{Equatorial Jets}
\acro{EKE}{eddy kinetic energy}
\acro{EMD}{Empirical Mode Decomposition}
\acro{EOF}{Empirical Orthogonal Function}
\acro{EOS}{Earth Observing System}
\acro{EOSDIS}{Earth Observing System Data Information System}
\acro{EPA}{Environmental Protection Agency}
\acro{EPSCoR}{Experimental Program to Stimulate Competitive Research}
\acro{EPR}{East Pacific Rise}
\acro{ERD}{Environmental Research Division}
\acro{ERS}{European Remote-sensing Satellite}
\acro{ESA}{European Space Agency}
\acro{ESDS}{Earth Science Data Systems}
%{ -- \it http://www.esdswg.org/}
\acro{ESDSWG}{Earth Science Data Systems Workign Group}
\acro{ESE}{Earth Science Enterprise}
\acro{ESG}{Earth System Grid}
%{ -- \it http://www.earthsystemgrid.org/}
\acro{ESG II}{Earth System Grid -- II}
%{ -- \it  http://www.earthsystemgrid.org}
\acro{ESIP}{Earth Science Information Partner}
%{ -- \it http://www.esipfed.org}
\acro{ESMF}{Earth System Modeling Framework}
%{ -- \texttt{http://www.esmf.ucar.edu}}
\acro{ESML}{Earth System Markup Language}
%{ -- \it http://esml.itsc.uah.edu/index.jsp}
\acro{ESP}{eastern South Pacfic}
\acro{ESRI}{Environmental Systems Research Institute}
%{-- \it http://www.esri.com}
\acro{ESR}{Earth and Space Research}
\acro{ETOPO}{Earth Topography}
\acro{EUC}{Equatorial Undercurrent}
\acro{EUMETSAT}{European Organisation for the Exploitation of Meteorological Satellites}
\acro{Ferret}{}
%{ -- \it http://ferret.pmel.noaa.gov/Ferret}

\acro{FASINEX}{Frontal Air-Sea Interaction Experiment}
\acro{FDS}{Ferret Data Server}
\acro{FFT}{Fast Fourier Transform}
\acro{FGDC}{Federal Geographic Data Committee}
%{ -- \it http://www.fgdc.gov}
\acro{FITS}{Flexible Image (or Interchange) Transport System}
\acro{FLOPS}{FLoating point Operations Per Second} 
\acro{FRTG}{Flow Rate Task Group}
\acro{FreeForm}{}
%{ -- \it http://www.ngdc.noaa.gov/seg/freeform/freeform.shtml}
\acro{FNMOC}{Fleet Numerical Meteorology and Oceanography Center}
\acro{FSU}{Florida State University}
\acro{FTE}{Full Time Equivalent}
\acro{ftp}[\normalsize  ftp]{File Transport Protocol}
\acro{FTP}[\normalsize  ftp]{File Transport Protocol}

\acro{GAC}{Global Area Coverage}
\acro{GAN}{Generative Adversarial Network}
\acro{GB}{GigaByte - $10^{9}$ bytes}
\acro{GCMD}{Global Change Master Directory}
%{ -- \it http://gcmd.nasa.gov}
\acro{GCM}{general circulation model}
\acro{GCOM-W1}{Global Change Observing Mission - Water}
\acro{GCOS}{Global Climate Observing System}
\acro{GDAC}{Global Data Assembly Center}
\acro{GDS}{GrADS Data Server}
\acro{GDS2}{GHRSST Data Processing Specification v2.0}
%{ -- \it http://grads.iges.org/grads/gds}
\acro{GEBCO}{General Bathymetric Charts of the Oceans}
\acro{GeoTIFF}{Georeferenced Tag Image File Format}
\acro{GEO-IDE}{Global Earth Observation Integrated Data Environment}
\acro{GES DIS}{Goddard Earth Sciences Data and Information Services Center}
\acro{GEMPACK}{General Equilibrium Modelling PACKage}
\acro{GEOSS}{Global Earth Observing System of Systems}
\acro{GFDL}{Geophysical Fluid Dynamics Laboratory}
\acro{GFD}{Geophysical Fluid Dynamics}
\acro{GHRSST}{Group for High Resolution Sea Surface Temperature}
\acro{GHRSST-PP}{GODAE High Resolution Sea Surface Temperature Pilot Project}
%{ -- \it  http://www.ghrsst-pp.org}
%\acro{GHRSST-PP}{\acs{GODAE}High Resolution Sea Surface Temperature Pilot Project}
%{ -- \it  http://www.ghrsst-pp.org}
\acro{GINI}{GOES Ingest and NOAA/PORT Interface}
\acro{GIS}{Geographic Information Systems} %{ -- \it http://www.gis.com}
\acro{Globus}{} %{ -- {http://www.globus.org}}
\acro{GMAO}{Global Modeling and Assimilation Office}
\acro{GML}{Geography Markup Language}
\acro{GMT}{Generic Mapping Tool}
\acro{GODAE}{Global Ocean Data Assimilation Experiment} %{ -- \it http://www.bom.gov.au/bmrc/ocean/GODAE}
\acro{GOES}{Geostationary Operational Environmental Satellites} %{ -- \it http://www.oso.noaa.gov/goes}
\acro{GOFS}{Global Ocean Forecasting System}
\acro{GoMOOS}{Gulf of Maine Ocean Observing System}
\acro{GOOS}{Global Ocean Observing System}
\acro{GOSUD}{Global Ocean Surface Underway Data}
\acro{GPFS}{ General Parallel File System}
\acro{GPU}{Graphics Processing Unit}
\acro{GRACE}{Gravity Recovery and Climate Experiment} %{ -- \it http://www.csr.utexas.edu/grace/spacecraft/config.html}
\acro{GRIB}{GRid In Binary} %{ -- \it http://www.wmo.ch/web/www/WDM/Guides/Guide-on-DataMgt-1.htm}
\acro{GrADS}{Grid Analysis and Display System} %{ -- \it http://grads.iges.org/grads/index.html}
\acro{GridFTP}{FTP with GRID enhancements}
\acro{GRIB}{GRid in Binary} %{ -- \it http://www.wmo.ch/web/www/DPS/grib-2.html}
\acro{GPS}{Global Positioning System}
\acro{GSFC}{Goddard Space Flight Center} %{ -- \it http://www.gsfc.nasa.gov}
\acro{GSI}{Grid Security Infrastructure}
\acro{GSO}{Graduate School of Oceanography}
\acro{GTSPP}{Global Temperature and Salinity Profile Program}
\acro{GUI}{Graphical User Interface}
\acro{GS}{Gulf Stream}

\acro{HAO}{High Altitude Observatory} %{ -- \it http://www.hao.ucar.edu/public/inside/data.html}
\acro{HLCC}{Hawaiian Lee Countercurrent}
\acro{HCMM}{Heat Capacity Mapping Mission}
\acro{HDF}{Hierarchical Data Format}
\acro{HDF-EOS}{Hierarchical Data Format - EOS} %{ -- \it http://hdfeos.gsfc.nasa.gov}
\acro{HEC}{High-End Computing}
\acro{HEALPix}{Hierarchical Equal Area isoLatitude Pixelation}
\acro{HF}{High Frequency}
\acro{HGE}{High Gradient Event}
\acro{HPC}{High Performance Computing}
\acro{HPCMP}{High Performance Computing Modernization Program}
\acro{HPSS}{High Performance Storage System} %{ -- \it http://www.sdsc.edu/hpss/hpss1.html}
\acro{HR DDS}{High Resolution Diagnostic Data Set}
\acro{HRPT}{High Resolution Picture Transmission}
\acro{HTML}{Hyper Text Markup Language}
\acro{html}{Hyper Text Markup Language}
\acro{http}{the hypertext transport protocol}
\acro{HTTP}{Hyper Text Transfer Protocol}
\acro{HTTPS}{Secure Hyper Text Transfer Protocol} %\acro{http}{Hypertext Transport Protocol} It seems that the acronym is all caps.
\acro{HYCOM}{HYbrid Coordinate Ocean Model} %{ -- \it http://oceanmodeling.rsmas.miami.edu/hycom/}

\acro{I-band}{imagery resolution band}
\acro{IDD}{Internet Data Distribution}
\acro{IB}{Image Band}
\acro{IBL}{internal boundary layer}
\acro{IBM}{Internation Business Machines}
%{ -- \it http://www.ibm.com/us}
\acro{ICCs}{Intermediate Countercurrents}
\acro{IDE}{Integrated Development Environment}
\acro{IDL}{Interactive Display Language}
%{ -- \it  http://www.rsinc.com/idl/index.asp}
\acro{IDLastro}{IDL Astronomy User's Library}
\acro{IDV}{Integrated Data Viewer}
%{ -- \it http://my.unidata.ucar.edu/content/software/metapps/index.html}
\acro{IEA}{Integrated Ecosystem Assessment}
\acro{IEEE}{Institute (of) Electrical (and) Electronic Engineers}
%{ -- \it http://www.ieee.org/portal/index.jsp}
\acro{IETF}{Internet Engineering Task Force}
\acro{IFREMER}{Institut Fran\c{c}ais de Recherche pour l'Exploitation de la MER}
%\acro{IMF}{Interplanetary Magnetic Field}
\acro{IMAPRE}{El Instituto del Mar del Per\'u}
\acro{IMF}{Intrinsic Mode Function}
\acro{IOOS}{Integrated Ocean Observing System}
\acro{ISAR}{Infrared Sea surface temperature Autonomous Radiometer}
\acro{ISO}{International Organization for Standardization}
\acro{ISSTST}{Interim Sea Surface Temperature Science Team}
\acro{IT}{Information Technology}
\acro{ITCZ}{Intertropical Convergence Zone} 
\acro{IP}{Internet Provider}
\acro{IPCC}{Intergovernmental Panel on Climate Change}
\acro{IPRC}{International Pacific Research Center}
\acro{IR}{Infrared}
\acro{IRI}{International Research Institute for Climate and Society}
\acro{ISO}{International Standards Organization}

\acro{JASON}{JASON Foundation for Education}
%\acro{JASON}{NASA Altimeter}
%{ -- \it http://www.jason.org}
\acro{JDBC}{Java Database Connectivity}
\acro{JFR}{Juan Fern\'andez Ridge}
\acro{JGOFS}{Joint Global Ocean Flux Experiment}
%{ -- \it http://puddle.mit.edu/datasys/jgsys.html}
\acro{JHU}{Johns Hopkins University}
\acro{JPL}{Jet Propulsion Laboratory}
\acro{JPSS}{Joint Polar Satellite System}
%{ -- \it http://www.jpl.nasa.gov}

\acro{KDE}{Kernel Density Estimation}
\acro{KVL}{Keyword-Value List}
\acro{KML}{Keyhole Markup Language}
\acro{KPP}{K-Profile Parameterization}

\acro{LAC}{Local Area Coverage}
\acro{LAN}{Local Area Network}
\acro{LAS}{Live Access Server}
%{ -- \it http://www.ferret.noaa.gov/nopp/main.pl? }
\acro{LASCO}{Large Angle and Spectrometric Coronagraph Experiment}
\acro{LatMIX}{Scalable Lateral Mixing and Coherent Turbulence}
\acro{LDAP}{Lightweight Directory Access Protocol}
\acro{LDEO}{Lamont Doherty Earth Observatory}
\acro{LEAD}{Linked Environments for Atmospheric Discovery}
\acro{LEIC}{Lower Equatorial Intermediate Current}
\acro{LES}{Large Eddy Simulation}
\acro{L1}{Level-1}
\acro{L2}{Level-2}
\acro{L2P}{Level-2P}
\acro{L3}{Level-3}
\acro{L4}{Level-4}
\acro{LL}{Log-Likelihood}
\acro{LLC}{Latitude/Longitude/polar-Cap}
\acro{LLC4320}[LLC4320]{\acl{LLC}4320}
\acro{LLC2160}[LLC2160]{\ac{LLC}2160}
\acro{LLC1080}[LLC1080]{\ac{LLC}1080}
\acro{LHF}{Latent Heat Flux}
\acro{LST}{local sun time}
\acro{LTER}{Long Term Ecological Research Network}
\acro{LTSRF}{Long Term Stewardship and Reanalysis Facility}
\acro{LUT}{Look Up Table}

\acro{M-band}{moderate resolution band}
\acro{MABL}{marine atmospheric boundary layer}
\acro{MADT}{Maps of Absolute Dynamic Topography}
\acro{MapServer}{MapServer}
%{ -- \it http://mapserver.gis.umn.edu/}
\acro{MAT}{Metadata Acquisition Toolkit}
\acro{MATLAB}{}
%{ -- \it http://www.mathworks.com/products/}
\acro{MARCOOS}{Mid-Atlantic Coastal Ocean Observing System}
\acro{MARCOORA}{Mid-Atlantic Coastal Ocean Observing Regional Association}
\acro{MB}{MegaByte - $10^{6}$ bytes}
\acro{MCC}{Maximum Cross-Correlation}
\acro{MCR}{MATLAB Component Runtime}
\acro{MCSST}{Multi-Channel Sea Surface Temperature}
\acro{MDT}{mean dynamic topography}
\acro{MDB}{Match-up Data Base}
\acro{MDOT}{mean dynamic ocean topography}
\acro{MEaSUREs}{Making Earth System data records for Use in Research Environments}
\acro{MERRA}{Modern Era Retrospective-Analysis for Research and Applications}
\acro{MERSEA}{Marine Environment and Security for the European Area}
\acro{MTF}{Modulation Transfer Function}
\acro{MICOM}{Miami Isopycnal Coordinate Ocean Model}
\acro{MIRAS}{Microwave Imaging Radiometer with Aperture Synthesis}
\acro{MITgcm}{MIT general circulation model}
\acro{MIT}{Massachusetts Institute of Technology}
\acro{mks}{meters, kilograms, seconds}
\acro{MLP}{Multilayer Perceptron}
\acro{ML}{machine learning}
\acro{MLSO}{Mauna Loa Solar Observatory}
%{ -- \it http://mslo.\ac{HAO}ucar.edu/}
\acro{MM5}{Mesoscale Model}
\acro{MMI}{Marine Metadata Initiative}
\acro{MMS}{Minerals Management Service}
\acro{MODAS}{Modular Ocean Data Assimilation System}
\acro{MODIS}{MODerate-resolution Imaging Spectroradiometer}
%{ -- \it http://modis.gsfc.nasa.gov}
\acro{MOU}{Memorandum of Understanding}
\acro{MPARWG}{Metrics Planning and Reporting Working Group}
\acro{MSE}{mean square error}
\acro{MSG}{Meteosat Second Generation}
\acro{MTPE}{Mission To Planet Earth}
\acro{MUR}{Multi-sensor Ultra-high Resolution}
\acro{MV}{Motor Vessel}

\acro{NAML}{National Association of Marine Laboratories}
\acro{NAHDO}{National Association of Health Data Organizations}
%{ -- \it http://www.nahdo.org}
\acro{NAS}{Network Attached Storage}
\acro{NASA}{National Aeronautics and Space Administration}
%{ -- \it http://www.nasa.gov}
\acro{NCAR}{National Center for Atmospheric Research}
%{ -- \it http://www.ncar.ucar.edu/ncar/index.html}
\acro{NCEI}{National Centers for Environmental Information}
\acro{NCEP}{National Centers for Environmental Prediction}
\acro{NCDC}{National Climatic Data Center}
\acro{NCDDC}{National Coastal Data Development Center}
\acro{NCL}{NCAR Command Language}
\acro{ncBrowse}{}
%{ -- \it http://www.epic.noaa.gov/java/ncBrowse}
\acro{NcML}{netCDF Markup Language}
\acro{NCO}{netCDF Operator}
\acro{NCODA}{Navy Coupled Ocean Data Assimilation}
\acro{NCSA}{National Center for Supercomputing Applications}
\acro{NDBC}{National Data Buoy Center}
\acro{NDVI}{Normalized Difference Vegetation Index}
\acro{NEC}{North Equatorial Current}
\acro{NECC}{North Equatorial Countercurrent}
\acro{NEFSC}{Northeast Fisheries Science Center}
\acro{NEIC}{North Equatorial Intermediate Current}
\acro{netCDF}{NETwork Common Data Format}
\acro{NEUC}{North Equatorial Undercurrent}
\acro{NGDC}{National Geophysical Data Center}
%{ -- \it http://www.ngdc.noaa.gov}
\acro{NICC}{North Intermediate Countercurrent}
\acro{NIST}{National Institute of Standards and Technology}
\acro{NLSST}{Non-Linear Sea Surface Temperature}
\acro{NMFS}{National Marine Fisheries Service}
\acro{NMS}{New Media Studio}
%{ -- \it http://newmediastudio.org/Homepage/TNMSHomeFramset.htm}
\acro{NN}{Neural Network}
\acro{NOAA}{National Oceanic and Atmospheric Administration}
%{ -- \it http://www.noaa.gov}
\acro{NODC}{National Oceanographic Data Center}
\acro{NOGAPS}{Navy Operational Global Atmospheric Prediction System}
\acro{NOMADS}{NOAA Operational Model Archive Distribution System}
%{ -- \it http://www.ncdc.noaa.gov/oa/climate/nomads/nomads.html}
\acro{NOPP}{National Oceanographic Parternership Program}
%{ -- \it http://www.nopp.org}
\acro{NOS}{National Ocean Service}
\acro{NPP}{National Polar-orbiting Partnership}
\acro{NPOESS}{National Polar-orbiting Operational Environmental Satellite System}
\acro{NSCAT}{NASA SCATterometer}
%{ -- \it http://winds.jpl.nasa.gov}
%\acro{NSEN}{\acs{NASA}Science and Engineering Network}
\acro{NSEN}{NASA Science and Engineering Network}
\acro{NSF}{National Science Foundation}
%\acro{NSIPP}{\acs{NASA}Seasonal-to-Interannual Prediction Project}
\acro{NSIPP}{NASA Seasonal-to-Interannual Prediction Project}
\acro{NRA}{NASA Research Announcement}
\acro{NRC}{National Research Council}
\acro{NRL}{Naval Research Laboratory}
\acro{NSCC}{North Subsurface Countercurrent}
\acro{NSF}{National Science Foundation}
\acro{NSIDC}{National Snow and Ice Data Center}
\acro{NSPIRES}{NASA Solicitation and Proposal Integrated Review and Evaluation System}
\acro{NSSDC}{National Space Science Data Center}
\acro{NVODS}{National Virtual Ocean Data System}
%{ -- \it http://nvods.org}
\acro{NWP}{Numerical Weather Prediction}
\acro{NWS}{National Weather Service}

\acro{OBPG}{Ocean Biology Processing Group}
\acro{OB.DAAC}{Ocean Biology DAAC}
\acro{ODC}{OPeNDAP Data Connector}
\acro{OC}{ocean color}
\acro{OCAPI}{OPeNDAP C API}
\acro{ODSIP}{Open Data Services Invocation Protocol}
\acro{OFES}{Ocean Model for the Earth Simulator}
\acro{OCCA}{OCean Comprehensive Atlas}
\acro{OGC}{Open Geospatial Consortium}
%{ -- \it http://www.opengis.org}
\acro{OGCM}{Ocean General Circulation Model}
\acro{OISSTv1}{Optimally Interpolated SST Version 1}
\acro{ONR}{Office of Naval Research}
\acro{OLCI}{Ocean Land Colour Instrument}
\acro{OLFS}{OPeNDAP Lightweight Front-end Server}
\acro{OOD}{out-of-distribution}
\acro{OOPC}{Ocean Observation Panel for Climate}
\acro{OPeNDAP}{Open source Project for a Network Data Access Protocol}
%{ -- \it http://opendap.org}
\acro{OPeNDAPg}{GRID-enabled OPeNDAP tools}
\acro{OpenGIS}{OpenGIS}
\acro{OSCAR}{Ocean Surface Current Analyses Real-time}
\acro{OSI SAF}{Ocean and Sea Ice Satellite Application Facility}
\acro{OSS}{Office of Space Science}
\acro{OSTM}{Ocean Surface Topography Mission }
\acro{OSU}{Oregon State University}
\acro{OS X}{}
\acro{OWL}{Web Ontology Language}
\acro{OWASP}{Open Web Application Security Project}

\acro{PAE}{Probabilistic AutoEncoder}
\acro{PBL}{planetary boundary layer}
\acro{PCA}{Principal Components Analysis}
\acro{PDistF}[PDF]{probability distribution function}
\acro{PDF}{probability density function}
\acro{PF}{Polar Front}
\acro{PFEL}{Pacific Fisheries Environmental Laboratory}
\acro{PI}{Principal Investigator}
\acro{PIV}{Particle Image Velocimetry}
\acro{PL}{Project Leader}
\acro{PM}{Project Member}
\acro{PMEL}{Pacific Marine Environmental Laboratory}
%{ -- \it http://pmel.noaa.gov}
\acro{POC}{particulate organic carbon}
\acro{PO.DAAC}{Physical Oceanography Distributed Active Archive Center}
%{ -- \it http://podaac.jpl.nasa.gov}
\acro{POP}{Parallel Ocean Program }
\acro{PSD}{Power Spectral Density}
\acro{PSPT}{Precision Solar Photometric Telescope}
\acro{PSU}{Pennsylvania State University}
\acro{PyDAP}{Python Data Access Protocol}
\acro{PV}{potential vorticity}

\acro{QC}{quality control}
\acro{QG}{quasi-geostrophic}
\acro{QuikSCAT}{Quick Scatterometer}
%{ -- \it http://winds.jpl.nasa.gov/missions/quikscat/quikindex.html}
\acro{QZJ}{quasi-zonal jet}

\acro{R2HA2}{Rescue \& Reprocessing of Historical AVHRR Archives }
\acro{RAFOS}{SOFAR, SOund Fixing And Ranging, spelled backward}
\acro{RAID}{Redundant Array of Independent Disks}
\acro{RAL}{Rutherford Appleton Laboratory}
\acro{RAN2}{$2^{nd}$ full-mission reanalysis}
\acro{RAN3}{$3^{rd}$ full-mission reanalysis}
\acro{RDF}{Resource Description Language}
\acro{REASoN}{Research, Education and Applications Solutions Network}
\acro{REAP}{Realtime Environment for Analytical Processing}
\acro{ReLU}{Rectified Linear Unit}
\acro{REU}{Research Experiences for Undergraduates}
\acro{RFA}{Research Focus Area}
\acro{RFI}{Radio Frequency Interference}
\acro{RFC}{Request For Comments}
\acro{R/GTS}{Regional/Global Task Sharing}
\acro{RSI}{Research Systems Inc.}
\acro{RISE}{Radiative Inputs from Sun to Earth}
\acro{rms}{root mean square}
\acro{RMI}{Remote Method Invocation}
\acro{ROMS}{Regional Ocean Modeling System}
\acro{ROSES}{Research Opportunities in Space and Earth Sciences}
\acro{RSMAS}{Rosenstiel School of Marine and Atmospheric Science}
\acro{RSS}{Remote Sensing Sytems}
\acro{RTM}{radiative transfer model}

%\acro{SACCF}{Southern \acs{ACC} Front}
\acro{SACCF}{Southern ACC Front}
\acro{SAC-D}[(SAC)-D]{Sat\'elite de Aplicaciones Cient\'ificas-D}
%\acro{SAF}{Satellite Application Facility}
\acro{SAF}{Subantarctic Front}
\acro{SAIC}{Science Applications International Corporation}
\acro{SANS}{SysAdmin, Audit, Networking, and Security}
\acro{SAR}{synthetic aperature radar}
\acro{SATMOS}{Service d'Archivage et de Traitement M\'et\'eorologique des Observations Spatiales}
\acro{SBE}{Sea-Bird Electronics}
\acro{SciDAC}{Scientific Discovery through Advanced Computing}
\acro{SCC}{Subsurface Countercurrent}
\acro{SCCWRP}{Southern California Coastal Water Research Project}
\acro{SDAC}{Solar Physics Data Analysis Center}
%\acro{SeaWinds}{NASA Scatterometer}
\acro{SDS}{Scientific Data Set}
\acro{SDSC}{San Diego Supercomputer Center}
\acro{SeaDAS}{SeaWiFS Data Analysis System}
\acro{SeaWiFS}{Sea-viewing Wide Field-of-view Sensor}
%\acro{SEC}{Sun Earth Connection}
\acro{SEC}{South Equatorial Current}
\acro{SECC}{South Equatorial Countercurrent}
\acro{SECDDS}{Sun Earth Connection Distributed Data Services}
\acro{SEEDS}{Strategic Evolution of ESE Data Systems}
%{ -- \it http://lennier.gsfc.nasa.gov/seeds}
\acro{SEIC}{South Equatorial Intermediate Current}
\acro{SEUC}{Southern Equatorial Undercurrent}
\acro{SEVIRI}{Spinning Enhanced Visible and Infra-Red Imager}
\acro{SeRQL}{SeRQL}
\acro{SGI}{Silican Graphics Incorporated}
%{ -- \it http://www.sgi.com}
\acro{SHF}{Sensible Heat Flux}
\acro{SICC}{South Intermediate Countercurrent}
\acro{SIED}{single image edge detection}
\acro{SIPS}{Science Investigator--led Processing System}
\acro{SIR}{Scatterometer Image Reconstruction}
\acro{SISTeR}{Scanning Infrared Sea Surface Temperature Radiometer}
\acro{SLA}{sea level anomaly}
\acro{SLSTR}{Sea and Land Surface Temperature Radiometer}
\acro{SMAP}{Soil Moisture Active Passive}
\acro{SMMR}{Scanning Multichannel Microwave Radiometer}
\acro{SMOS}{Soil Moisture and Ocean Salinity}
\acro{SMTP}{Simple Mail Transfer Protocol}
\acro{SOAP}{Simple Object Access Protocol}
\acro{SOEST}{School of Ocean and Earth Science and Technology}
\acro{SOFAR}{SOund Fixing And Ranging}
\acro{SOFINE}{Southern Ocean Finescale Mixing Experiment}
\acro{SOHO}{Solar and Heliospheric Observatory}
\acro{SPARC}{Space Physics and Aeronomy Research Collaboratory}
%{ -- \it http://intel.si.umich.edu/SPARC/}
\acro{SPARQL}{Simple Protocol and RDF Query Language}
\acro{SPASE}{Space Physics Archive Search Engine}
\acro{SPCZ}{South Pacific Convergence Zone}  
\acro{SPDF}{Space Physics Data Facility}
\acro{SPDML}{Space Physics Data Markup Language}
%{ -- \it http://sd-www.jhuapl.edu/SPDML/}
\acro{SPG}{Standards Process Group}
%{ -- \it http://www.esdswg.org/spg/}
\acro{SQL}{Structured Query Language}
\acro{SSCC}{South Subsurface Countercurrent}
%\acro{SSL}{Secure Sockets Layer}
\acro{SSL}{self-supervised learning}
\acro{SSO}{Single sign-on}
\acro{SSES}{Single Sensor Error Statistics}
\acro{SSH}{sea surface height}
%\acro{SSHA}{\ac{SSH} anomaly}
\acro{SSHA}{sea surface height anomaly}
\acro{SSMI}{Special Sensor Microwave/Imager}
\acro{SST}{Sea Surface Temperature}
\acro{SSTa}{sea surface temperature anomaly}
\acro{SSS}{sea surface salinity}
\acro{SSTST}{Sea Surface Temperature Science Team}
\acro{STEM}{science, technology, engineering and mathematics}
\acro{STL}{Standard Template Library}
\acro{STCZ}{Subtropical Convergence Zone}
\acro{Suomi-NPP}[Suomi-NPP]{Suomi-\acl{NPP}}
\acro{SWEET}{Semantic Web for Earth and Environmental Terminology}
\acro{SWFSC}{Southwest Fisheries Science Center}
\acro{SWOT}{Surface Water and Ocean Topography}
\acro{SWRL}{Semantic Web Rule Language}
\acro{SubEx}{Submesoscale Experiment}
\acro{SURA}{Southeastern Universities Research Association}
\acro{SURFO}{Summer Undergraduate Research Fellowship Program in Oceanography}
\acro{SuperDARN}{Super Dual Auroral Radar Network}

\acro{TAMU}{Texas A\&M University}
\acro{TB}{TeraByte - $10^{12}$ bytes}
\acro{TCASCV}{Technology Center for Advanced Scientific Computing and Visualization}
%{ -- \newline \it http://www.cascv.brown.edu/aboutus.html}
\acro{TCP}{Transmission Control Protocol}
\acro{TCP/IP}{Transmission Control Protocol/Internet Protocol}
\acro{TDS}{THREDDS Data Server}
\acro{TEX}{external temperature or T-External}
\acro{THREDDS}{Thematic Realtime Environmental Data Distributed Services}
%{ -- \newline \it http://my.unidata.ucar.edu/content/projects/THREDDS/index.html}
\acro{TIDI}{TIMED Doppler Interferometer}
\acro{TIFF}{Tag Image File Format}
\acro{TIMED}{Thermosphere, Ionosphere, Mesosphere, Energetics and Dynamics}
\acro{TLS}{Transport Layer Security}
\acro{TRL}{Technology Readiness Level}
\acro{TMI}{TRMM Microwave Imager}
\acro{TOPEX/Poseidon}{TOPography EXperiment for Ocean Circulation/Poseidon}
\acro{TRMM}{Tropical Rainfall Measuring Mission}
%{ -- \it http://trmm.gsfc.nasa.gov/}
\acro{TSG}{thermosalinograph}

\acro{UCAR}{University Corporation for Atmospheric Research}
%{ -- \it http://www.ucar.edu}
\acro{UCSB}{University of California, Santa Barbara}
\acro{UCSD}{University of California, San Diego}
\acro{uCTD}{Underway Conductivity, Temperature and Salinity or Underway \acs{CTD}}
\acro{UDDI}{Universal Description, Discovery and Integration}
\acro{UMAP}{Uniform Manifold Approximation and Projection}
\acro{UMiami}{University of Miami}
\acro{Unidata}{}
%{ -- \it http://unidata.ucar.edu}
\acro{URI}{University of Rhode Island}
%{ -- \it http://www.uri.edu}
\acro{UPC}{Unidata Program Committee}
\acro{URL}{Uniform Resource Locator}
\acro{USGS}{United States Geological Survey}
\acro{UTC}{Coordinated Universal Time}
\acro{UW}{University of Washington}

\acro{VCDAT}{Visual Climate Data Analysis Tools}
\acro{VIIRS}{Visible Infrared Imaging Radiometer Suite}
\acro{VR}{Virtual Reality}
\acro{VSTO}{Virtual Solar-Terrestrial Observatory}

\acro{WCR}{Warm Core Ring}
\acro{WCS}{Web Coverage Service}
%{ -- \it http://www.ogcnetwork.org}
\acro{WCRP}{World Climate Research Program}
\acro{WFS}{Web Feature Service}
%{ -- \it http://www.ogcnetwork.org}
\acro{WMS}{Web Map Service}
%{ -- \it http://www.ogcnetwork.org}
\acro{W3C}{World Wide Web Consortium}
\acro{WJ}{Wyrtki Jets}
\acro{WHOI}{Woods Hole Oceanographic Institution}
\acro{WKB}{Well Known Binaries}
\acro{WIMP}{Windows, Icons, Menus, and Pointers}
\acro{WIS}{World Meteorological Organisation Information System}
\acro{WOA05}{World Ocean Atlas 2005}
\acro{WOCE}{World Ocean Circulation Experiment}
\acro{WP-ESIP}{Working Prototype Earth Science Information Partner}
\acro{WRF}{Weather \& Research Forecasting Model}
\acro{WSDL}{Web Services Description Language}
\acro{WSP}{western South Pacfic}
\acro{WWW}{World Wide Web}

\acro{XBT}{Expendable BathyThermograph}
\acro{XML}{Extensible Markup Language}
%{ -- \it http://www.w3.org/XML}
\acro{XRAC}{eXtreme Digital Request Allocation Committee}
\acro{XSEDE}{Extreme Science and Engineering Discovery Environment}

\acro{YAG}{yttrium aluminium garnet}

\end{acronym}
}

%\clearpage

%% REFERENCES

%% The reference list is compiled as follows:

%\begin{thebibliography}{}
%
%\bibitem[AUTHOR(YEAR)]{LABEL1}
%REFERENCE 1
%
%\bibitem[AUTHOR(YEAR)]{LABEL2}
%REFERENCE 2
%
%\end{thebibliography}

%% Since the Copernicus LaTeX package includes the BibTeX style file copernicus.bst,
%% authors experienced with BibTeX only have to include the following two lines:
%%
\bibliographystyle{copernicus}
\bibliography{bibli.bib}

\end{document}